\documentclass[fleqn,usenatbib]{mnras}

\usepackage{newtxtext,newtxmath}

\usepackage[T1]{fontenc}

\DeclareRobustCommand{\VAN}[3]{#2}
\let\VANthebibliography\thebibliography
\def\thebibliography{\DeclareRobustCommand{\VAN}[3]{##3}\VANthebibliography}

\usepackage{graphicx}	
\usepackage{amsmath}	
\usepackage[normalem]{ulem}
\usepackage{float}
\usepackage{xcolor}

\newcommand{\ha}{H$\mathrm{\alpha}$}   
\newcommand{\hb}{H$\mathrm{\beta}$}  
\newcommand{\oiii}{[\mbox{O\,\textsc{iii}}]}
\newcommand{\nii}{[\mbox{N\,\textsc{ii}}]}
\newcommand{\sii}{[\mbox{S\,\textsc{ii}}]}
\newcommand{\oii}{[\mbox{O\,\textsc{ii}}]}
\newcommand{\oi}{[\mbox{O\,\textsc{i}}]}
\newcommand{\siii}{[\mbox{S\,\textsc{iii}}]}
\newcommand{\hii}{\mbox{H\,\textsc{ii}}}
\newcommand{\ncl}{$N_{\mathrm{cluster}}$}
\newcommand{\mcl}{$M_{\mathrm{cl}}$}
\newcommand{\msol}{M$_{\odot}$}
\newcommand{\hdenini}{$n_{\mathrm{H,0}}$}

\title[Emission-line diagnostics of \hii\ regions using cINNs]{Emission-line diagnostics of \hii\ regions using conditional Invertible Neural Networks}

\author[D. E. Kang et al.]{%
Da Eun Kang,$^{1}$\thanks{E-mail: kang@uni-heidelberg.de (DK)} 
Eric~W.~Pellegrini,$^{1}$ 
Lynton~Ardizzone,$^{2}$
Ralf~S.~Klessen,$^{1,3}$ \newauthor 
Ullrich Koethe,$^{2}$ 
Simon~C.~O.~Glover$^{1}$ 
and Victor~F.~Ksoll$^{1}$ 
\\
$^{1}$Universit\"{a}t Heidelberg, Zentrum f\"{u}r Astronomie, Institut f\"{u}r Theoretische Astrophysik, Albert-Ueberle-Stra{\ss}e 2, D-69120 Heidelberg, Germany\\
$^{2}$Universit\"{a}t Heidelberg, IWR, Computer Vision and Learning Lab, Berliner Stra{\ss}e 43, D-69121 Heidelberg, Germany\\
$^{3}$Universit\"{a}t Heidelberg, Interdisziplin\"{a}res Zentrum f\"{u}r Wissenschaftliches Rechnen, Im Neuenheimer Feld 205, D-69120 Heidelberg, Germany\\ 
}

\date{Accepted XXX. Received YYY; in original form ZZZ}

\pubyear{2022}

\begin{document}
\label{firstpage}
\pagerange{\pageref{firstpage}--\pageref{lastpage}}
\maketitle

\begin{abstract}
Young massive stars play an important role in the evolution of the interstellar medium (ISM) and the self-regulation of star formation in giant molecular clouds (GMCs) by injecting energy, momentum, and radiation (stellar feedback) into surrounding environments, disrupting the parental clouds, and regulating further star formation. Information of the stellar feedback inheres in the emission we observe, however inferring the physical properties from photometric and spectroscopic measurements is difficult, because stellar feedback is a highly complex and non-linear process, so that the observational data are highly degenerate.
On this account, we introduce a novel method that couples a conditional invertible neural network (cINN) with the WARPFIELD-emission predictor (WARPFIELD-EMP) to estimate the physical properties of star-forming regions from spectral observations. We present a cINN that predicts the posterior distribution of seven physical parameters (cloud mass, star formation efficiency, cloud density, cloud age which means age of the first generation stars, age of the youngest cluster, the number of clusters, and the evolutionary phase of the cloud) from the luminosity of 12 optical emission lines, and test our network with synthetic models that are not used during training. Our network is a powerful and time-efficient tool that can accurately predict each parameter, although degeneracy sometimes remains in the posterior estimates of the number of clusters. We validate the posteriors estimated by the network and confirm that they are consistent with the input observations. We also evaluate the influence of observational uncertainties on the network performance.

\end{abstract}




\section{Introduction}
\label{sec:introduction}

Stellar feedback, the interaction between young massive stars and their birthplace, plays a crucial role in the evolution of the interstellar medium (ISM) in galaxies. Once massive stars are formed in giant molecular clouds (GMCs), they inject a large amount of energy and momentum into the surrounding environment, which leads to the destruction of their birthplace and suppression of further star formation in the molecular cloud~\citep[for a review about stellar feedback, see][]{Krumholz+14, Klessen&Glover16}.

There are diverse stellar feedback mechanisms that can act against gravity such as supernovae, stellar winds, radiation pressure, or photoionization. Different feedback modes are coupled non-linearly and complicate feedback-influenced regions like \hii\ regions and photodissociation regions (PDRs) morphologically and dynamically. Recent studies theoretically describe stellar feedback models including several different feedback mechanisms~\citep[e.g.,][]{Dale+14, Rahner+17, Rahner+19, Kim+18, Ali&Harries19,Geen+20, Grudic+22} and apply their models to characterize observed star-forming regions~\citep[e.g.,][]{Pellegrini+11, Rahner+18, Rugel+19}. 
However, finding a best-fitting model for observations using classical forward modelling methods is difficult and limited in parameter space due to the high dimensionality, nonlinearity, and degeneracy of the complex stellar feedback process.

Artificial neural networks~\citep[NNs;][]{Goodfellow+16} link physical parameters to observational measurements through a statistical model rather than depending on a preselected physical model. NNs have been applied in various astronomical studies, e.g., to predict physical parameters~\citep[e.g.,][]{Fabbro+18, Ksoll+20, Olney+20}, for classification~\citep[e.g.,][]{Wu+19, Sharma+20, Wei+20}, and to identify structures in maps and images~\citep{Abraham+18}. In this study, we apply NNs to estimate the physical properties of stellar feedback and star-forming regions from observations. 

We adopt a supervised learning approach that trains the NNs based on a large database, which contains both the target physical parameters and corresponding observables, and is built from synthetic models or well-determined observations.
The forward process that translates parameters into observations is well-defined but involves a loss of information, such that different parameter sets are mapped on to identical observations, rendering the inverse process ambiguous and degenerate. For inverse inference, we need a full posterior distribution conditioned on the observed measurements to fully characterize the ambiguity.

The invertible Neural Network~\citep[INN;][]{Ardizzone+19a} is an architecture introduced to solve ambiguous inverse problems. Unlike classical neural networks which solve the inverse problem directly, INNs learn the forward process, using additional latent output variables to capture the information otherwise lost. Leveraging their invertible architecture, INNs then derive a solution for the inverse process for free. Conditioned on the observations and the latent variable distribution, INNs can predict full posterior distributions, which is advantageous to study multi-modality or correlations between parameters. In this paper, we use conditional invertible neural networks~\citep[cINNs;][]{Ardizzone+19b, Ardizzone+21}, an extended class of INN, which has been applied in \cite{Ksoll+20} to predict stellar parameters from photometric observations. cINN has also been applied in several studies outside of the astronomical field such as particle physics or medical science~\citep[e.g.,][]{Bellagente+20, Trofimova+20}. 

We present an application of a cINN to estimate the physical properties of star clusters and star-forming clouds from spectral observations of \hii\ regions. For simplicity, we regularly use the term \hii\ region as a synonym for the entire star-forming complex. This is justified, because we focus in this study on optical emission lines that mostly trace the ionized gas in the feedback-generated hot bubble and the surrounding dense expanding shell. As mentioned above, supervised learning requires a large amount of well-interpreted data to train the network. In this paper, we build a training database by using the WARPFIELD emission predictor~\citep[WARPFIELD-EMP;][]{Pellegrini+20} which allows us to collect both cloud properties and corresponding observable quantities (i.e., line luminosity). WARPFIELD-EMP describes the evolution of a cluster, expanding bubble, and the surrounding cloud using the 1D stellar feedback code WARPFIELD~\citep{Rahner+17, Rahner+19} and calculates detailed emission predictions based on the output from WARPFIELD with the help of CLOUDY, a photoionization code~\citep[see][]{Ferland+17}, and the radiative transfer code POLARIS~\citep[see][]{Reissl+16}. WARPFIELD takes into account several feedback mechanisms (i.e., stellar winds, radiation pressure, thermal gas pressure, supernovae, and gravity) self-consistently. Although 1D models cannot describe detailed structures as much as 3D simulations, they typically do provide very good approximations to the \hii\ region properties of interest here, and due to the reduced dimensionality, they are well suited for building the large numbers of models required for training the cINN.

In this study, we focus on examining how accurately the network understands the hidden rules in the given training data and how accurately it can estimate the parameters of test synthetic models, which share the same physical system as the training data. In this study, we do not yet apply our network to real observations, but in the latter part of the paper, we discuss how observational uncertainties affect the network performance using a statistical approach and show the change of individual posterior distributions with increasing observational errors.

The paper is structured as follows. First, in Section~\ref{sec:training data}, we introduce a new WARPFIELD-EMP database used for the network training. We also review the physics of our synthetic models and present details of the new database. In Section~\ref{sec:nn}, we introduce the structure of the cINN, our network setup, and the performance evaluation methods for the trained network. We present the prediction performance of our network in Section~\ref{sec:training results} and validate the predicted posteriors in Section~\ref{sec:validation}. Section~\ref{sec:degenerate} discusses the degeneracy in the posterior distributions. We explore the influence of observational uncertainties on the network performance in Section~\ref{sec:luminosity error}. In Section~\ref{sec:discussion}, we discuss the physical assumptions inherent in the training data, and training methods to improve the network prediction. Finally, we summarize our main results in Section~\ref{sec:summary}.

\section{Training data}
\label{sec:training data}

We need a large data set containing both the physical parameters we want to predict and the observable quantities of \hii\ regions that they give rise to in order to train the neural network. However, it is difficult to collect such a large amount of well-analysed \hii\ regions from observations. Instead, we use a database of numerous synthetic \hii\ region models created through a pipeline known as the WARPFIELD emission predictor~\citep[WARPFIELD-EMP;][]{Pellegrini+20} to train the network. WARPFIELD-EMP follows the evolution of massive star-forming clouds using a 1D stellar feedback model (WARPFIELD; \citealt{Rahner+17, Rahner+19}) and predicts the time-dependent continuum and line emission radiated from the evolving clouds by using CLOUDY and POLARIS. \cite{Pellegrini+20} presented the first WARPFIELD-EMP database composed of synthetic \hii\ region models evolved from 180 initial clouds with different combinations of cloud mass (\mcl), star formation efficiency (SFE), and density (\hdenini). This first database is sufficient to reproduce the BPT diagram of observed \hii\ regions in NGC628, but the sampling of initial parameters (i.e., \mcl\, SFE, and \hdenini) is not dense enough to be used for our network training. Therefore, in this paper, we present a new, extended database of synthetic models evolved from 10,000 initial WARPFIELD clouds, which is suitable for our network training.
We will first explain about the WARPFIELD-EMP pipeline and the inherent physical mechanisms of the synthetic model, and introduce the new database.

\subsection{WARPFIELD and WARPFIELD-EMP}
\label{subsec:warpfield}
WARPFIELD-EMP~\citep{Pellegrini+20} is a pipeline that predicts continuum and line emissions of evolving star-forming clouds by coupling these three codes: WARPFIELD, CLOUDY, and POLARIS. First, WARPFIELD~\citep{Rahner+17, Rahner+19} is a 1D spherical symmetric stellar feedback model which explains the evolution of the isolated massive star-forming cloud. The initial star-forming cloud, which we call WARPFIELD cloud in this paper, is determined by four initial parameters: \mcl, star formation efficiency, \hdenini, and metallicity ($Z$).

The evolution of a WARPFIELD cloud begins (at $t=0$) with the formation of the first star cluster at the centre with a mass of $M_{*} = {\rm SFE} M_{\text{cl}}$, where SFE  is the star formation efficiency. The cluster ionizes a large area called the Strömgren sphere and the stellar winds freely expand outwards from the centre. This initial expansion phase is very short and essentially covered in the first time step of the WARPFIELD evolution.
Quickly, the cloud can be separated into distinct regions around the cluster: an inner wind zone (i.e., diffuse central bubble) and the surrounding dense shell which consists of swept-up material affected by stellar feedback. WARPFIELD describes the evolution of this system by solving the equation of motion of the dense shell, considering the effects of the stellar wind, radiation pressure, thermal gas pressure, supernovae, and gravity. We assume that the ionized and neutral/molecular phases of the shell are in quasi-hydrostatic equilibrium and that the evolution of the cloud can be characterized by four distinct evolutionary phases depending on the dynamics of the shell.

In the earliest stage, Phase 1, the shocked wind material reaches a very hot temperature, resulting in a fast adiabatic expansion. The influence of gravity and radiation pressure on the shell is negligible in this phase. It ends when the bubble loses the hot gas either because the gas cools by radiative cooling ($t_{\text{cool}}$) or because the bubble bursts and the hot gas escapes. In the latter case, we assume that the bubble bursts only when the shell sweeps up the whole material in the cloud ($t_{\text{sweep}}$, i.e., when $R_{\text{shell}} = R_{\text{initial cloud}}$), because we do not consider the three-dimensional structures such as inhomogeneity or asymmetries. Therefore, Phase 1 ends after $t$ = min($t_{\text{cool}}$, $t_{\text{sweep}}$).

If the hot gas has cooled before the shell has swept away all of the cloud material ($t_{\text{cool}}$ < $t_{\text{sweep}}$), the cloud evolution enters Phase 2. After Phase 1 or Phase 2, when the shell eventually sweeps up all of the cloud material ($t$ > $t_{\text{sweep}}$), the shell enters Phase 3, expanding into a low density warm neutral medium outside the cloud. The shell expansion in Phase 2 or Phase 3 is dominated by the ram pressure exerted on the shell by stellar winds and supernovae, and by radiation pressure. The counteracting effect of gravity is now not negligible anymore. To take this into account, WARPFIELD includes at each time step both the gravity from the central cluster and the self-gravity of the shell in the calculation of shell dynamics.

The fate of the evolving cloud is divided into two cases depending on the effect of stellar feedback against gravity. When gravity becomes dominant, the shell stops expanding and begins to recollapse. Please note that, in this paper, we designate this recollapsing evolutionary phase as Phase 0 to conveniently feed it into the neural network. In \cite{Rahner+17}, the evolution terminated when the shell recollapsed to a radius of 1~pc, but in \cite{Pellegrini+20} and this study, we assume that recollapse triggers the birth of a new star cluster followed by a second expansion, and that the star formation efficiency of the new burst of star formation is the same as for the original burst (i.e., $M_{*, \text{second}} = {\rm SFE} (M_{\text{cl}} - M_{*, \text{first}})$).
If gravity never becomes dominant, the shell continues to expand into the low-density ambient ISM without a recollapse. The density of the shell gradually decreases and it eventually becomes indistinguishable from the ambient ISM. We terminate the calculation if the maximum density of the shell is smaller than 1 cm$^{-3}$ for a period of 1 Myr or more.

WARPFIELD accounts for the time-dependent stellar feedback effect exerted on the shell during the whole evolution period. We assume that stellar mass within the star cluster follows a Kroupa initial mass function~\citep{Kroupa01}. The evolution of the star cluster as a function of time is calculated with STARBURST99~\citep{Leitherer+99, Leitherer+14} by using Geneva stellar evolution tracks for rotating stars~\citep{Ekstrom+12, Georgy+12, Georgy+13}. The spectral energy distribution (SED) of the cluster and surrounding cloud
in the WARPFIELD model are time-dependent due to the evolving physical conditions and complicated because we may be looking at the combination of multiple clusters of different ages if recollapse occurred.

The information on the time-dependent physical conditions is then passed on to CLOUDY, a spectral synthesis code developed by \cite{Ferland+17}. WARPFIELD-EMP uses the most recent version of CLOUDY, C17. It calculates both continuum emission and a large set of line emissions as well as the corresponding opacities as a function of position within the shell and surrounding natal cloud. If the shell swept up the entire natal cloud, we run CLOUDY only once to determine the emission from the shell, whereas if the natal cloud remains, we need to run CLOUDY twice; first for the shell and second for the static natal cloud. The output of the first CLOUDY run for the shell is used as the incident flux on the static cloud in the second CLOUDY run.

In the last step, we use POLARIS~\citep{Reissl+16,Reissl+19} to solve the radiative transfer equation for rays passing through a 3D grid of the shell and the natal cloud. Based on the CLOUDY output, POLARIS calculates the absorption and emission from the dust so that we obtain the luminosity information about the continuum and the lines, taking into account the overall attenuation inside the shell and the cloud. As WARPFIELD is a 1D spherical symmetric model, we select a 3D spherical grid for POLARIS calculations. The output containing the three-dimensional attenuated luminosity information is then projected onto a 2D space. This process provides a 2D map of the velocity-integrated luminosity of any given line and the continuum, which is similar to spatially resolved observations. Finally, we spatially integrate the 2D map and obtain the 1D integrated luminosity of a series of emission lines that correspond to a given WARPFIELD model at a certain age.

WARPFIELD-EMP provides us with a vast amount of information on a \hii\ region from the fundamental physical parameters such as mass or age to the final observable quantities like emission-line luminosity. Although our synthetic model does not account for the 3D structures and small-scale instabilities, because WARPFIELD is a 1D spherical symmetric model, it has been confirmed that WARPFIELD well describes the observations of feedback-affected regions around the star clusters~\citep{Rahner+18, Rugel+19}.
Moreover, \cite{Pellegrini+20} demonstrates that a WARPFIELD-EMP model is close to mimic the real observations, finding a good agreement between the BPT diagram of WARPFIELD-EMP models and that of observed \hii\ regions in NGC628.

\subsection{Database}
\label{subsec:database}

\begin{figure*}
	\includegraphics[width=2\columnwidth]{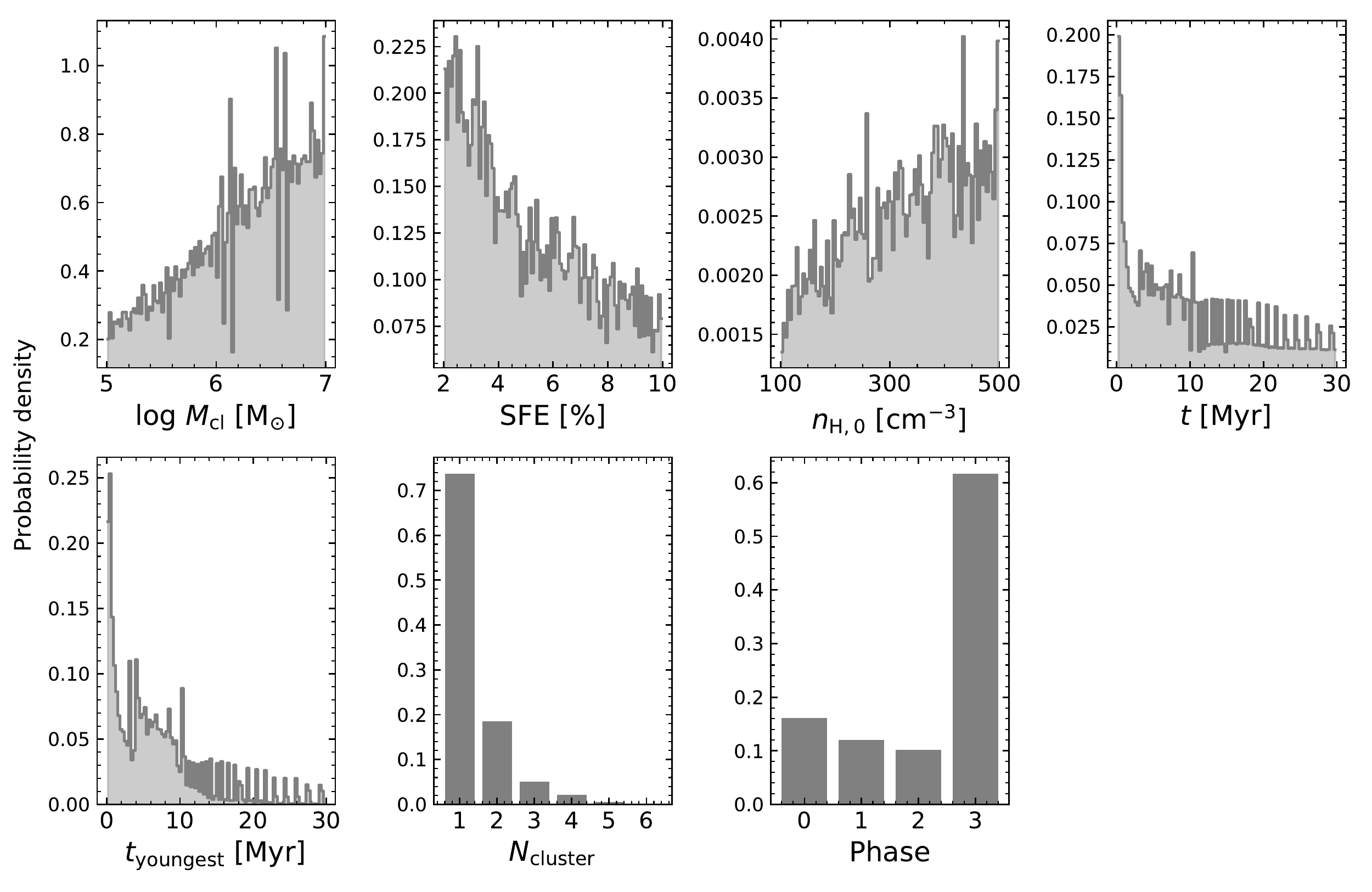}
    \caption{Distribution of seven physical parameters in the database used for the network training and evaluation. The database consists of 505,748 synthetic \hii\ region models simulated by WARPFIELD-EMP~\citep{Pellegrini+20}, which are evolved from 10,000 initial clouds. The first three parameters (cloud mass, star formation efficiency, and density) are initial conditions provided to WARPFIELD.  The fourth, fifth and sixth panels correspond to the cloud age (as measured from the onset of star formation),  the age of the youngest cluster in the cloud, and the number of clusters in the cloud, respectively. Phase is classified by the evolutionary state of the WARPFIELD cloud depending on its dynamics: Phase 1 is the energy-dominated phase, Phase 2 is momentum-dominated, Phase 3 is free-expansion, and Phase 0 denotes a recollapsing cloud (see the more detailed explanation in Section~\ref{subsec:warpfield} or in \citealt{Rahner+17, Pellegrini+20}).
    }  
    \label{fig:distr_param}
\end{figure*}

In this section, we introduce the new database of WARPFIELD-EMP synthetic \hii\ region models, which provides a considerable extension of the first database presented in \cite{Pellegrini+20}.
As mentioned in Section~\ref{subsec:warpfield}, the main initial parameters of WARPFIELD models are cloud mass \mcl, star formation efficiency SFE, initial cloud density \hdenini, and metallicity $Z$. However, we fix the metallicity at solar metallicity so that the initial WARPFIELD cloud in our database is determined only by a combination of the other three parameters. There are several parameters that control the physical condition of the cloud but we constrain all the other parameters to be fixed in the database. For example, we can add the effect of magnetic field or turbulent pressure in WARPFIELD, but in this database we turn off both of them.
As the cloud begins its evolution, age $t$ becomes the fourth independent parameter. So, each synthetic model, describing a certain evolution state of the cloud, is determined by the four parameters: \mcl, SFE, \hdenini, and $t$.

To create an evenly and densely populated database, we follow a total of 10,000 initial WARPFIELD clouds and randomly sample \mcl, star formation efficiency, and \hdenini\ within the range of 10$^{5-7}$\msol, 2-10\%, and 100-500 cm$^{-3}$, respectively. The ranges of the three parameters are similar or slightly narrower than those of the previous database. Then we evolve these uniformly distributed 10,000 initial clouds until an age of 30~Myr. 
Depending on the evolutionary condition, WARPFIELD may terminate the calculation earlier if the cloud already dissolved into the ambient ISM. We constrain the maximum age of cloud to be 30~Myr to prevent infinite expansion, and also because we expect some of the base assumptions of the model (that the cloud is isolated, or that it is unaffected by external feedback and large-scale galactic dynamics) to break down for very old clouds. 

WARPFIELD saves the evolution of the cloud in prescribed time intervals. For the new database, we adopt an interval of 0.1~Myr. However, we do not run CLOUDY and POLARIS for all of the saved models because of the limitations of data volume and computational time. Instead, as mentioned in \cite{Pellegrini+20}, we run CLOUDY and POLARIS whenever the physical properties of the cloud or star cluster have changed sufficiently to result in a considerable change in the emission. For example, we calculate the model whenever the shell density at the boundary of the inner shell, shell radius, shell mass, or the ionizing photon flux changes by 10\%. We also calculate the model if the evolutionary phase of WARPFIELD changes. Therefore, the cloud age and the youngest cluster age are not sampled in constant intervals. Especially, the time interval of the old cloud is sometimes wider because the physical properties mentioned above do not change as rapidly as during the early phases of evolution. This wide time interval is exhibited as periodic patterns at $t > 10$~Myr in Figure~\ref{fig:distr_param}.
For the new database, we sample the time adaptively but, if needed, we can sample intermediate times by interpolating the evolution track of each WARPFIELD cloud.
Our final sample consists of 505,748 synthetic \hii\ region models in total with different values of \mcl, SFE, \hdenini, and $t$.

\renewcommand{\arraystretch}{1.25}
\begin{table}
    \centering
    \caption{List of seven physical parameters of \hii\ region that our network predicts from the observation. \label{table:parameters}}
    \begin{tabular}{ l  r} 
        \hline
        \hline
        Parameter & Symbol \\
        \hline
        initial mass of star-forming cloud &  \mcl\ [M$_{\odot}$] \\
        initial star formation efficiency & SFE [\%] \\
        initial cloud density &  \hdenini\ [cm$^{-3}$] \\
        age of the cloud & $t$ [Myr]	\\
        age of the youngest cluster & $t_{\mathrm{youngest}}$ [Myr] \\
        number of star clusters & \ncl\ \\
        evolutionary phase of the cloud & Phase \\
        \hline
        \hline
    \end{tabular} 
\end{table}

Although the initial WARPFIELD clouds were uniformly distributed in our parameter space, the final database has a non-uniform parameter distribution because each cloud evolves differently. Figure~\ref{fig:distr_param} shows the non-uniform distributions of the seven physical parameters listed in Table~\ref{table:parameters} that we aim to predict with the cINN.
As seen in Figure~\ref{fig:distr_param}, our database includes more \hii\ region models corresponding to massive clouds or to clouds with smaller star formation efficiencies. This reflects the fact that in these models the power of stellar feedback against gravity is often not large enough to immediately disrupt the cloud, leading to one or more episodes of recollapse and hence a greater number of total output snapshots. On the other hand, more than 70\% of our \hii\ region models have only one stellar cluster, since in the majority of cases, the initial burst of feedback is enough to destroy the cloud and recollapse does not occur.
In the case of the phase distribution, about 60\% of the outputs correspond to clouds in Phase 3. The reason for this is that clouds typically remain in Phase 3 for a long period of time until the \hii\ region dissolves into the ambient ISM, whereas clouds in the other evolutionary phases usually evolve rapidly and change into different phases.

In terms of training the network, we need to construct the training data to be as evenly distributed as possible, because over- or less-populated regions may introduce a bias to the trained network. In this case, the network might provide poor predictions for clouds with less popular characteristics in Figure~\ref{fig:distr_param}. To remedy this problem, one could potentially post-process the database and augment it where needed to even out the parameter distributions. However, in this study, there are some difficulties in applying such an approach. First, the evolution of each cloud is a complex result as a function of four independent parameters. So, if we oversample less populated characteristics to achieve an even distribution for one parameter, this can lead in turn to a more biased distribution in the other parameters. For the same reason, it is also difficult to plan additional sampling measures, because it is not easy to predict the evolution of the cloud from a given initial condition. Moreover, WARPFIELD-EMP is not able to interpolate between different \mcl, star formation efficiency, or \hdenini, so it is not easy to fill less-populated regions by interpolation from the given database.
Although the parameter distributions in the current database are not entirely optimal for training, we decide to train the network without any further processing. We discuss the influence of the bias in the training data on the network performance in Section~\ref{subsec:overall performance} and \ref{sec:validation}.

We divide the database into two parts: a training set and a test set. 80\% of the database is used to train the network (training set) and the remaining 20\% is a set of held-out models (test set), which is used to evaluate the network training and the performance of the trained network. Training and test set have the same distribution because they are randomly selected. To sum up, our training set includes 404,599 \hii\ region models. Each synthetic model has information about 7 physical parameters (i.e., \mcl, SFE, \hdenini, age, age of the youngest cluster, \ncl, and phase) as well as the luminosity of 10 optical emission lines within the wavelength range 3700\AA\ -- 9600\AA. These lines are [\mbox{O\,\textsc{ii}}] 3726\AA,
[\mbox{O\,\textsc{ii}}] 3729\AA,  \hb\ 4861\AA, [\mbox{O\,\textsc{iii}}] 5007\AA, [\mbox{O\,\textsc{i}}] 6300\AA, \ha\ 6563\AA, [\mbox{N\,\textsc{ii}}] 6583\AA, [\mbox{S\,\textsc{ii}}] 6716\AA,  [\mbox{S\,\textsc{ii}}] 6731\AA, and [\mbox{S\,\textsc{iii}}] 9531\AA). We also store the total strength of the [\mbox{O\,\textsc{ii}}] and [\mbox{S\,\textsc{ii}}] doublets, referred to hereafter as [\mbox{O\,\textsc{ii}}] 3727\AA\ (blend) and [\mbox{S\,\textsc{ii}}] 6720\AA\ (blend), since at low spectral resolution these lines will not always be separately resolved. Our choice of emission lines is motivated by their strength and the fact that they will be targeted in the forthcoming SDSS-V LVM survey of ionized gas in the Milky Way and other Local Group galaxies~\citep{Kollmeier+17}. Most of these lines are also included in the recent PHANGS-MUSE survey\footnote{The exception is the [\mbox{O\,\textsc{ii}}] doublet, which lies outside of the frequency range of ESO's MUSE integral field unit.}  of \hii\ regions in nearby spiral galaxies \citep{Emsellem+21} or the SIGNALS survey using the SITELLE spectrograph at the CFHT \citep{Rousseau-Nepton+19}. Please note that WARPFIELD-EMP provides information on many additional emission lines at frequencies ranging from the optical to the radio (see Table D1 and D2 in \citealt{Pellegrini+20}), but we restrict ourselves here to the optical emission lines most relevant for the above-mentioned surveys.

\renewcommand{\arraystretch}{1.25}
\begin{table}
    \centering
    \caption{List of 12 emission lines whose luminosities are used as a condition for our network to predict the physical parameters listed in Table~\ref{table:parameters}. \label{table:emission lines}}
    \begin{tabular}{ l  c} 
        \hline
        \hline
        Line & Wavelength \\
        \hline
        \oii\ &  3726\AA  \\
        \oii\ (blend) & 3727\AA \\
        \oii\ &  3729\AA  \\
        \hb\ & 4861\AA \\
        \oiii\ & 5007\AA \\
        \oi\ & 6300\AA \\
        \ha\ & 6563\AA \\
        \nii\ & 6583\AA \\
        \sii\ & 6716\AA \\
        \sii\ (blend) & 6720\AA \\
        \sii\ & 6731\AA \\
        \siii\ & 9531\AA \\
        \hline
        \hline
    \end{tabular} 
\end{table}

\section{Neural Network}
\label{sec:nn}

\subsection{cINN}
\label{subsec:cINN}

\begin{figure*}
	\includegraphics[width=2\columnwidth]{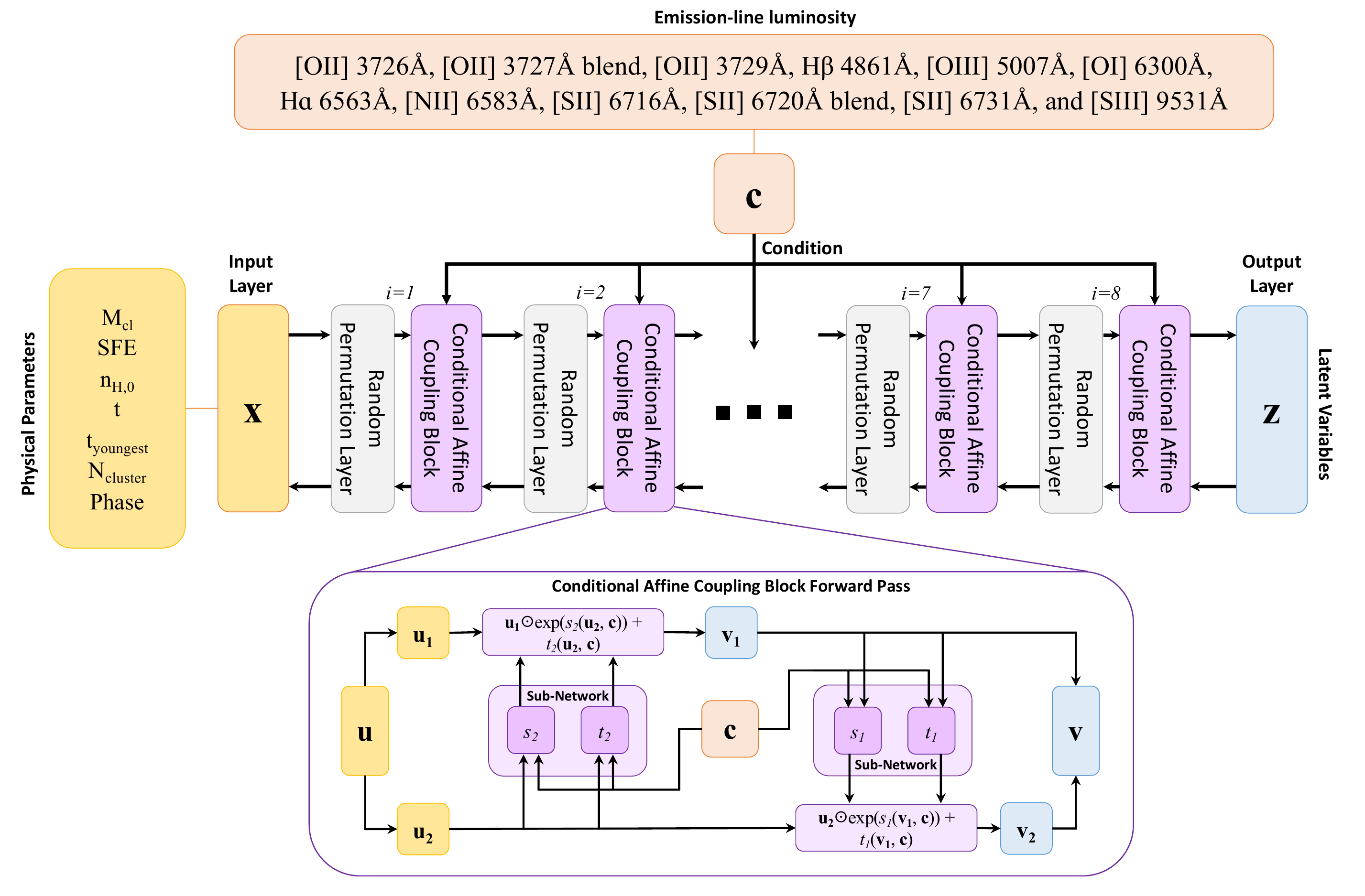}
    \caption{Schematic overview of our cINN architecture with the information of input parameters and observations for a condition. Our network consists of 8 affine coupling blocks interchanged with permutation layers. 
    The zoom-in panel of the conditional affine coupling block shows how the information is passed through the block in the forward direction. For each affine coupling block, we use the GLOW configuration where the internal transformations $s_{i}()$ and $t_{i}()$ are described by a single sub-network. } 
    \label{fig:cinn_structure}
\end{figure*}

In this study, we apply a conditional invertible neural network \cite[cINN;][]{Ardizzone+21} to predict physical parameters of \hii\ regions from their spectral observations. The cINN is an extension of the INN architecture described in \cite{Ardizzone+19a}. Both INN and cINN are able to solve inverse problems and provide posterior distribution of desired physical parameters on a given observation. \cite{Ksoll+20} used the same cINN architecture and demonstrated that the cINN is able to successfully estimate stellar parameters such as stellar age and mass from HST photometry observations for individual stars. The advantage of these invertible architectures is that the network can automatically learn the inverse process when it is trained to approximate a known forward process.

To apply the cINN or INN in a given inverse problem we assume that information loss occurs during the forward process such that different sets of physical parameters (\textbf{x}) are mapped onto identical observations (\textbf{y}). Consequently, the degenerate \textbf{y} cannot uniquely explain the corresponding \textbf{x}. By introducing the latent variables (\textbf{z}), which capture the information loss during the forward process, we can make a bijective mapping that could not be achieved with \textbf{x} and \textbf{y} alone. The original INN architecture links \textbf{x} and a unique pair of [\textbf{y}, \textbf{z}], making a bijective forward mapping $f(\textbf{x}) = [\textbf{y}, \textbf{z}]$ and a inverse mapping $\textbf{x} =  f^{-1}(\textbf{y}, \textbf{z}) = g(\textbf{y}, \textbf{z})$. A schematic overview of INN architecture is described in \citet{Ardizzone+19a} and \citet{Ksoll+20}. However, this method does not apply to all types of inverse problems: the forward process has to be deterministic and there are some requirements towards the intrinsic dimensionalities of $\textbf{x}$ and $\textbf{y}$.
On the other hand, the cINN, used in this paper, avoids these problems. While it also introduces latent variables for the same purpose, it uses a different mapping system by inputting the observations \textbf{y} in both the forward and inverse process as a condition \textbf{c}: $f(\textbf{x}; \textbf{c}=\textbf{y}) = \textbf{z}$, $\textbf{x} = g(\textbf{z}; \textbf{c} = \textbf{y})$ \citep{Ardizzone+21}.
This has the advantage that there are no assumptions or restrictions about the intrinsic dimensionalities of $\textbf{x}$ and $\textbf{y}$, meaning that effects such as stochastic modelling noise or measurement noise can also be accounted for.

The posterior distribution of physical parameters, $p(\textbf{x}|\textbf{y})$, is estimated on the basis of the inverse mapping $f^{-1} = g$. During training, we prescribe the latent variables to have a standard normal probability distribution $p(\textbf{z}) = N(\textbf{z}, 0, \textbf{I})$ with zero mean and unit width, where \textbf{I} is the identity matrix with a dimension of $\mathrm{dim}(\textbf{z}) \times \mathrm{dim}(\textbf{z})$. Following the inverse process $\textbf{x}=g(\textbf{z}; \textbf{c})$, the posterior distribution is a transformation of the known distribution $p(\textbf{z})$ to \textbf{x}-space, conditioned on the observation. So, the posterior distribution for a given observation $p(\textbf{x}|\textbf{y})$ is determined by sampling the latent variable following the prior distribution and using the inverse process $g$.

The INN and cINN are very similar overall, but unlike the INN, the cINN has the advantage of being free to choose the dimensions of \textbf{x} and \textbf{y}. In the case of the INN which links \textbf{x} and a pair of [\textbf{y}, \textbf{z}], zero padding is necessary if the dimension of \textbf{x} is smaller than the dimension of [\textbf{y}, \textbf{z}]~\citep{Ardizzone+19a}. However, in the cINN, \textbf{y} can have an arbitrarily large dimension regardless of the dimension of \textbf{x} because the cINN uses observation \textbf{y} as a condition and connects \textbf{x} and \textbf{z}, matching the dimension of each other.

According to \cite{Ardizzone+19a, Ardizzone+21}, cINNs and INNs consist of a series of affine coupling blocks following an architecture proposed by \cite{Dinh+16}. The schematic structure of our conditional affine coupling block is described in Figure~\ref{fig:cinn_structure}. Each coupling block splits the input \textbf{u} into two parts $[\textbf{u}_{\textbf{1}}, \textbf{u}_{\textbf{2}}]$ and passes them through two invertible affine transformations. The output of the coupling block \textbf{v}  is the concatenation of outputs from each affine transformation $\textbf{v}_{\textbf{1}}$ and  $\textbf{v}_{\textbf{2}}$. The invertibility of the cINN and INN architecture is based on each reversible affine coupling block. The difference between cINN and INN is that the cINN uses the observation as an additional input in each affine transformation as follows: 
\begin{equation} 
\label{eq:forward affine}
\begin{split}
\mathbf{v_{1}} &= \mathbf{u_{1}} \odot \mathrm{exp}(s_{2}(\mathbf{u_{2}}, \mathbf{c})) + t_{2}(\mathbf{u_{2}}, \mathbf{c}), \\
\mathbf{v_{2}} &= \mathbf{u_{2}} \odot \mathrm{exp}(s_{1}(\mathbf{v_{1}}, \mathbf{c})) + t_{1}(\mathbf{v_{1}}, \mathbf{c}).
\end{split}
\end{equation}
The internal transformations $s_{i}$ and $t_{i}$ are only evaluated in the forward direction even when the network implements the inverse process:
\begin{equation} 
\label{eq:inverse affine}
\begin{split}
\mathbf{u_{2}} &= (\mathbf{v_{2}} - t_{1}(\mathbf{v_{1}}, \mathbf{c})) \odot \mathrm{exp}(-s_{1}(\mathbf{v_{1}}, \mathbf{c})), \\
\mathbf{u_{1}} &= (\mathbf{v_{1}} - t_{2}(\mathbf{u_{2}}, \mathbf{c})) \odot \mathrm{exp}(-s_{2}(\mathbf{u_{2}}, \mathbf{c})).
\end{split}
\end{equation}
Therefore, an arbitrary neural network that does not need to be invertible itself can represent the internal transformations $s_{i}$ and $t_{i}$. In this study, we use a single sub-network for the internal transformations in each coupling block, adapting the GLOW (Generative Flow; \citealt{Kingma&Dhariwal18}) configuration.

\subsection{Network setup}
\label{subsec:network setup}
\subsubsection{Architecture}
\label{subsubsec:architecture}

To construct our network with a cINN architecture we use the `Framework for Easily Invertible Architectures' (FrEIA) for Python \citep{Ardizzone+19a, Ardizzone+21} which is based on the `pytorch' library~\citep{Paszke+19}, just as \cite{Ksoll+20} did in their study. As described in Figure~\ref{fig:cinn_structure}, seven physical parameters of the \hii\ region are the input \textbf{x} for our network, while the condition \textbf{y} is given by the twelve emission line luminosities that we can obtain from observations. Following the structure of the cINN architecture, the dimension of \textbf{z} matches that of \textbf{x} and we have seven latent variables in our network. 

We build our network with 8 conditional affine coupling blocks, and for the sub-networks in the affine coupling blocks, we adopt a simple three-layer, fully-connected architecture with a width of 256, using rectified linear units (ReLU) as the activation functions. Additionally, we apply soft clamping on the sub-network output $s_{i}()$, which is introduced in \cite{Ardizzone+21} to prevent instability from the exponential component in Eq.~\ref{eq:forward affine}.

To mix the information stream $\textbf{u}_{\textbf{1}}$ and $\textbf{u}_{\textbf{2}}$, we add permutation layers with a random orthogonal matrix after each coupling layer. The permutation layer is invertible and fixed during the training. Therefore, the final network is made up of 8 invertible blocks where each block is the combination of an affine coupling block and a permutation layer.

We train the network to minimize the maximum likelihood loss, as described in \cite{Ardizzone+21} and \cite{Ksoll+20}. The cINN model is trained until both the train-loss curve calculated by the training set and the test-loss curve calculated by the test set converge and deviation between the two curves is small enough. 
The training time varies depending on the batch size and the number of training epochs, even in the same network setting. 
Training our network for 300 epochs took about 2 hours with a batch size of 1024 and about 6 hours with a batch size of 256 when we used an NVIDIA GeForce RTX 2080 Ti graphic card. The GPU memory used for each case was 892~MB and 976~MB respectively.

\subsubsection{Data Pre-processing}
\label{subsubsec:preprocess}

When training a network or using the network in practice, we apply pre-processed physical parameters and observations based on the procedure described by \cite{Ksoll+20}. 
First, we transform the variables that have a relatively broad range of values in linear space to logarithmic space. In our case, we transform the emission line luminosities ($y_{i}$) into log-scale.
We retain most of the physical parameters ($x_{i}$) in linear space because when generating the WARPFIELD-EMP database, we sampled each physical parameter within reasonable ranges. The exception is the cloud mass (\mcl), which we sampled in logarithmic space in the beginning.

Next, we add artificial noise to relatively discretized parameters such as the \ncl\ and phase, which have a sampling interval of 1 (see Figure~\ref{fig:distr_param}). According to \cite{Ardizzone+21}, smoothing out the distribution using a small amount of Gaussian noise helps the network converge in training. We smooth out the distribution of \ncl\ and phase by adding Gaussian noise with a standard deviation of 0.05. The noise augmentation also improves the prediction performance of our network. A more detailed explanation about the effect of this smoothing process on the network performance is given in Section~\ref{subsec:smoothing}, where we compare the performance of our network with and without noise augmentation.

Last, we re-scale the distribution of physical parameters and observations by using linear transformations. We transform the physical parameter, $x_{i}$, following 
\begin{equation} 
\label{eq:x-rescale}
\hat{x}_{i} = (x_{i} - \mu_{x_{i}}) \cdot \frac{1}{\sigma_{x_{i}}},
\end{equation}
where $\mu_{x_{i}}$ and $\sigma_{x_{i}}$ are the mean and standard deviation of the physical parameter $x_{i}$ in the database so that the re-scaled distribution of each parameter has zero mean and unit standard deviation. 
In the case of $y_{i}$, we first centre the observable ($\tilde{y}_{i} = y_{i} - \mu_{y_{i}}$) and whiten the observable matrix ($\mathbf{\hat{Y}} = \mathbf{W_{\tilde{Y}}} \mathbf{\tilde{Y}} $), following Equation 35 in \cite{Hyvarinen&Oja00}, so that the variance of each emission line is unity and the covariance of the matrix $\mathbf{\hat{Y}}$ is an identity matrix. The values of
$\mu$, $\sigma$, and $\mathbf{W}$ used in the linear transformations are calculated based on the whole database including both training set and test set. We use the same values when training the network and utilizing the trained network. When we predict physical parameters through the inverse process of the cINN model, we transform $y_{i}$ to $\hat y_{i}$ for the condition of the network and transform the output $\hat x_{i}$ to $x_{i}$.

\subsection{Network evaluation methods}
\label{subsec:evaluation methods}
In this section, we describe how we evaluate the trained network using the held-out 101,149 models of the test set. As mentioned in Section~\ref{subsec:database}, we split the original database and only use 80\%\ of it for training and retain the rest for network evaluation.

We evaluate the network in the following five ways.
First, we need to verify the network by using the latent variables of the test set, $\mathbf{Z_{test}}$. If the network is converged to a good solution, latent variables should follow the prescribed Gaussian normal distribution. This can be confirmed by checking whether the covariance matrix of $\mathbf{Z_{test}}$ is close to an identity matrix and whether the distribution of each latent variable follows the standard normal distribution.

The next three measures allow us to evaluate how accurately and precisely the network predicts each physical parameter with respect to the true value. To quantify the accuracy and precision, we compute the median calibration error ($e\mathrm{_{cal}^{med}}$), median uncertainty at 68\%\ confidence interval ($u\mathrm{_{68}^{med}}$), and the root mean square error (RMSE).
The calibration error, our second criterion, evaluates the shape of the posterior distribution. At a given confidence interval $q$ it is defined as 
\begin{equation}
e_{\mathrm{cal}} = q_{\mathrm{inliers}} - q, 
\end{equation}
where $q_{\mathrm{inliers}}$ is the fraction of test models where the true value falls within the given confidence interval of the posterior distribution. The calibration error is an important evaluation index of the network because it represents the correctness of the shape of the posterior distribution~\citep{Ardizzone+19a}. A negative calibration error indicates an overconfident network, which means that the predicted posterior distribution is too narrow, whereas a positive value means the opposite (under-confident network). We calculate $e\mathrm{_{cal}^{med}}$, the median of the absolute value of calibration error over the confidence range from 0.01 to 0.99 in 0.01 confidence level interval for each physical parameter.

The third quantity is a median uncertainty interval at a 68\% confidence level. The uncertainty interval is the width of the posterior distribution corresponding to the given confidence interval. We chose a confidence level of 68\%, close to the width of $\pm1\sigma$, and take a median value over the whole test set.

Fourth, we determine the root mean square error (RMSE) of the maximum a posteriori (MAP) point estimates, with respect to the ground true value ($x^{*}$). The RMSE of each parameter,
\begin{equation}
    \mathrm{RMSE} = \sqrt{\frac{\Sigma_{i=1}^{N}(x^{\mathrm{MAP}}_{i} - x^{*}_{i})^{2}    }{N}  },
\end{equation}
indicates how accurately our network can predict the true value.

We use a similar method as \cite{Ksoll+20} did to determine the MAP value from each posterior distribution. We perform a Gaussian kernel density estimation on a posterior distribution and find the point where the probability density becomes maximum. To find the suitable bandwidth of the kernel, we generally apply Silverman’s rule of thumb~\citep{Silverman86} but, we also apply the Improved Sheather-Jones (ISJ) algorithm~\citep{Botev+10} which works better for a multi-modal distribution in some cases. For simplicity, we did not check the multi-modality of the distribution for all seven parameters, and only investigate the multi-modality in the \ncl\ posterior distribution. The reason is that the multi-modality of the posteriors in our network usually results from the degeneracy in the \ncl\ prediction, as we explain in Section~\ref{sec:degenerate}. When the posterior distribution of \ncl\ shows multi-modality, we apply a factor of five narrower width than that of Silverman’s rule of thumb except for the posterior distribution of age. We apply the ISJ algorithm to the age posterior distribution which needs more careful kernel density estimation because it usually has both very narrow mode and wide mode in one distribution.

The final method we use to validate the network is to examine whether or not our cINN model constrains the physical parameters correctly. This is different from the previous three approaches of evaluating how accurately or precisely the network predicts the true values. Different \hii\ regions can have similar emission line strengths, so our cINN model is designed to find all possible physical models $\mathbf{x'}$ conditioned on a given observation \textbf{y} (= \textbf{c}). Therefore, even if the predicted value of parameters is different from the true value, it does not demonstrate the incapacity of our network. What we need to examine is whether the predicted model has the same emission line luminosity as the conditioned observation. Observational properties \textbf{y} cannot be obtained through the forward process of the cINN because \textbf{y} is used as a condition in both the forward and backward process. For that reason, we re-simulate the emission line strengths of predicted models using WARPFIELD-EMP in the same way as we created the database.

As explained in Section~\ref{subsec:database}, one synthetic \hii\ region is determined by four independent physical parameters (\mcl, star formation efficiency, \hdenini, and the age of the cloud). We first evolve the cloud using WARPFIELD, taking as initial conditions the values of \mcl, the star formation efficiency, and \hdenini\ predicted by the network. We stop the evolution of the cloud at the age predicted by the network and then use CLOUDY and POLARIS to compute the line emission produced at this time. In practice, we generally cannot collect the cloud information at the very age that we want because WARPFIELD records the evolution with a finite time interval. In this re-simulation process for network validation, we adopt a time interval of 0.05~Myr so that the age difference between the re-simulated model and the predicted model is certain to be less than 0.025~Myr.

\section{Training Results}
\label{sec:training results}

Once trained, the network is able to sample posterior distributions very efficiently. It takes less than 10 minutes to generate posterior distributions for the whole 101,149 observations in our test set, sampling 4096 times for each observation with an NVIDIA GeForce RTX 2080 Ti graphic card (the same graphic card used to measure training time in Section~\ref{subsubsec:architecture}). On average, we can obtain posterior distributions of 170 observations per second through our network.

\subsection{Training evaluation}
\label{subsec:training evaluation}

\renewcommand{\arraystretch}{1.25}
\begin{table*}
    \caption{Overview of our network performance using all of the 101,149 \hii\ region models in the test set. For each parameter, we present median calibration error, median width of posterior distributions (uncertainty at 68\%\ confidence interval), and mean accuracy of the MAP estimates (RMSE). We divide median width and mean accuracy into two types depending on whether we use re-scaled parameters ($\hat{x}_{i}$) or original parameters ($x_{i}$). The one using the re-scaled parameters is dimensionless whereas the one using the original parameters has the same unit as the parameters. \label{table:overview}}
    \begin{tabular}{ l  c  c  c  c  c  c   c }
        \hline
        \hline
        Performance measure & log \mcl\ & SFE & \hdenini\ & $t$ & $t_{\text{youngest}}$ & \ncl\ & Phase \\
             & log [M$_{\odot}$] & [\%]	& [cm$^{-3}$]	& [Myr]	& [Myr]	\\
        \hline
        Median calibration error [\%] & 0.44 & 0.26 & 0.87 & 1.27 & 1.05 & 2.34 & 0.12   \\
        Median uncertainty at 68\%\ confidence ($\hat{x}$) & 0.0284 & 0.0945 & 0.1694 & 0.0176 & 0.0097 & 0.1512 & 0.0855 \\
        Median uncertainty at 68\%\ confidence (${x}$)& 0.0154 & 0.2173 & 19.0633 & 0.1491 & 0.0637 & 0.1108 & 0.0995 \\
          
        RMSE ($\hat{x}$) & 0.0823 & 0.2506 & 0.2846 & 0.5589 & 0.1210 & 0.6385 & 0.1724 \\
        RMSE ($x$) & 0.0448 & 0.5760 & 32.0280 & 4.7285 & 0.7918 & 0.4681 & 0.2006 \\
            
        \hline
        \hline
    \end{tabular} 
\end{table*}

In this section, we evaluate our trained network by using four of the five methods presented in Section~\ref{subsec:evaluation methods}: $\mathbf{Z_{test}}$, $e\mathrm{_{cal}^{med}}$, $u\mathrm{_{68}^{med}}$, and RMSE of MAP estimates. First, we confirm whether the latent variables follow the prescribed Gaussian normal distribution on the test set ($\mathbf{Z_{test}}$) or not. In Figure~\ref{fig:z_cov_pdf}, we present the covariance matrix of $\mathbf{Z_{test}}$ and the probability distribution of each latent variable. The covariance matrix is close to the unit matrix and distributions of latent variables are almost following the Gaussian normal distribution with a residual of less than 0.03. These results confirm that our network is trained well.

The other three evaluation indices ($e\mathrm{_{cal}^{med}}$, $u\mathrm{_{68}^{med}}$, and RMSE) are presented in Table~\ref{table:overview}. The median calibration errors of our cINN model are very low, around 1\% or less in most cases. This means that the expected accuracy of our model is well calibrated to the confidence. The largest value is 2.34\%\ for \ncl. This means that the network is the least calibrated for the \ncl\ prediction among the seven parameters, but the error value of 2.34\%\ is still very low and highly acceptable~\citep{Guo+17}.

The second and third rows in Table~\ref{table:overview} show the median uncertainty at 68\% confidence interval (i.e., $u\mathrm{_{68}^{med}}$) for each parameter in the re-scaled parameter space ($\hat{x}$-space) and original parameter space ($x$-space), respectively. For the original parameter space, $u\mathrm{_{68}^{med}}$ has the same units as the physical parameter, whereas for the re-scaled parameter space $u\mathrm{_{68}^{med}}$ is dimensionless, since it is the width of the posterior distribution with respect to the range of the parameter in the database. According to the $u\mathrm{_{68}^{med}}$ for the $\hat{x}$-space, \hdenini\ and \ncl\ have on average wider predicted posterior distributions compared to the other parameters. Nevertheless, when it is transformed to real physical space, the uncertainty interval of \hdenini\ is around 20~cm$^{-3}$ and that of \ncl\ is 0.11 which is small compared to its sampling size of 1.

The RMSE of each parameter is also calculated for both the re-scaled parameter space and the physical parameter space (the fourth and fifth rows in Table~\ref{table:overview}). Age and \ncl\ have relatively large RMSEs in the re-scaled space, implying again that the age of the cloud and \ncl\ are difficult to predict. On the other hand, \mcl\ has a low RMSE in both $x$-space and $\hat{x}$-space; in $x$-space, the RMSE for the cloud mass is around 0.04~dex, corresponding to an uncertainty of less than 10\%.

Considering the overall results in Table~\ref{table:overview}, our cINN model is well-calibrated and is able to predict each parameter accurately and precisely in general. However, some parameters such as \ncl\ and the age of the cloud are relatively difficult to predict. On the other hand, \mcl\ and phase have small values in all three indices, meaning that our network predicts \mcl\ and phase stably.

\subsection{Posterior probability distribution}
\label{subsec:showcase}

\begin{figure*}
	\includegraphics[width=1.53\columnwidth]{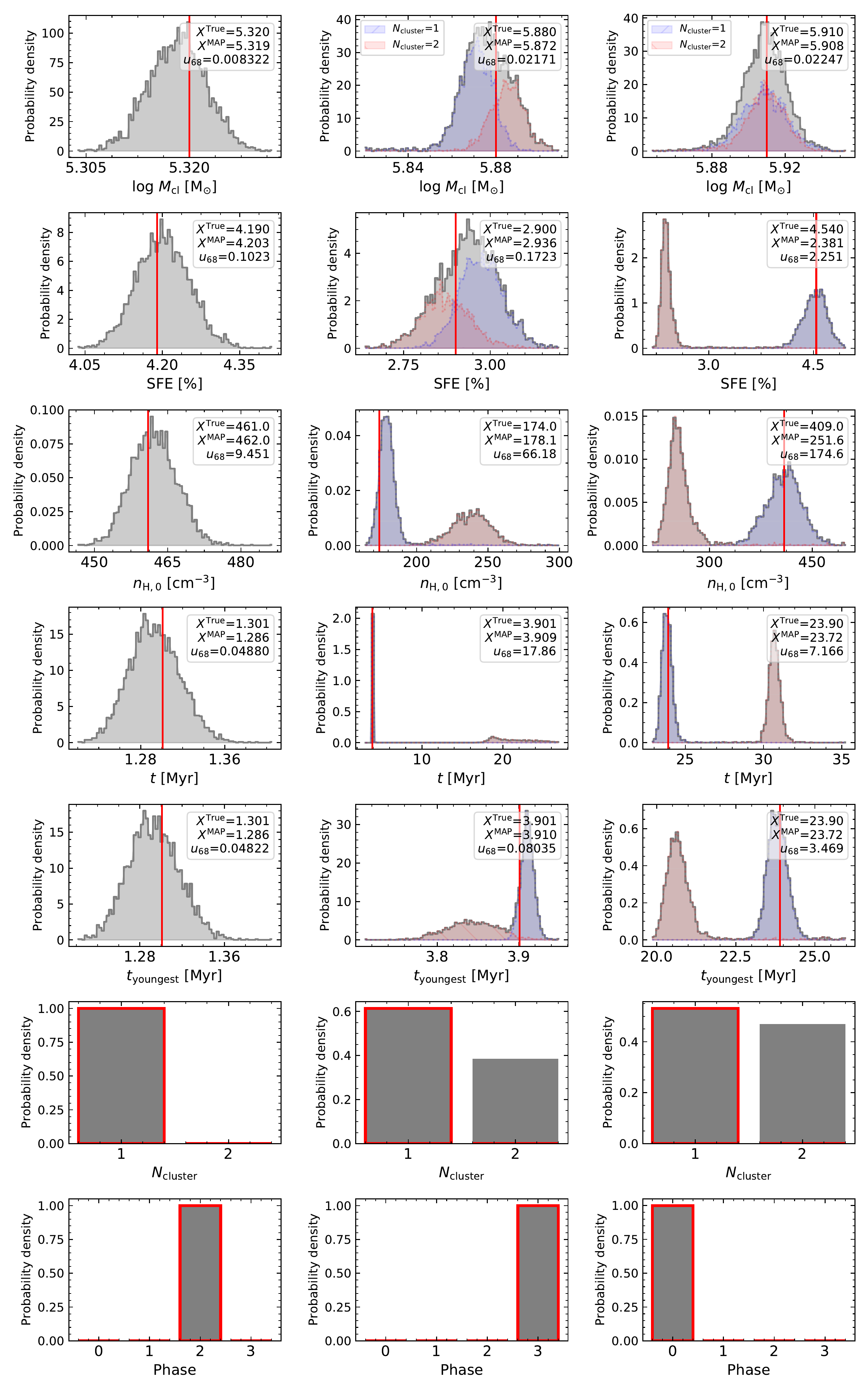}
    \caption{Posterior probability distributions (grey) of seven physical parameters. Each column corresponds to a different model selected from the test set. Red vertical lines (or red edges for bar-shaped histograms) denote the true parameter values of the model. The true value, the MAP estimate, and uncertainty at 68\%\ confidence interval of the posterior distribution are presented in the upper right corner of each panel except for \ncl\ and phase. 
    The left column shows an example of an extremely well-predicted case. The other columns represent degenerate cases where the posterior of some parameters have multi-modal distribution. The middle column has a bi-modal distribution where one mode has a higher probability than the other, whereas the model in the right column represents a degenerate case where different solutions have similar probabilities. For the two degenerate cases, we divide the posteriors into two groups depending on the \ncl\ posterior: blue distributions indicate posteriors that predict \ncl\ as 1, and red distributions indicate posteriors that predict \ncl\ as 2.} %
    \label{fig:showcase_hist}
\end{figure*}

In this section, we show representative posterior distributions conditioned on individual observations. We select three \hii\ region models from the test set that exhibit typical shapes of the posterior distribution. Figure~\ref{fig:showcase_hist} shows the one-dimensional posterior distributions of each parameter for these examples.

The first model (left column) represents extremely well-predicted cases. As shown in the figure, posteriors of this model have a clear unimodal distribution with narrow width, and the MAP estimate is very close to the true value marked by red vertical lines or red-edged bars for \ncl\ and phase. In the case of age, for which our cINN model showed a less good performance in the previous section, the difference between the MAP estimate and the true value is less than 0.01~Myr, and the width of the posterior distribution represented by $u_{68}$ is about 0.05~Myr. Though this model is one of the best examples, the accurate and precise unimodal posterior distribution is the most common characteristic found in our test set.

The other two models are examples of degenerate cases, for which posterior distributions usually have two or more peaks as shown in Figure~\ref{fig:showcase_hist}. The second case in the middle column shows two different solutions, one with a higher probability and the other with a lower probability, whereas the last example has two solutions with similar probabilities. 
In the second example, \mcl, star formation efficiency, and phase have a unimodal distribution (grey histograms) and the other four parameters have a bimodal distribution where true values always fall within the first mode with a higher probability density. The posterior distribution of the youngest cluster age has an overlapping area between the first mode and the secondary mode, suggesting that the unimodal distribution such as in \mcl\ could be decomposed into two different modes that are considerably overlapped with each other.
The third case shows two distinct peaks with similar amplitudes without overlap in all parameters excluding \mcl\ and phase. The true value lies on the first mode in the case of cloud age and the youngest cluster age but in the case of star formation efficiency and \hdenini, the true value falls in the secondary mode. Nevertheless, it is clear that the true value is within the range of the posterior distribution.

\subsubsection{Degeneracy}
\label{subsubsec:showcase_deg}

For these degenerate examples, we divide the posteriors into two groups based on the \ncl\ prediction: the first group predicting \ncl\ as 1 and the second group predicting \ncl\ as 2. Then we decompose the posterior distribution of the remaining parameters except for the phase into each group and present them in Figure~\ref{fig:showcase_hist} using different colours: blue for the first group (\ncl=1) and red for the second group (\ncl=2). The red and blue histograms in Figure~\ref{fig:showcase_hist} clearly show that the two groups classified by the \ncl\ prediction correspond to the two different modes exhibited in the posterior distributions. Moreover, in the second example, the unimodal posterior distribution for \mcl\ and star formation efficiency is divided into two modes albeit with a wide range of overlap. 
As shown in these cases, most of the degeneracy revealed in the posterior distribution reflects the degeneracy in the \ncl\ prediction. In other words, this kind of degeneracy suggests how other parameters should change to produce the same amount of emission line luminosity with more or fewer star clusters.

Both degenerate examples actually have only one cluster, or more precisely one stellar generation in the central cluster, so the red mode in Figure~\ref{fig:showcase_hist} suggests another solution with one more star cluster (stellar generation) which generates the same emission. 
In the case of the second example, the initial clouds of the red mode are on average more massive and denser but have smaller star formation efficiency compared to those of the blue mode (i.e., true model). The red mode shows that to emit the same amount of emissions as the true model, the cloud of the red mode should live longer till the second-star cluster is a bit younger than that of the blue mode. 
In the third example, the second solution requires a similar mass of cloud but with smaller star formation efficiency, smaller density, older cloud age, and younger age for the youngest cluster.

\begin{figure}
    \includegraphics[width=0.99\columnwidth]{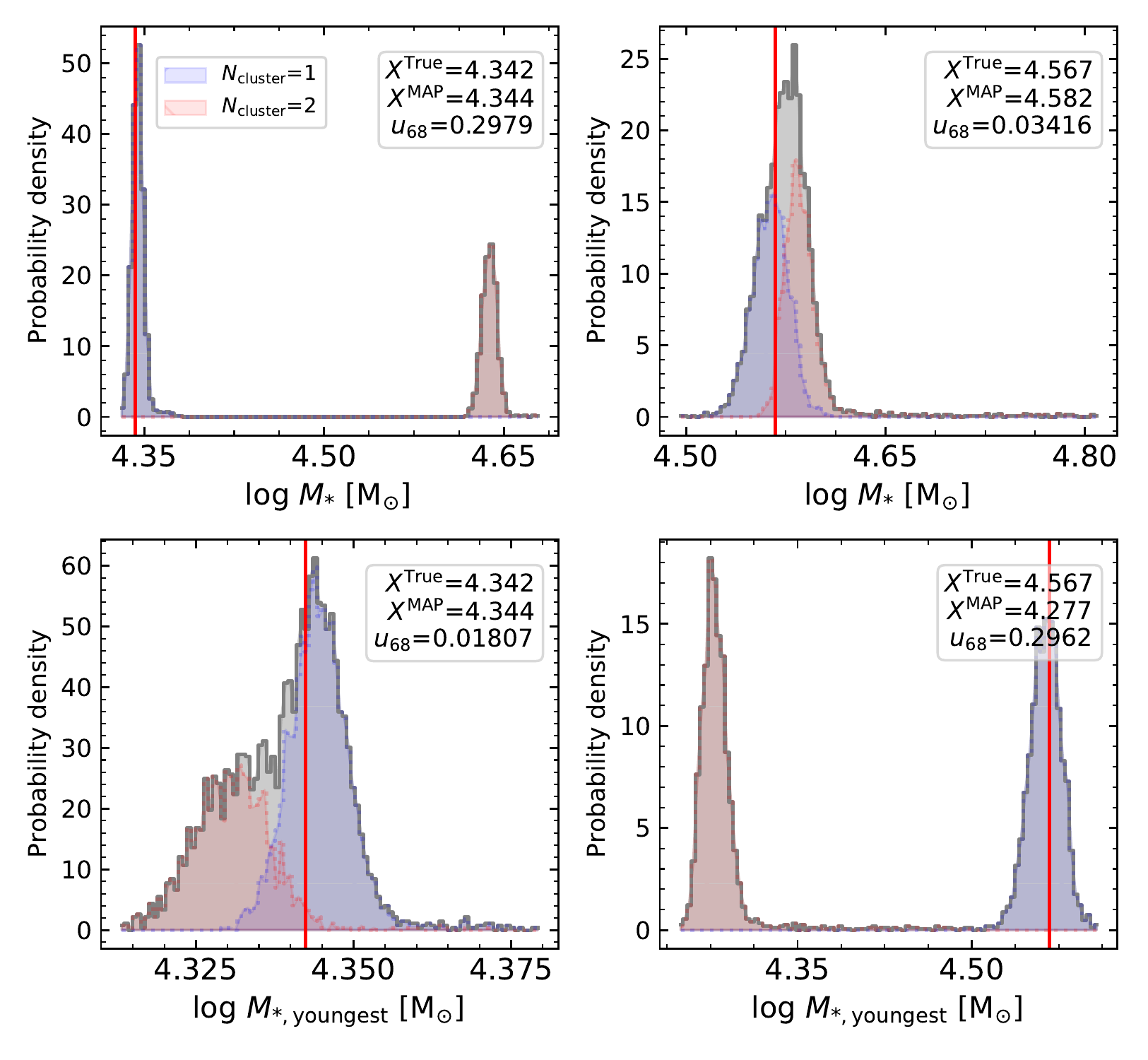}
    \caption{Posterior distributions (grey) of total cluster mass and the youngest cluster mass. The first and second columns correspond to the second example and the third example in Figure~\ref{fig:showcase_hist} respectively. Lines and legends are the same as in Figure~\ref{fig:showcase_hist}.
    Posteriors are divided into two groups depending on the \ncl\ posterior as shown in Figure~\ref{fig:showcase_hist}: the blue mode whose \ncl\ prediction is 1 and the red mode whose \ncl\ prediction is 2. }
    \label{fig:showcase_Ms}
\end{figure}

The degenerate solution demonstrates that our network understands the correlation between parameters that affect the emission line luminosity. Based on the posterior of cloud mass, star formation efficiency, and the number of clusters, we additionally calculate the total stellar mass of clusters in the cloud because stellar mass is a more direct index of the ionizing power. As the youngest cluster dominates the ionizing luminosity, we also calculate the mass of the youngest cluster. In Figure~\ref{fig:showcase_Ms}, we present the posterior distribution of these two parameters for the second example and the third example and divide the distribution into two groups based on the \ncl\ prediction in the same way as in Figure~\ref{fig:showcase_hist}.
In the case of the second example, the total stellar mass of the red mode with two clusters is a factor of two larger than that of the blue mode, whereas the youngest cluster mass is almost identical with only a 0.015~dex difference. The second cluster of the red mode and the cluster of the blue mode, which are the main luminosity sources of each mode, may have similar luminosity because they have similar mass and similar age around 3.8$\sim$4~Myr. The small difference in the cluster age offsets the small difference in the cluster mass. The first cluster of the red mode is less luminous than the others in spite of its similar stellar mass because it is older than the others (~$\sim$20~Myr) so that all of its bright O type stars and many of its brightest B type stars are already dead. However, the first cluster slightly increases the total luminosity of the red mode. This additional luminosity is cancelled out by the larger amount of dust attenuation due to the higher cloud density of the red mode.

In Figure~\ref{fig:showcase_hist}, the red mode of the third example shows a similar trend to the red mode of the second example, except for the density. However, Figure~\ref{fig:showcase_Ms} shows that the situation is different from the second example. The total mass of the clusters in the red and blue modes are almost the same, but the mass of each of the two clusters in the red mode is half of that in the blue mode. The age of the clusters is also different from the second example. All three clusters are older than 20~Myr, so they have already lost their brightest stars. In this case, the age difference between the individual clusters is not significant. The total luminosity of the blue and red mode is similar because it is determined by their total stellar mass.
However, the stellar feedback of the red mode is weaker than that of the blue mode because the mass of the individual cluster is about a factor of two smaller. In order for the red mode to have two clusters and to evolve to a phase similar to that of the blue mode, the gravitational potential must be smaller to balance the weak stellar feedback. As the two modes have similar cloud mass, the red mode has a smaller density to reduce the gravitational potential. 
The results of these two degenerate examples reveal that our cINN model successfully learned the hidden rules in the training data and provides the physically reasonable alternatives as well as the posteriors close to the true solution.

\begin{figure*}
	\includegraphics[width=2\columnwidth]{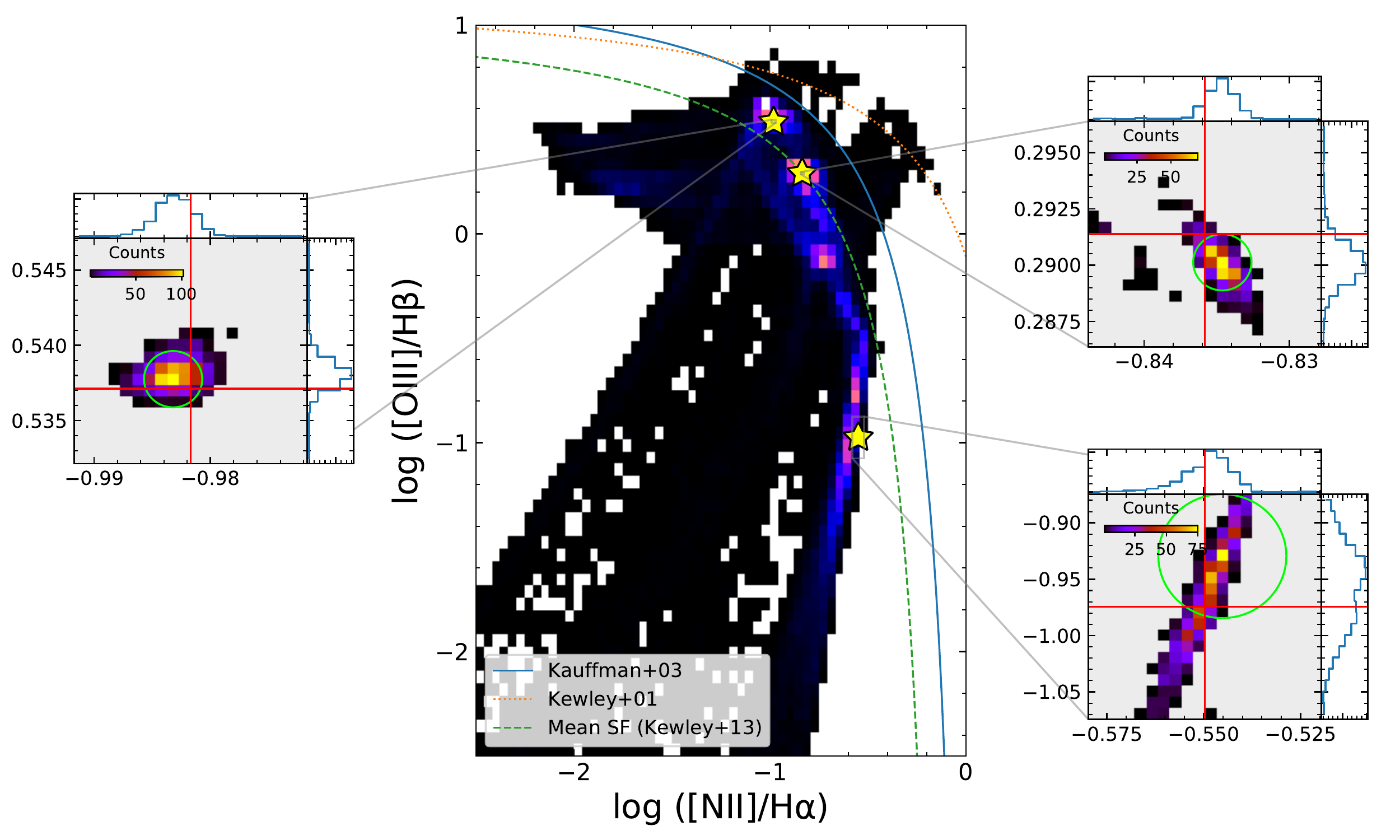}
    \caption{Middle panel: 2D histogram showing the \nii/\ha\ and \oiii/\hb\ line ratios for all of the models in the test set, where    brighter colour indicates a higher number of models. Overlaid as yellow stars are the corresponding values of the models presented in Figure~\ref{fig:showcase_hist}. Also shown are demarcation curves distinguishing star-forming galaxies and AGN, taken from \citet{Kauffmann+03} (blue solid line) and \citet{Kewley+01} (orange dotted line) as well as the mean line of star-forming galaxies from \citet{Kewley+13b} (green dashed line). Star-forming galaxies are expected to sit below and to the left of the \citeauthor{Kauffmann+03} and \citeauthor{Kewley+01} lines.
    Zoom-in panels show the distribution of the line ratios that we recover if we sample the posterior distribution for each example model 1024 times and use the resulting values as input for new WARPFIELD-EMP calculations, as described in Section~\ref{subsec:evaluation methods}. The true line ratio values for each example model are represented by red lines in these zoom-in panels.
    The left panel corresponds to the first model, whereas the upper right and lower right panels correspond to the second and third models, respectively. Green circles in each zoom-in panel indicate the area in which 68\%\ of the models are included from the centre of the distribution.
    }
    \label{fig:showcase_bpt}
\end{figure*}

\subsubsection{Re-simulating posterior models}
\label{subsubsec:showcase_vd}
We now investigate the emission line luminosity of the predicted posteriors to validate our network performance. As explained in Section~\ref{subsec:evaluation methods}, we reproduce the synthetic \hii\ region model of the posterior and calculate the luminosity of 12 emission lines through WARPFIELD-EMP based on the four parameters that determine a unique star-forming cloud: \mcl\, SFE, \hdenini, and age. The posterior distribution of each example model in Figure~\ref{fig:showcase_hist} consists of 4096 posteriors, but we only simulate 1024 posteriors of them. 

In the middle panel of Figure~\ref{fig:showcase_bpt}, we present a 2D comparison of the \oiii/\hb\ and \nii/\ha\ line ratios -- an example of a so-called BPT diagram~\citep{Baldwin+81} -- for the entire 101,149 test set models. Here the colour indicates the number of different models at each point in this 2D histogram. The true locations of our three example models are represented by yellow stars.
In each zoom-in panel of Figure~\ref{fig:showcase_bpt}, we compare the \nii/\ha\ and \oiii/\hb\ line ratios produced by the models sampled from the predicted posteriors with the true line ratio values (red lines) of each example model. The green circles show the area in which 68\%\ of the posteriors are included from the centre of the distribution.
The left panel corresponds to the first example model. We find that the line ratio distribution of the posterior samples is very narrow and lies close to the true values. The peak of the distribution is just shifted by about 0.002~dex from the expected point and the overall width of the distribution is around 0.01~dex. The green circle shows that 1$\sigma$ of the distribution is within the 0.003~dex radius.
Even though we only present two line ratios in the figure, we have confirmed that the re-simulated luminosity of all 12 emission lines also shows a good agreement with the true values with an average error of 0.013~dex in logarithmic scale.

The second example (upper right panel in Figure~\ref{fig:showcase_bpt}) also indicates a good result. The reproduced line ratios are very narrowly distributed within a box of size 0.01~dex. The centre of the distribution is shifted by 0.0025~dex from the true values and 68\%\ of the entire models are located within a radius of 0.002~dex. 
On the other hand, the last example (lower right panel in Figure~\ref{fig:showcase_bpt}) exhibits a wider distribution and a larger offset compared to the other two. The line ratio distribution is elongated along the y-axis over a range of 0.2~dex and its peak is shifted by about 0.04~dex. Although the overall results are worse than for the other two examples, still 68\%\ of the posterior samples are within a 0.05~dex radius of the true values. We have verified that the luminosities of the 12 emission lines are in good agreement with the true values with an average error of 0.12~dex in log units, or around 30\%.

Overall, the results of the reproduced emission lines of the posterior samples shown in Figure~\ref{fig:showcase_bpt} demonstrate that our network is understanding the hidden rules in the training data and is able to provide reliable posteriors, especially in that the secondary solutions in two degenerate examples also match the observation successfully.

\subsection{Overall performance}
\label{subsec:overall performance}

\begin{figure*}
	\includegraphics[width=1.97\columnwidth]{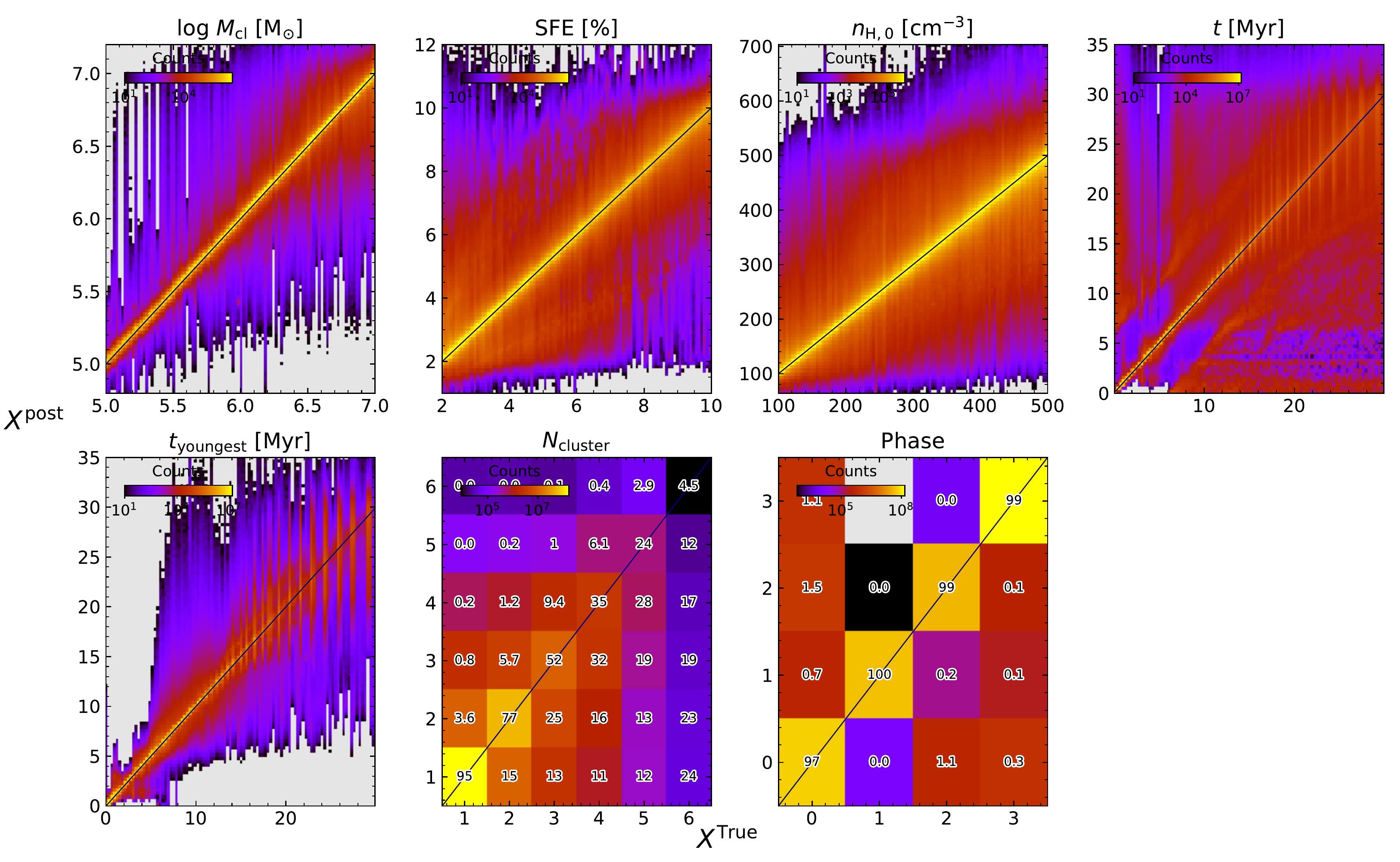}
    \caption{2D-histogram comparing all posterior estimates predicted by our network with the true values of each parameter for all 101,149 models in the test set. We sampled 4096 posteriors for each test model. Colours indicate the number of models at each point in the two-dimensional histograms. Please note that we only plot areas with more than 10 models, leaving the otherwise part in grey. For the two discretized parameters (\ncl\ and phase), we additionally present the confusion matrices on the 2D-histograms that show the number of models in each point divided by the number of all models over the column in percent.
    } 
    \label{fig:all_post}
\end{figure*}

\begin{figure*}
	\includegraphics[width=1.97\columnwidth]{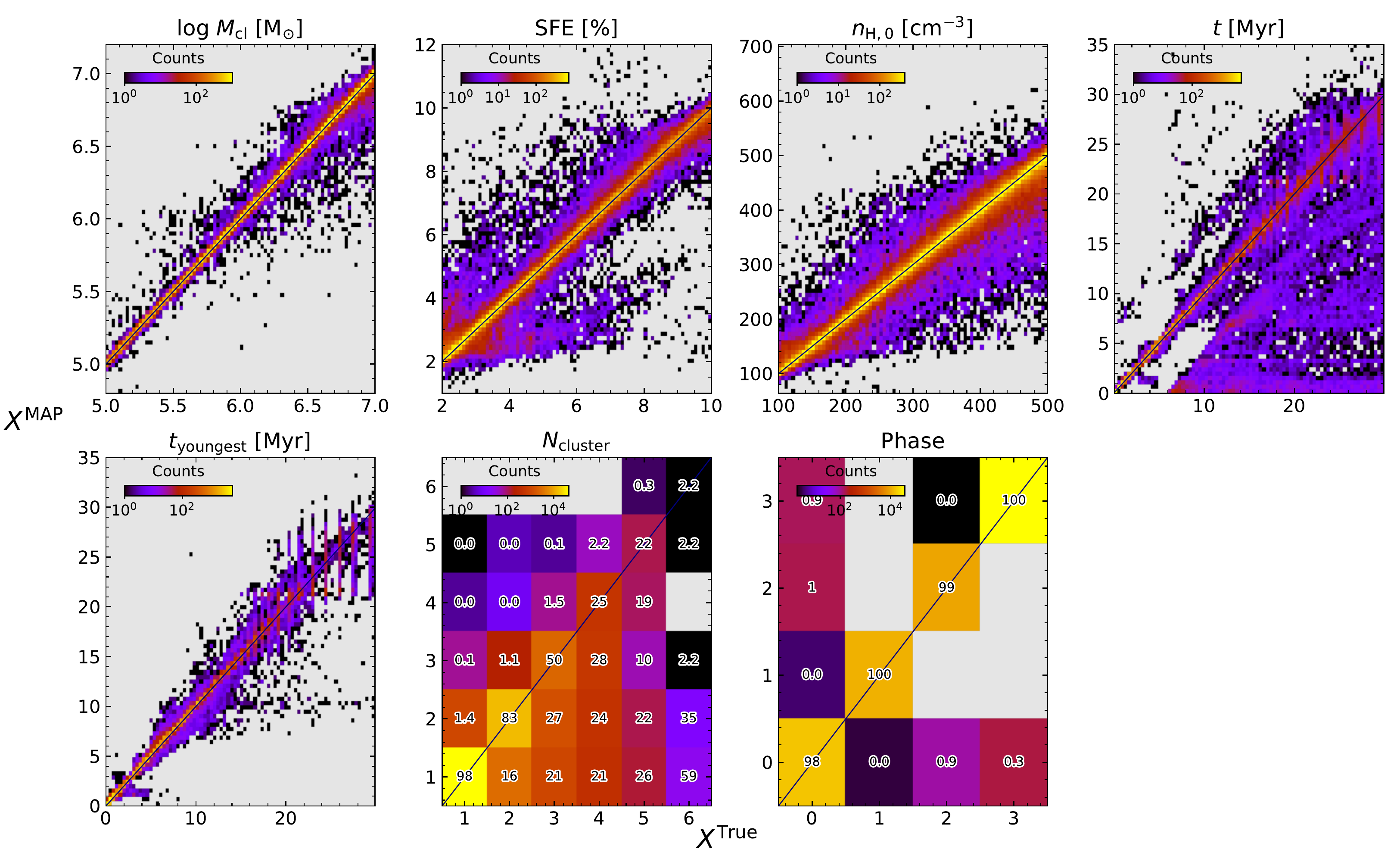}
    \caption{2D histogram comparing the MAP estimates with the true values of each parameter for all 101,149 test models. Colour code is the same as in Figure~\ref{fig:all_post} but the grey colour indicates the area without any posterior model.} 
    \label{fig:all_map}
\end{figure*}

Now we extend the target to all 101,149 test models to probe the overall predictive performance of our network. We sample sets of latent variables 4096 times for each test model and measure the MAP estimates as the representative of each posterior distribution. We compare the true values of the seven parameters with all the obtained posterior estimates in Figure~\ref{fig:all_post} or only with the MAP estimates in Figure~\ref{fig:all_map}. Please note that colours representing the number of models in the 2D histogram in Figures~\ref{fig:all_post} and \ref{fig:all_map} are in logarithmic scale. For \ncl\ and phase, we additionally present the confusion matrix of the 2D histogram, showing the percent of the number of models in each point to the number of all models in the same column, to visualize the prediction performance depending on the true values. Most of the predictions are in excellent agreement with the true values albeit with large scatters around the one-to-one correspondence, and the scatter is much smaller when we compare MAP estimates with the true values.

Both figures indicate that it is difficult to predict the cloud age and the number of clusters, especially when the number of clusters is larger than three. 
In the age prediction in Figure~\ref{fig:all_post}, we find two characteristic features. The first one is that the older the true age, the wider the scatter around the one-to-one correspondence. We can find a similar feature in the age MAP estimates in Figure~\ref{fig:all_map} and the result of the youngest cluster age in both Figures as well. 
Analysing the posterior distribution of various models, we notice that young \hii\ regions, less than a few million years old, usually have very narrow age posterior distributions compared to old \hii\ regions. The $u_{68}$ values of the first model and the third model in Figure~\ref{fig:showcase_hist} clearly show this feature. Larger scatter around the true value for old \hii\ regions reflects this general feature exhibited in most posterior distributions. 
The second feature in the age prediction is two parallel lines shifted by about 6~Myr above and below the one-to-one correspondence line. These parallel lines reflect solutions that are younger or older than the true age values due to the degenerate prediction of the number of clusters as shown in Figure~\ref{fig:showcase_hist}.

In Figure~\ref{fig:all_map}, the MAP estimates of the cloud age are usually distributed around the true values or at younger ages, which is different to the case in Figure~\ref{fig:all_post} using all posteriors estimates where outliers are scattered in both older and younger age range. The parallel lines shown in Figure~\ref{fig:all_post} using all posterior estimates remain in Figure~\ref{fig:all_map} using MAP estimates, but only the bottom line is clear in Figure~\ref{fig:all_map}. This is related to the first characteristic mentioned above that the age posterior distributions of young \hii\ regions are narrower than those of old \hii\ regions. 
When the posterior distribution has multi-modality, the width of the younger mode is  by and large narrower than that of the older mode. This is because the probability density of the younger mode is higher even if the number of posteriors corresponding to the younger mode is similar to that of the older mode.
Sometimes, even if most of the posterior samples belong to the older mode, the peak of the younger mode is determined as a MAP estimate because of its extremely narrow width compared to that of the older mode. Numerous outliers below the one-to-one line in Figure~\ref{fig:all_map} originates from this feature.

The reason why our network predicts the cloud age more precisely for younger \hii\ regions is related to the training data. For both cloud age and the youngest cluster age, the fraction of young models in the database is larger as shown in Figure~\ref{fig:distr_param} so that the network may have learned better about younger models. Moreover, as explained in Section~\ref{subsec:warpfield}, cloud age and the youngest cluster age are not sampled in a constant interval, unlike the other five parameters. Age intervals are sometimes wider in the old period of the cloud because the cloud does not evolve as dramatically as in the younger period.

In the case of \ncl, the more star clusters (more precisely stellar generations) contribute to the observed luminosity, the harder it is for the cINN to predict the correct number of clusters (generations). When this number is one or two, the network predicts the posterior accurately close to the true value but it shows poor performance when the number is larger than three. The poor performance for models with many clusters is also attributed to the distribution of \ncl\ in the training data. Our network is trained mostly on the single cluster \hii\ regions which occupy more than 70\%\ of the training set as shown in Figure~\ref{fig:distr_param}. 
Apart from the bias of the training data, it is difficult to accurately predict the number of clusters because of the intrinsic physical degeneracy of the \hii\ region models with respect to the twelve optical emission lines that we use to predict the parameters.
The strengths of these lines are dominated mostly by the ionizing luminosity of the youngest stellar generation of the \hii\ region model. So adding one or more old clusters with low ionizing power does not have much impact on the overall strength of most of the lines, making it hard for our network to identify the accurate number of clusters.

In the star formation efficiency case, we discover an arrow-shaped structure pointing towards the lower left side in both Figures~\ref{fig:all_post} and \ref{fig:all_map}. These are similar to a one-to-two line and two-to-one line respectively, which implies the degeneracy revealed in the posterior distribution. As most of the degeneracy occurring in our network is because of the degenerate \ncl\ prediction, this feature is more distinct in the range of low star formation efficiency. As shown in the \ncl\ predictions, the degeneracy occurs more frequently for \hii\ regions with many clusters. These \hii\ region models usually have small star formation efficiencies, which makes star-forming clouds easily collapse again as stellar feedback is too weak to destroy the cloud.

Prediction of the cloud initial density and the cloud mass is well constrained to the true values. The scatter in the cloud mass distribution is larger for higher cloud masses. We consider that this feature is also related to the degeneracy of \ncl\ because the massive clouds are more likely to recollapse. Phase prediction is accurate and less degenerate compared to the other parameters. We notice that even when the \ncl\ prediction is degenerate and the other parameters have multi-modal posterior distributions, the posteriors of the phase are usually accurate and have unimodal distributions, as shown in Figure~\ref{fig:showcase_hist}.

\section{Validation of the network}
\label{sec:validation}

\begin{figure}
	\includegraphics[width=0.99\columnwidth]{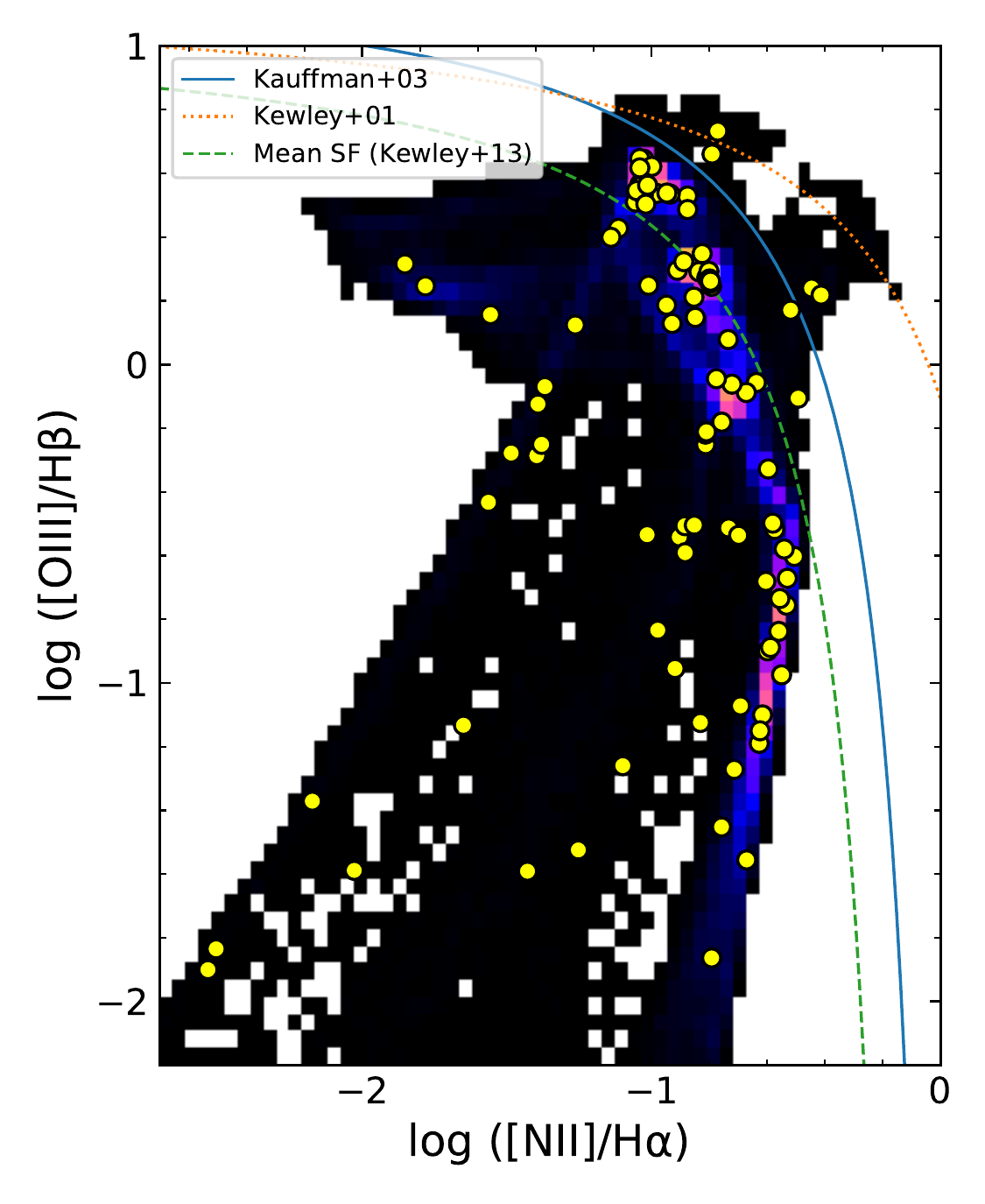}
    \caption{BPT diagram showing the locations of the 100 test models (yellow circles) used for the network validation in Section~\ref{sec:validation} and all of the models in the test set. The three demarcation lines and the colour code of the 2D histogram are the same as in Figure~\ref{fig:showcase_bpt}. } 
    \label{fig:vtest_sample_bpt}
\end{figure}

\begin{figure}
	\includegraphics[width=1.\columnwidth]{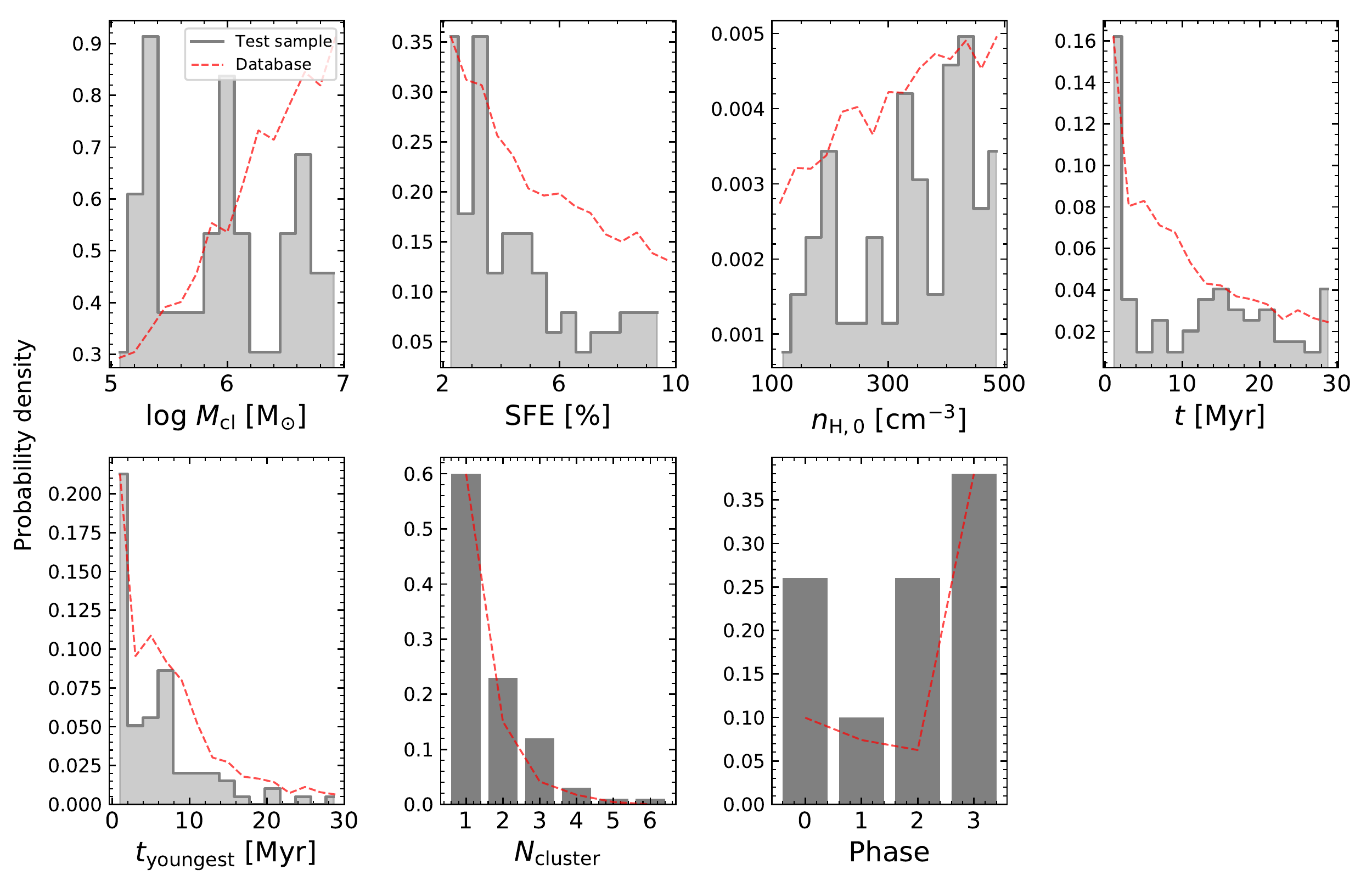}
    \caption{Distribution of the seven physical parameters of the 100 randomly selected test models for the network validation in Section~\ref{sec:validation}. The grey histogram shows the distribution of the 100 models and the dashed red line shows the distribution of the whole database shown in Figure~\ref{fig:distr_param}, where the amplitudes of the distributions are re-scaled.
    } 
    \label{fig:vtest_sample_parameter}
\end{figure}

\begin{figure*}
	\includegraphics[width=2\columnwidth]{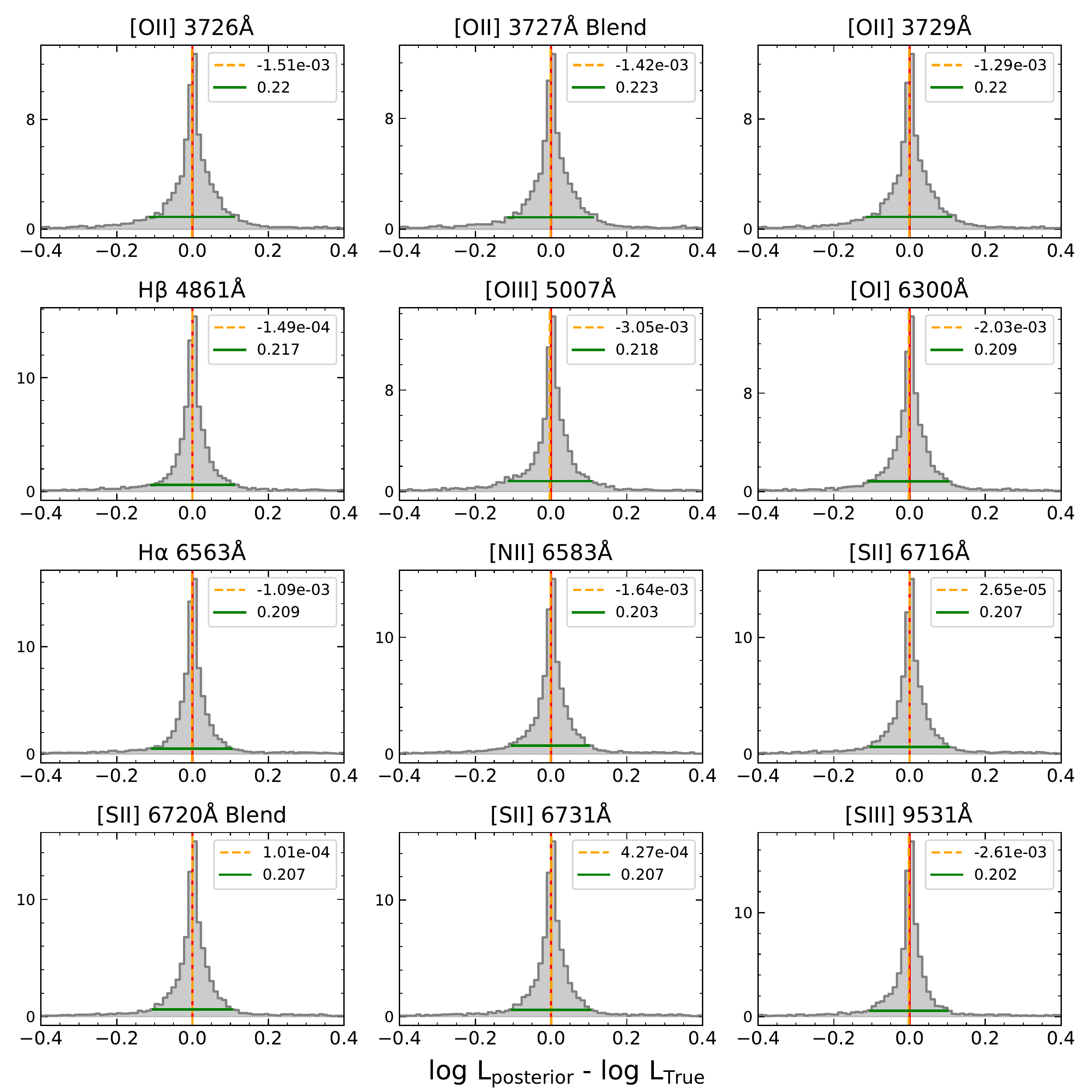}
    \caption{Density histograms of the logarithmic difference between the re-simulated emission-line luminosity of the posterior samples and true emission-line luminosity values of the test models. 
    Twelve panels show the distribution of different emission lines required for our cINN model. 
    For our 100 test models given in Figure~\ref{fig:vtest_sample_bpt}, we draw 100 posterior samples per test model, so histograms are based on the luminosity of 10,000 posterior samples in total.
    We calculate the density-weighted first moment and second moment of each distribution. The vertical dashed orange line indicates the first moment and the horizontal green bar denotes twice the second moment, representing the offset and width of the distribution respectively. The corresponding values are presented by the legend in each panel. 
     } \label{fig:vtest_all}
\end{figure*}

In Section~\ref{subsubsec:showcase_vd}, we validated our network prediction for three example models based on the re-simulation method described in Section~\ref{subsec:evaluation methods}. Now we assess the general performance of our tool with more test models by comparing the luminosity of all twelve emission lines used in the network with the true values. To complete the validation of our tool, we have to check two things. First, we need to confirm whether the network correctly learned the physical rules hidden in the training data. Second, we need to examine how well the database used for training reproduces real nature. In this paper, we focus on the first validation step, evaluating the machine learning aspects of the problem, since the underlying physical input models from WARPFIELD-EMP have already been probed by previous studies~\citep[e.g.,][]{Rahner+18, Rugel+19, Pellegrini+20}.  Nevertheless, we will address some of the limitations of the WARPFIELD-EMP synthetic model and our training data in Section~\ref{subsec:assumptions}.

Altogether, we select 100 models from the test set for the network validation and draw 100 samples for the posterior of each model. We randomly select these 100 models, subject to the following criteria. First, we only use models with \nii/\ha $\geq$ -3 and \oiii/\hb $\geq$ -2 because the line ratios of most of the observed star-forming galaxies~\citep{Kauffmann+03, Kewley+06} or \hii\ regions~\citep{Sanchez+15} are larger than these minimum values. Second, four of the 100 models are randomly selected among extreme cases, which are located
in the BPT diagram beyond the revised demarcation curve between starburst galaxies and active galactic nuclei (AGNs) of \cite{Kauffmann+03}. One of our four extreme cases is located even beyond the demarcation curve of \cite{Kewley+01}. Third, for the other 96 models, we set the fractions of models with relatively uncommon characteristics in order to prevent too many models from being selected close to the mean of the distribution. Considering the parameter distribution of our database (Figure~\ref{fig:distr_param}), we limited the fraction of single cluster models to 60\%\ and the fraction of Phase 3 models to 40\%. The former is to ensure a sufficient number of multicluster models in the validation-test sample and the latter is to prevent too many Phase 3 models from being selected. 
We present the location of the 100 selected models in the BPT diagram in Figure~\ref{fig:vtest_sample_bpt}. As shown in the figure, test models are selected from both high density and low-density regions with respect to the BPT locations of all held-out models. 
In Figure~\ref{fig:vtest_sample_parameter}, grey histograms represent the distribution of parameters for the selected test sample, whereas the red dashed lines show the distributions of our entire database, which are the same as in Figure~\ref{fig:distr_param}. As we set selection criteria, some parameters have slightly different distributions to those of the database.

As indicated above, for our 100 test models, we draw 100 posterior samples per model, giving us a total of 10,000 sets of parameters (\mcl, SFE, \hdenini, and age). We then re-run WARPFIELD-EMP for each set of values, giving us 100 sets of emission line luminosities for each of our test models. For each test model, we then investigate whether the luminosities derived from the posterior samples are equal to the input-condition luminosity (i.e., the true luminosity of each model). 
In Figure~\ref{fig:vtest_all}, we present a density histogram of the logarithmic difference between the luminosities from the posterior samples and the true luminosity values. All twelve emission lines used in our network reveal similar distributions in Figure~\ref{fig:vtest_all}. We quantify the accuracy and precision of the luminosities derived from the posterior samples with respect to the true luminosity values by calculating the first and the second moment of each distribution. In each panel in Figure~\ref{fig:vtest_all}, the vertical dashed orange line and the horizontal green bar represent the first moment and twice the second moment respectively, the legend indicating their corresponding values.
All 12 emission lines have very small first moment values. The smallest absolute value is $\sim 2.65 \times 10^{-5}$~dex for the \sii\ 6716\AA\ distribution and the largest one is around $3\times10^{-3}$~dex in \oiii. The width of the distribution is denoted by the green bar, twice the second moments, and all twelve distributions have a similar width within the range of 0.2~dex $\sim$ 0.22~dex. As shown in Figure~\ref{fig:vtest_all}, the posterior samples predicted by the network have emission line luminosities very close to the true luminosities of the test models, although there is a $\pm$0.11~dex error, which is equivalent to a factor of 1.3. This demonstrates that the posteriors predicted by our network are reliable and correctly conditioned on the given observation apart from the result of parameter prediction.

\begin{figure}
	\includegraphics[width=1\columnwidth]{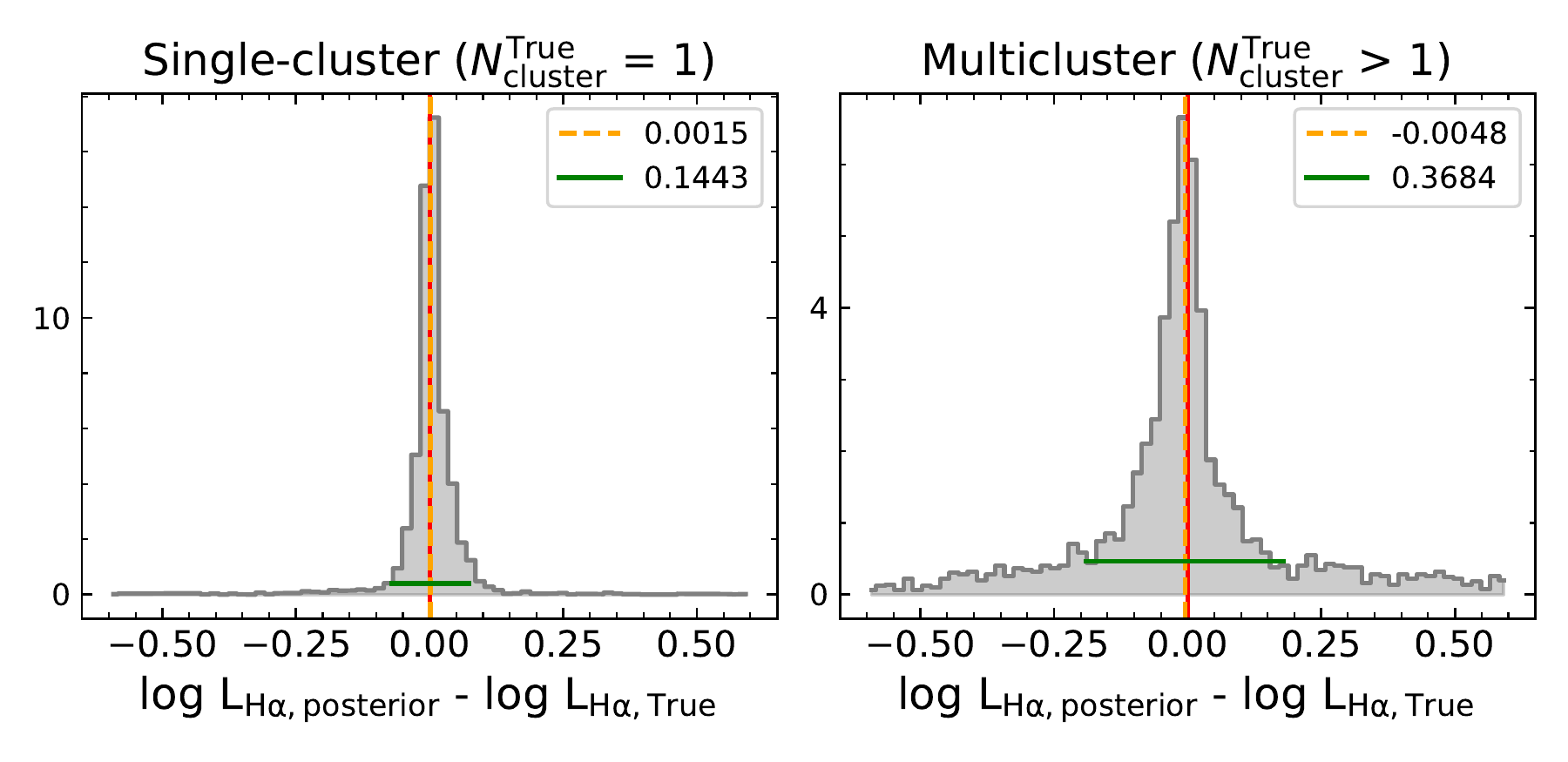}
    \caption{Density histograms of the logarithmic difference between the \ha\ re-simulated luminosity from the posterior samples and the true \ha\ luminosity of the test models. We divide the 100 test models into two groups depending on the true value of \ncl. 
    The left panel shows the result of 6,000 posterior samples from the 60 test models, which have only one cluster and the right panel shows the result of 4,000 posterior samples from the 40 test models that contain more than one cluster.
    The colours and lines are the same as in Figure~\ref{fig:vtest_all}.
    } \label{fig:vtest_comparison1}
\end{figure}

The performance of the network varies depending on the characteristic of the observed target that we want to analyse. For example, there is a large scatter or degeneracy in parameter predictions especially when the target \hii\ region is very old or has many star clusters in the cloud. Since we confirmed in Figure~\ref{fig:vtest_all} that the distribution of the luminosity difference is similar regardless of the emission line, we now select the \ha\ as the representative emission line and investigate how the luminosity distribution shifted with respect to the true luminosity differs according to the characteristics of the conditioned test models.
First, we compare the case when the selected test object actually has only one star cluster (i.e., a single-cluster model) with the case when the object has two or more star clusters (i.e., a multicluster model). As mentioned above, 40 of our 100 validation test models are single-cluster models, whereas the remaining 60 are multicluster models. Figure~\ref{fig:vtest_comparison1} shows the distribution of the \ha\ luminosity difference between the predicted posterior samples and conditioned test models of each case. 
The distribution of single-cluster models (left panel) exhibits a better result considering the smaller offset and width compared to the multicluster models (right panel). The offset of multicluster models ($\sim$ 0.005~dex) is slightly larger than that of single-cluster models or that of all models combined in Figure~\ref{fig:vtest_all}, but it is still a very small value.
Compared to the \ha\ distribution of entire models in Figure~\ref{fig:vtest_all}, the width of the single-cluster case is narrower by 0.06~dex, whereas the width of multicluster case is wider by $\sim$0.16~dex. 
It is expected that our network performs better for single-cluster models because, in the previous section, we find more accurate parameter prediction for single-cluster models. Figure~\ref{fig:vtest_comparison1} reflects that even when our network shows relatively poor performance for multi-clusters, our network provides reliable predictions with the luminosity difference of -0.0048$\pm$0.1842~dex on average with respect to the conditioned observations.

Additionally, we divide the test samples according to various other features and compare the distribution of the \ha\ luminosity difference. For example, we split the samples according to the cloud age, where the cloud age of the first group is less than 10~Myr and that of the other group is older than 10~Myr. We confirm that the distribution of the \ha\ luminosity difference for the younger group has a better result than that of the older group. When the observed \hii\ region is older than 10~Myr, the luminosity difference of the posterior samples with respect to the true luminosity value is -0.0001$\pm$0.157~dex on average. 
On the other hand, we divide samples into bright and dark models, where bright models have luminosities larger than $10^{34}$~erg/s in all twelve emission lines. This criterion serves to probe the influence of uncommon characteristics in the observation space because more than 60\%\ of our training data is classified as bright models. We confirm that bright models have a luminosity error of 0.003$\pm$0.086~dex, while dark models have an average luminosity difference of -0.01$\pm$0.183. In conclusion, the performance of our network varies depending on the characteristics of the target we want to analyse. However, even in the poorly performing cases, the re-simulated emission line luminosity of the posterior samples predicted by our network is very similar to the conditioned luminosity with an average error of less than $\pm$0.2~dex.

\section{Degenerate Prediction}
\label{sec:degenerate}

The shapes of the 1D posterior distributions as shown in Figure~\ref{fig:showcase_hist} mainly have the following three characteristics. First,  a unimodal distribution is common but some posterior distributions have two or more distinguishable modes. Second, for a given observation, the number of modes in the posterior distribution is different for each parameter. Third, as mentioned in Section~\ref{subsec:showcase}, the multi-modality in the posterior distribution of \mcl, star formation efficiency, \hdenini, cloud age, or the age of the youngest cluster is often due to the degeneracy in the \ncl\ prediction.
In this section, we perform a statistical analysis of the modality of the posterior distribution. Rather than interpreting physical meanings of the degeneracy of the network prediction, we focus on the shape of the 1D posterior distributions from a statistical point of view, such as how many modes are in the posterior distribution in general, and how the number of modes differs depending on the parameters.

\begin{figure*}
	\includegraphics[width=2\columnwidth]{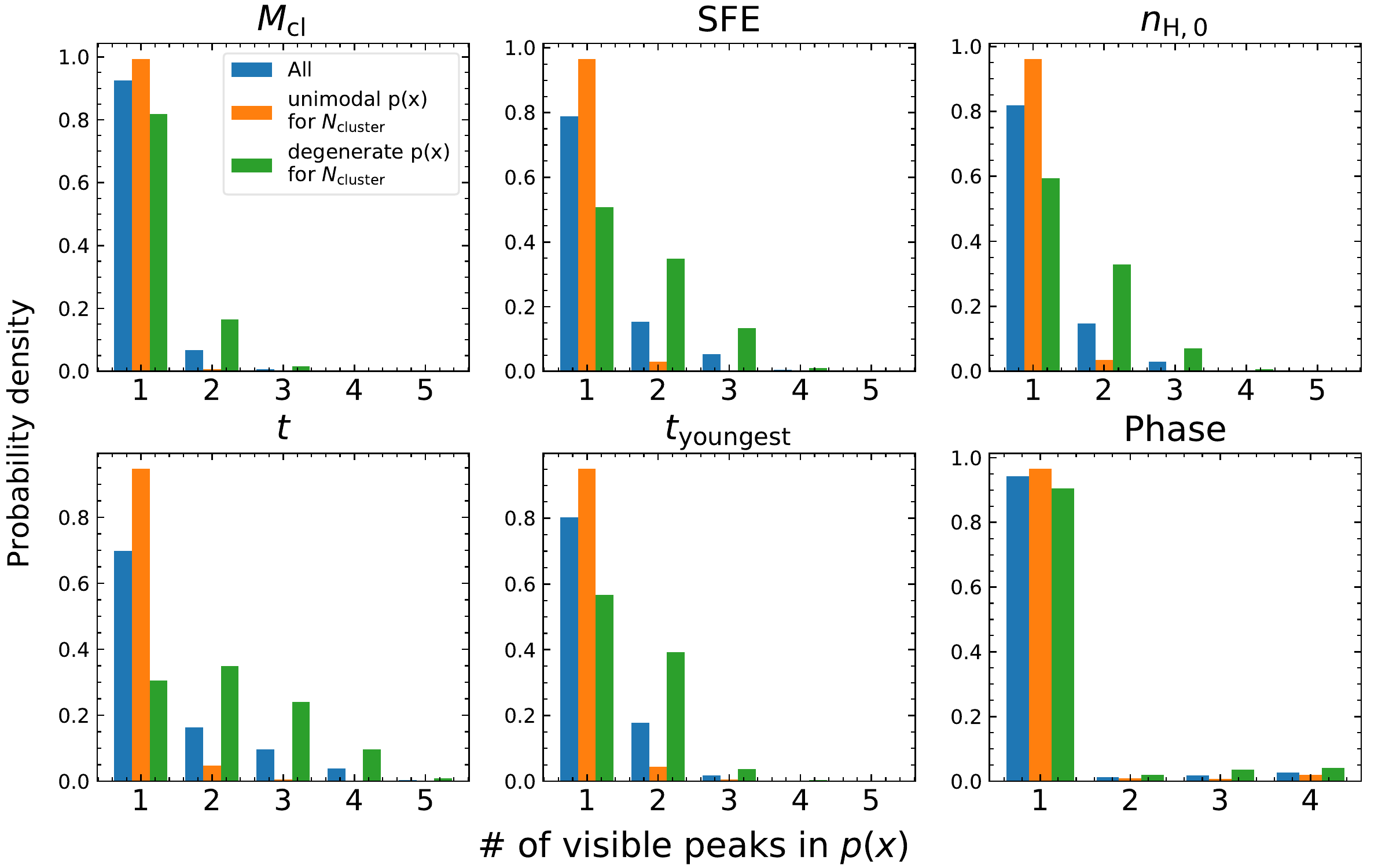}
    \caption{Density histograms of the number of visible peaks in the posterior distributions for six parameters. The blue histogram shows the results for all models in the test set. The orange histogram represents models whose posterior distribution of \ncl\ are unimodal (61.4\%\ of test set), whereas the green histogram represents the rest of models with multimodal posterior distribution in \ncl.} 
    \label{fig:vpeak_hist}
\end{figure*}

\subsection{Method of counting the number of modes}

For discretized parameters such as \ncl\ and phase, we simply measure the number of modes in the posterior distribution by rounding the posteriors and count the number of unique values.
For non-discretized parameters, we analyse the number of modes in the posterior distribution in 3 steps. First, we bin the posteriors of each parameter to make a probability distribution. Next, we fit the posterior distribution with multiple Gaussians and in the last step, we measure the number of modes by counting the number of peaks in the posterior distribution according to three different definitions of peaks. The reasoning behind the binning method in the first step is that different binning can affect the fitting result.

We sample the posterior 2048 times per model in the test set and divide the predicted posteriors into bins with regular intervals to obtain a posterior probability distribution for the non-discretized parameters. We first bin with a number equal to the square root of the number of posterior samples. If it fails to fit the posterior distribution, we bin again according to the Freedman-Diaconis rule (i.e., FD rule: \citealt{Freedman&Diaconis1981}) and retry the fitting. The FD rule takes into account the interquartile range (IQR) of the data ($x$) to determine the binning width ($h$):
\begin{equation}
    \begin{split}
        h &= 2 \frac{\mathrm{IQR}(x)}{^{3}\sqrt{n}}, \\
        n_{\mathrm{bin}} &= \frac{\mathrm{max}(x)-\mathrm{min}(x)}{h},
    \end{split}
\end{equation}
where $n$ is the number of data.

We start to fit the posterior distribution with one Gaussian model and increase the number of Gaussian components if the integral of root mean square of the residual is larger than 3\%\ of the full integral of the posterior distribution. The maximum number of Gaussian components per posterior distribution is limited to six. For each model, we fit the posterior distribution for the five parameters (\mcl, SFE, \hdenini, $t$, and $t_{\rm youngest}$) independently.

In order to determine the number of modes in one posterior distribution, we employ three different definitions for the peak of the mode and count the number of peaks according to each definition. First, we treat the number of Gaussian components fitted in the posterior distribution as the number of fitted peaks. In the second definition, we determine the number of visible peaks, which are defined as the local maxima of the fitted function. We use the fitted curve and the first derivative of it to find visible peaks. For the third one, we compute the number of separated peaks, where "separated" means that the distance between the centres of two Gaussian components is larger than the L2 norm of the two dispersions. If the distance between the centres is smaller than the L2 norm, we regard the pair of Gaussian components as part of one extended peak. In the previous two definitions, the position of the peak is accurately identified on the fitted posterior distribution, whereas the extended peak does not indicate a certain position but provides the range of possible peak positions.

\subsection{Visible modes in posterior distributions}

For the entire 101,149 test models, we count the number of peaks in the posterior distribution following the above steps and according to the three different definitions. As mentioned, we simply count the number of peaks for \ncl\ and phase, so the number of peaks for these two parameters does not depend on the definition of the peak.
In this section, we only discuss the number of visible peaks (i.e., $n_{\mathrm{vpeak}}$), which is the most conservative and straightforward among the three definitions. We provide an investigation of the other two approaches in Appendix \ref{appendix-B}. The blue histogram in Figure~\ref{fig:vpeak_hist} represents the distribution of the number of visible peaks in the posterior distributions. In the previous section (Section~\ref{subsubsec:showcase_deg}), we have mentioned that degenerate predictions (i.e., multi-modality in the posterior distribution) in the non-discretized five parameters are attributed to the degeneracy in the \ncl\ prediction. So, we divide the test set into two groups: one with a unimodal \ncl\ posterior distribution (orange histogram in Figure~\ref{fig:vpeak_hist}) and the other one with multimodal \ncl\ posterior distribution (green histogram). Please note that the criterion for dividing models into two groups is not based on the true \ncl\ value or the posterior value of \ncl, but the number of modes in the posterior distribution of \ncl.

From Figure~\ref{fig:vpeak_hist}, we find that the distribution of $n_{\mathrm{vpeak}}$ differs for each parameter. However, in all parameters, about 95\%\ of models have one visible peak if the posterior distribution of \ncl\ is not degenerate. This shows that the multi-modality in the posterior distribution of \ncl\ obviously affects the number of visible peaks in the posterior distribution of other parameters. However, the remaining 5\%\ of the models whose posterior distribution is non-degenerate for \ncl\ but degenerate for other parameters implies that the number of clusters is not the only cause of degeneracy.
As expected from the results in the previous section, the posterior distribution of the phase is unimodal in most cases. The fraction of models with only one visible peak slightly decreases if the \ncl\ posterior distribution is degenerate but it is still larger than 90\%. We confirm that degeneracy rarely occurs in phase posteriors so we focus more on the result of the other five parameters.

Among the five parameters, the fraction of models with one visible peak is the highest in \mcl\ and the lowest in age.
First, in the case of \mcl, more than 90\%\ of all models have only one visible peak. When the posterior distribution of \ncl\ is degenerate, about 20\%\ of models have two or more visible peaks, which is smaller compared to those of the other parameters.
In the case of age, about 70\%\ of models have unimodal posterior distributions. If the \ncl\ posterior distribution is not degenerate, more than 90\%\ of models have one visible peak. On the other hand, if the \ncl\ prediction is degenerate, 70\%\ of models have two or more visible peaks, which implies that multi-modality of the age posterior distribution based on the visible peak is significantly influenced by the degeneracy in the \ncl\ posterior distribution. 

For all parameters, it is more probable to have multiple visible peaks in the posterior distribution if the posterior distribution of \ncl\ is degenerate.
Even when the \ncl\ prediction is degenerate, the number of visible peaks is two in most cases. However, in the case of age, half of the models with multiple visible peaks have three or more peaks. This means that, when the \ncl\ prediction is degenerate, the age posterior distributions have more distinct and not blended multi-modes compared to other parameters.

In conclusion, the \ncl\ prediction is not the only source of degeneracy in the posterior but it is the most influential one. As mentioned in the previous section (Section~\ref{subsec:overall performance}), it is difficult to break out the degeneracy remaining in the \ncl\ posterior because the information of the number of clusters is not well reflected in the selected 12 emission lines used for the input. 
The amounts of emissions of these lines are contributed mostly by the youngest cluster with bright O and B types stars and they do not depend much on the other old stellar generations.
These results suggest that we need an additional observable which is sensitive to the number of clusters such as the high-resolution photometry observations obtained with the Hubble Space Telescope (HST) to enhance the performance of our network by breaking out the degeneracy aroused from \ncl\ prediction.

\section{Posterior distributions considering luminosity errors} 
\label{sec:luminosity error}

So far, we presented posteriors conditioned on the luminosity of 12 emission lines, but we did not take into account the fact that in real observations, these luminosities cannot be measured with arbitrarily high precision. Here, we explain how to obtain the posterior distribution by taking into account the luminosity error, and how the posterior distribution changes depending on the amount of the error of observations.

In this study, we use a Monte Carlo approach to sample posteriors while accounting for luminosity errors arising from  different signal to noise levels of the observations. Suppose that one set of luminosities (\textbf{y}) has a 1$\sigma$ error in percent unit for each of the 12 emission lines we consider ($\sigma_{1}$, ...,  $\sigma_{12}$). The first step is to generate a sufficient number $N_{\mathrm{MC}}$ of mock luminosity sets ($\textbf{y}'_{1}, ...,\textbf{y}'_{N_{\mathrm{MC}}} $) by adding a small amount of random noise to each emission line luminosity considering the corresponding error. The second step is to sample the posterior $N_{\mathrm{cINN}}$ times conditioned on each mock luminosity set so that we draw $N_{\mathrm{MC}} \times N_{\mathrm{cINN}}$ posterior samples in total. We found that producing a sufficient number of mock luminosity sets ($N_{\mathrm{MC}}$) at least around 3000$\sim$5000 is important to obtain a smooth posterior distribution. In this study, we decided to sample the posterior 300,000 times in total for one observation by making 3000 mock luminosity sets and sampling the posterior 100 times per mock luminosity set. However, our network sometimes returns extremely extrapolated posterior samples that are physically incorrect (e.g., negative age or negative star formation efficiency) especially when the luminosity errors are large, and so we exclude all physically unrealistic posterior samples.

As we use synthetic \hii\ region models, which by definition are fully accurate, we add a simple noise model to make mock errors of the emission line luminosity. In real observations, we have to consider errors from diverse sources such as the Poisson error, calibration error, or systematic errors, e.g., arising from incomplete sky subtraction. However, in this study, we ignore error sources other than Poisson noise. We also ignore any covariances between different emission lines including the physical relations between blended lines and their individual components. When the 1$\sigma$ percent error of the brightest emission line of one observation is given, the error of the other 11 emission lines are automatically determined according to
\begin{equation} 
    \label{eq:luminosity noise model}
    \sigma_{\mathrm{line}} = \sigma_{\mathrm{min}} \times \sqrt{\frac{L_{\mathrm{brightest}}}{L_{\mathrm{line}}}},
\end{equation}
where $\sigma_{\mathrm{min}}$ is the error of the brightest emission line, which is the smallest among the luminosity errors of the 12 emission lines. In real observations, this is typically the \ha\ line.

\subsection{Statistical analysis}

\begin{figure*}
	\includegraphics[width=2\columnwidth]{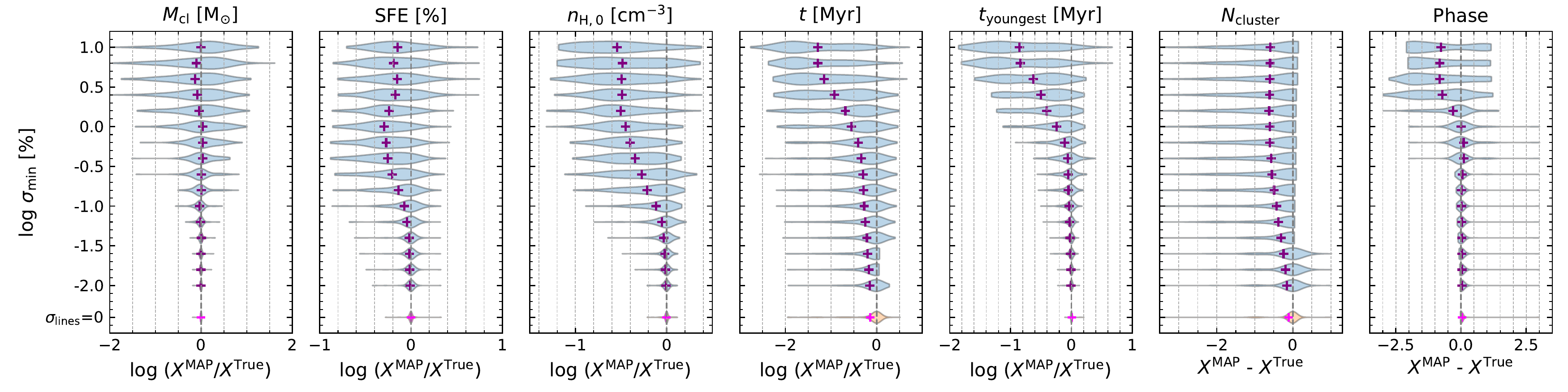}
	\includegraphics[width=2\columnwidth]{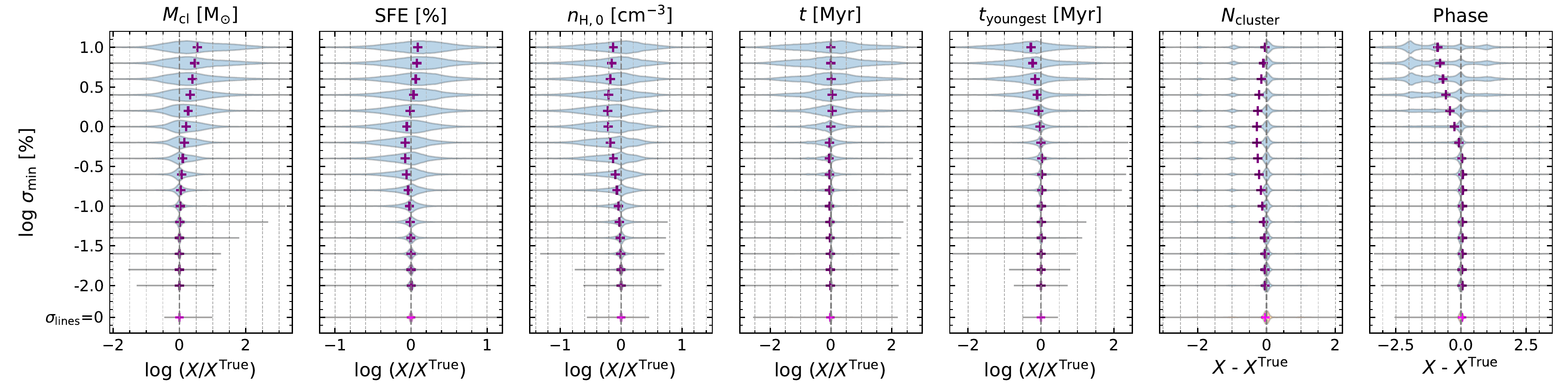}
    \caption{Using 1600 posterior distributions (100 test models and 16 different luminosity errors of the brightest emission line for each model), we present histograms of the logarithmic difference between MAP estimates and true values of each parameter (blue histograms in the first row). Please note that for \ncl\ and phase, we use difference in linear scale. The yellow histogram at the bottom in each panel is the distribution obtained from the posterior without luminosity error. The cross symbol in each histogram represents the average of the distribution. In the same manner, the lower panels show distributions of the logarithmic difference between all posterior estimates and true values instead of MAP estimates.
     } \label{fig:violin}
\end{figure*}

\begin{figure*}
	\includegraphics[width=2\columnwidth]{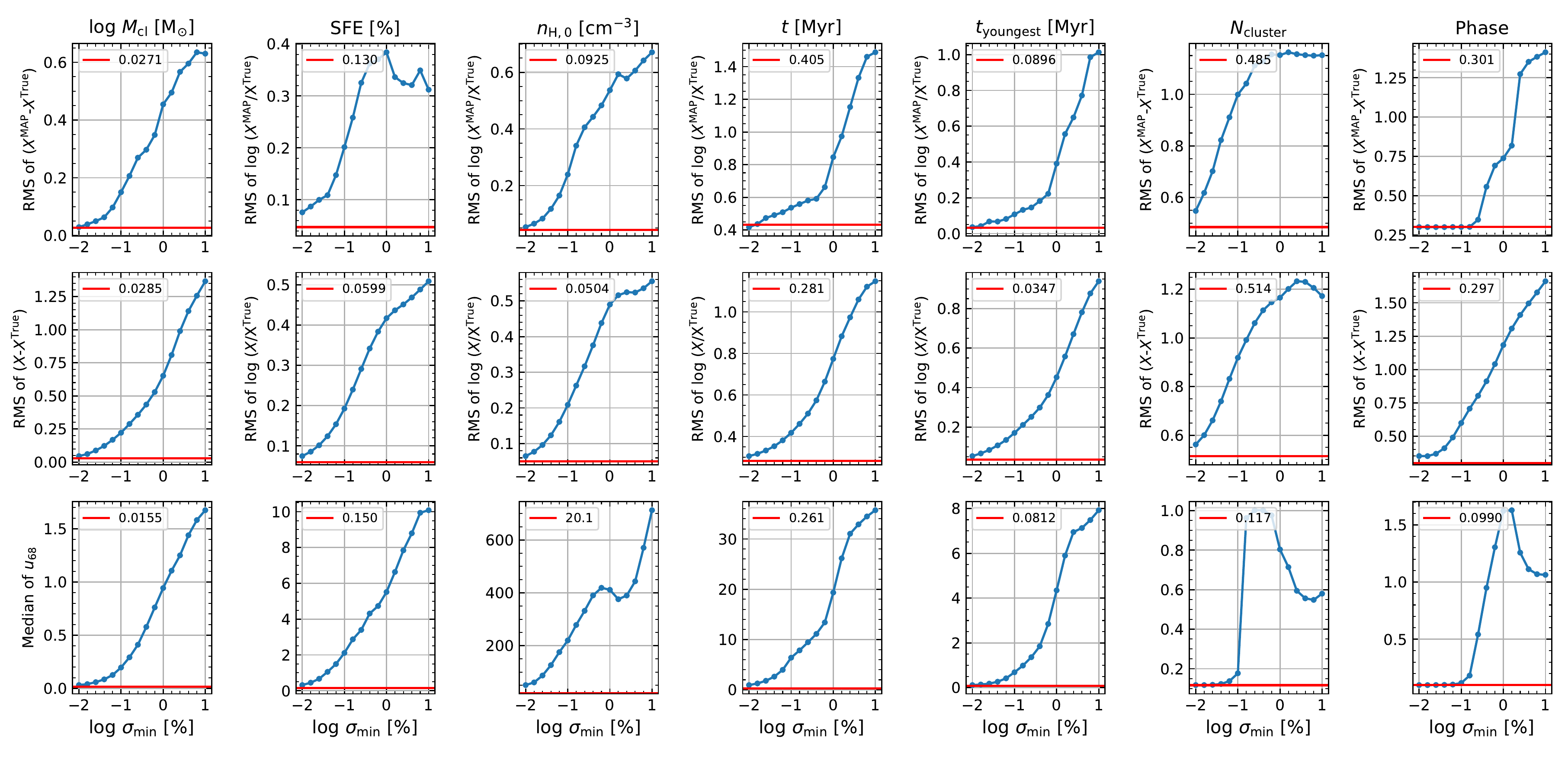}
    \caption{Accuracy and precision of our network as a function of the luminosity error of the brightest emission line ($\sigma_{\mathrm{min}}$) using 100 test models. Please note that the luminosity error is given in percent. We use the RMS of MAP accuracy in the first row and the RMS of accuracy using all posterior estimates in the second row. 
    The last row shows the median uncertainty at 68\%\ confidence interval ($u_{68}$) which represents the precision of our network. 
    The red horizontal lines and corresponding values on the upper left corner in each panel indicate the value obtained from posterior distributions without luminosity error ($\sigma_{\mathrm{lines}}=0$). 
     } \label{fig:rmse_sigma}
\end{figure*}

We expect the posterior distribution to change depending on the characteristics of the observed target as well as the magnitude of the luminosity error. So, we utilize the same test sample of 100 models used for network validation in Section~\ref{sec:validation}. For each model, we obtain 16 posterior distributions for each parameter by varying the luminosity error of the brightest emission line in a logarithmic scale from 0.01\%\ to 10\%\ with 0.2~dex intervals. As mentioned above, we draw 300,000 posterior samples to construct the full posterior distribution for each of these 1600 cases.

To evaluate the accuracy and precision of the posterior, we use the MAP estimate and uncertainty at a 68\%\ confidence interval ($u_{68}$) in the same way as we did to evaluate the posterior without considering luminosity errors (i.e., $p_{0}(x)$).
In this section, we take the logarithm of the ratio between the posterior estimate and the true value (i.e., log deviation, $\mathrm{log}\frac{X}{X^{*}}$) as a proxy of the network accuracy. We define accuracy in two ways either by using only the MAP estimate as a representative value of the posterior distribution (i.e., MAP accuracy), or by using all of the posterior estimates. However, in the case of the \ncl\ and phase, we use a linear deviation (i.e., $X-X^{*}$) instead. The precision of the posterior distribution is represented by $u_{68}$.

Using the 1600 posterior distributions, we first examine how the accuracy of the posterior depends on the luminosity error. The violin plots in Figure~\ref{fig:violin} show how the accuracy distribution changes with increasing minimum luminosity error. We present the accuracy using only MAP estimates in the first row of Figure~\ref{fig:violin} and accuracy using all posterior estimates in the second row. Blue histograms in the panel show the accuracy of the posterior distribution for 100 models with the same minimum luminosity error. Please note that we apply the same luminosity error only for the brightest emission line so each model has different luminosity errors for the other emission lines. The yellow histogram located at the bottom of each panel represents the accuracy of posterior distributions without luminosity errors ($p_{0}(x)$). In order to maintain the same sampling size, we draw 300,000 posterior samples per test model to make a posterior distribution without luminosity errors. The average of each distribution in the figure is denoted by a cross symbol.

From Figure~\ref{fig:violin}, we find that the accuracy distribution using all posterior estimates and that using only MAP estimates are significantly different.
First, in the upper panels, the MAP accuracy distributions are on average shifted to negative values, except for \mcl, and the magnitude of the shift increases with increasing luminosity error. This feature is especially evident in the distribution of \hdenini, cloud age, and the youngest cluster age. The star formation efficiency exhibits a similar trend, but when the luminosity error ($\sigma_{\mathrm{min}}$) is within a range of 1$\sim$10\%\ the distributions become less biased in comparison to those with smaller luminosity error. 
On the other hand, when using all posterior estimates, the accuracy distributions show different trends. In the case of \mcl, the distribution gradually shifts to the positive direction on average as the luminosity error increases ($\sim$0.1~dex). The distributions of star formation efficiency, \hdenini, cloud age, and the youngest cluster age become wider but do not shift to one side even when the luminosity error is large. The density distribution shows a negative shift around 0.2 dex but this is smaller than the amount of shift found in the MAP accuracy distributions. This implies that the posterior distributions of these four parameters are skewed towards the range below the true values, but have a long tail-like distribution with a lower probability in the range above the true values so that the distributions are not shifted on average. In particular, it is expected that density, cloud age, and the youngest cluster age will clearly show this characteristic when the minimum luminosity error is 1\%\ or more. In the case of \mcl\ showing the opposite trend, it is expected that a large number of posterior estimates have larger values than the true one from the positive shifts in accuracy distributions. From the distribution of MAP accuracy, we infer that the posteriors of \mcl\ have multi-modal or spiky distributions so that the peak of the distribution does not have a systematic bias to a certain direction on average.

To measure the average accuracy as a function of the minimum luminosity error, we calculate the root mean square (RMS) value of the accuracy shown in Figure~\ref{fig:violin}. The first and second row in Figure~\ref{fig:rmse_sigma} present the RMS of the MAP accuracy and the RMS of the accuracy using all posteriors estimates, respectively. Additionally, in the third row, we examine the median of the $u_{68}$ values from 100 models as a function of the minimum luminosity error to quantify the average width of the posterior distribution. The red horizontal line in each panel represents the values from the posterior without luminosity errors ($p_{0}(x)$). 
Figure~\ref{fig:rmse_sigma} quantitatively shows how much our network performance decreases with increasing luminosity error.
In the case of \mcl, star formation efficiency, and \hdenini, the RMS of MAP accuracy gradually increases up to 0.4$\sim$0.6~dex. The RMS value of the star formation efficiency slightly decreases when the minimum luminosity error is larger than 1\%, which is also confirmed in Figure~\ref{fig:violin}.
In the case of cloud age and the youngest cluster age, the slope of the RMS curve becomes larger around a luminosity error of 1\%. The RMS value of the youngest cluster age is smaller than other parameters when the minimum luminosity error is smaller than 1\%, but thereafter the RMS value rapidly increases to 1 dex. This indicates that the network has more difficulty in finding accurate MAP estimates of the age and age of the youngest cluster than other parameters when the minimum luminosity error is larger than 1\%.
In the case of the number of clusters, the RMS value increases steadily up to the minimum luminosity error of 1\%, and thereafter, it is almost constant. In the case of phase, on the contrary, there is no change up to the minimum luminosity error of 0.1\%, but after that, the RMS value increases significantly.

There is no significant difference in the RMS curve when the whole posterior estimates are considered (second row in Figure~\ref{fig:rmse_sigma}) or when only the MAP estimates are used (first row in Figure~\ref{fig:rmse_sigma}). Although the increment is different, the RMS values of the seven parameters gradually increase with increasing luminosity error. This represents the increase in the overall width of the distribution shown in Figure~\ref{fig:violin}. The median of $u_{68}$ also shows the change in the width of the posterior distribution. The unit of the $u_{68}$ in the third row of Figure~\ref{fig:rmse_sigma} is the same as the physical unit of the corresponding parameter. Compared to the width of the posterior distribution without luminosity errors, $p_{0}(x)$, represented by red lines, the posterior distributions become much wider with increasing luminosity error. When the minimum luminosity error is around 10\%\, the average width of the posterior distribution is almost the same as the entire parameter range of the training data (see Figure~\ref{fig:distr_param} to compare with the parameter ranges of the training data).

\subsection{Change of the posterior distribution for individual models}

\begin{figure*}
	\includegraphics[width=2.1\columnwidth]{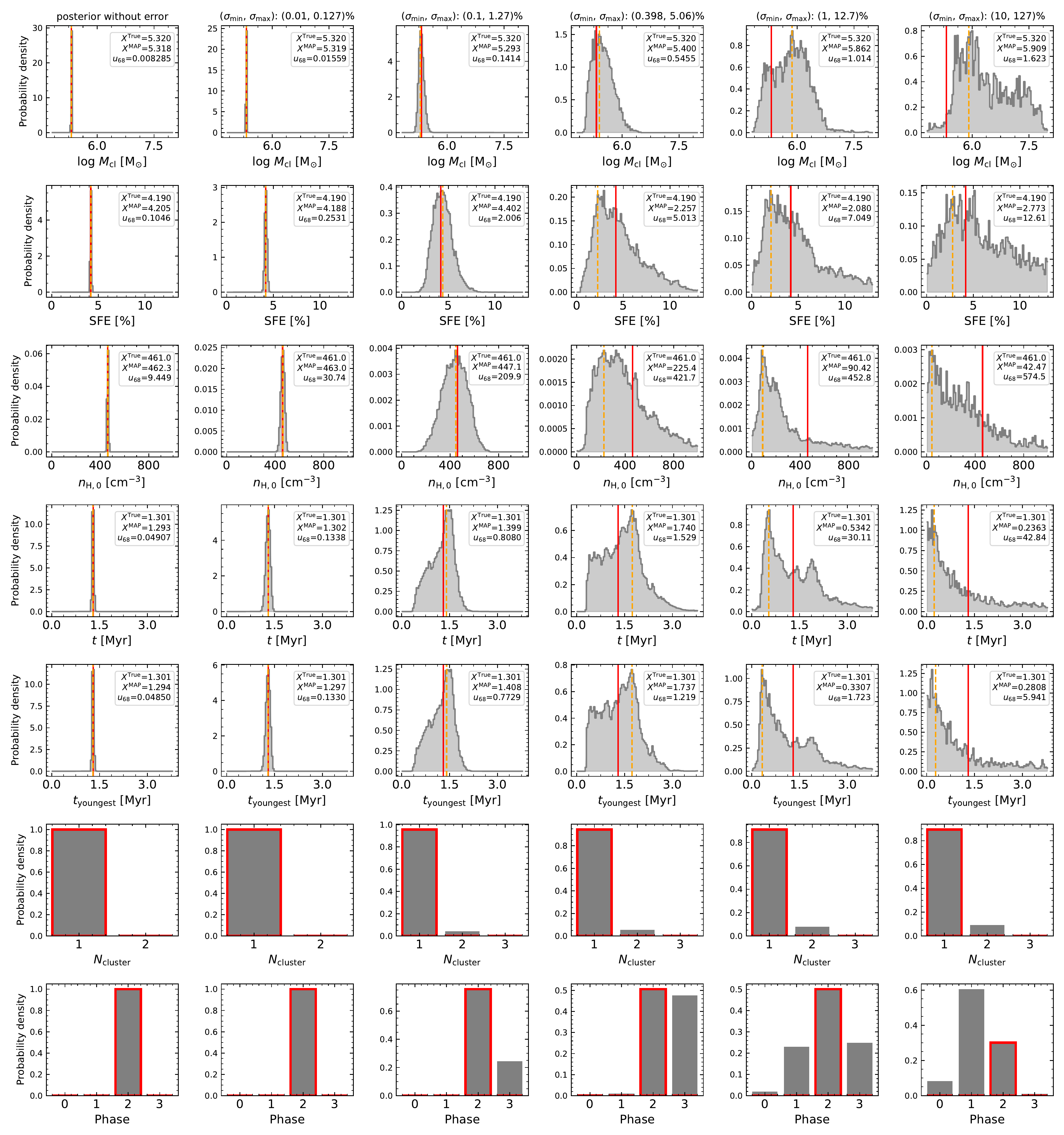} 
    \caption{Posterior probability distributions (grey) of the first example model. This model is the same as the model in the first column of Figure~\ref{fig:showcase_hist}. The first column shows the posterior distribution when no luminosity error is applied. From the second column to the last column, we present the posterior distribution with luminosity errors of the brightest emission-line of 0.01, 0.1, 10$^{-0.4}$, 1, and 10\%\, respectively. The luminosity errors of the brightest and faintest emission lines are indicated at the top of each column. Red vertical lines or red edges in the bar-shaped histograms indicate the true values of the model. On the upper right corner of each panel, we present true values, MAP estimates and their 1$\sigma$ uncertainties, and u$_{68}$ values of each posterior distribution. 
    } \label{fig:post_lerr1}
\end{figure*}

\begin{figure*}
	\includegraphics[width=2.1\columnwidth]{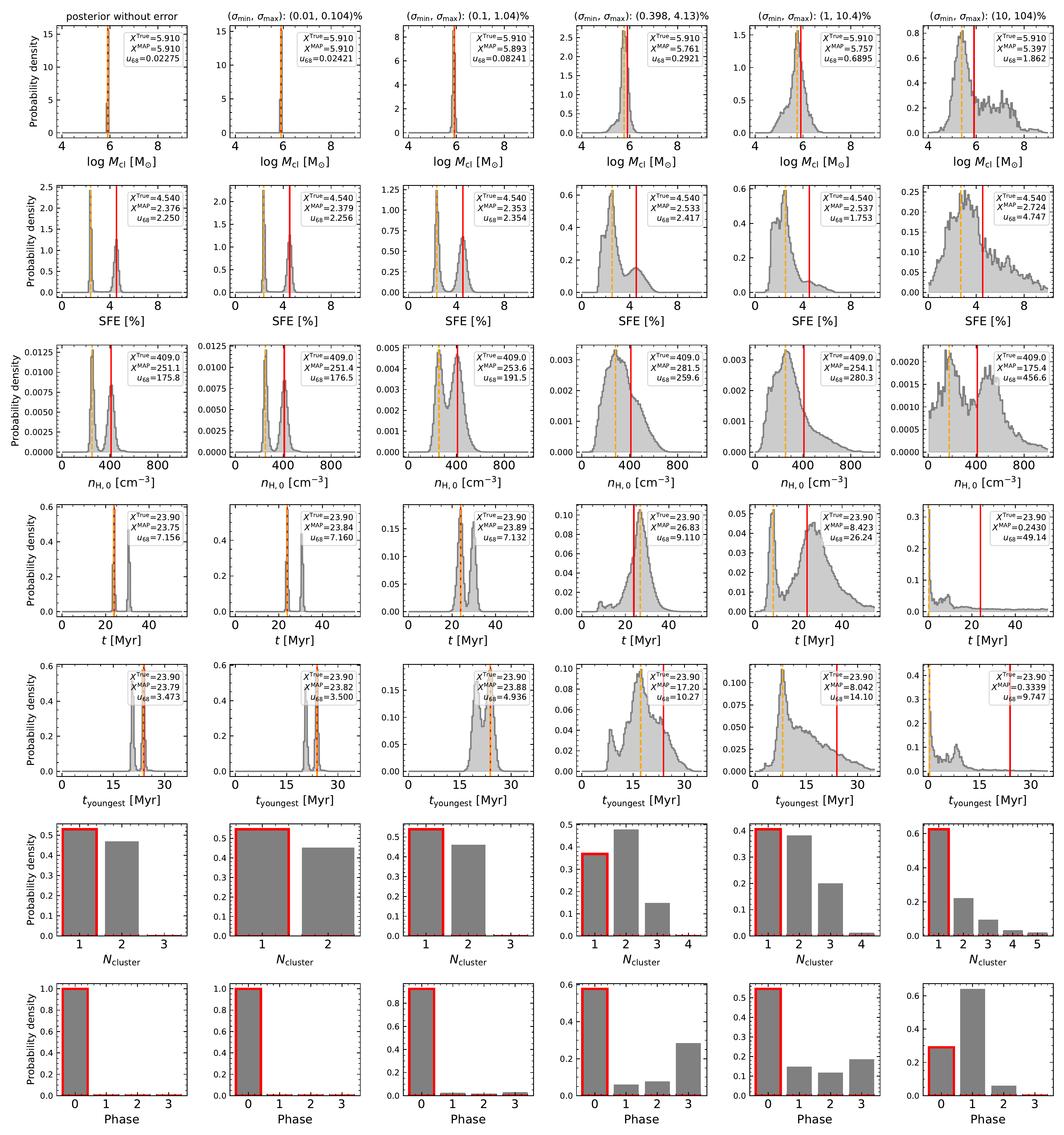}
    \caption{Posterior distributions of the second example model corresponding to the model in the third column of Figure~\ref{fig:showcase_hist}. Colour codes and lines are the same as in Figure~\ref{fig:post_lerr1}.
     } \label{fig:post_lerr2}
\end{figure*}

\begin{figure*}
	\includegraphics[width=2.1\columnwidth]{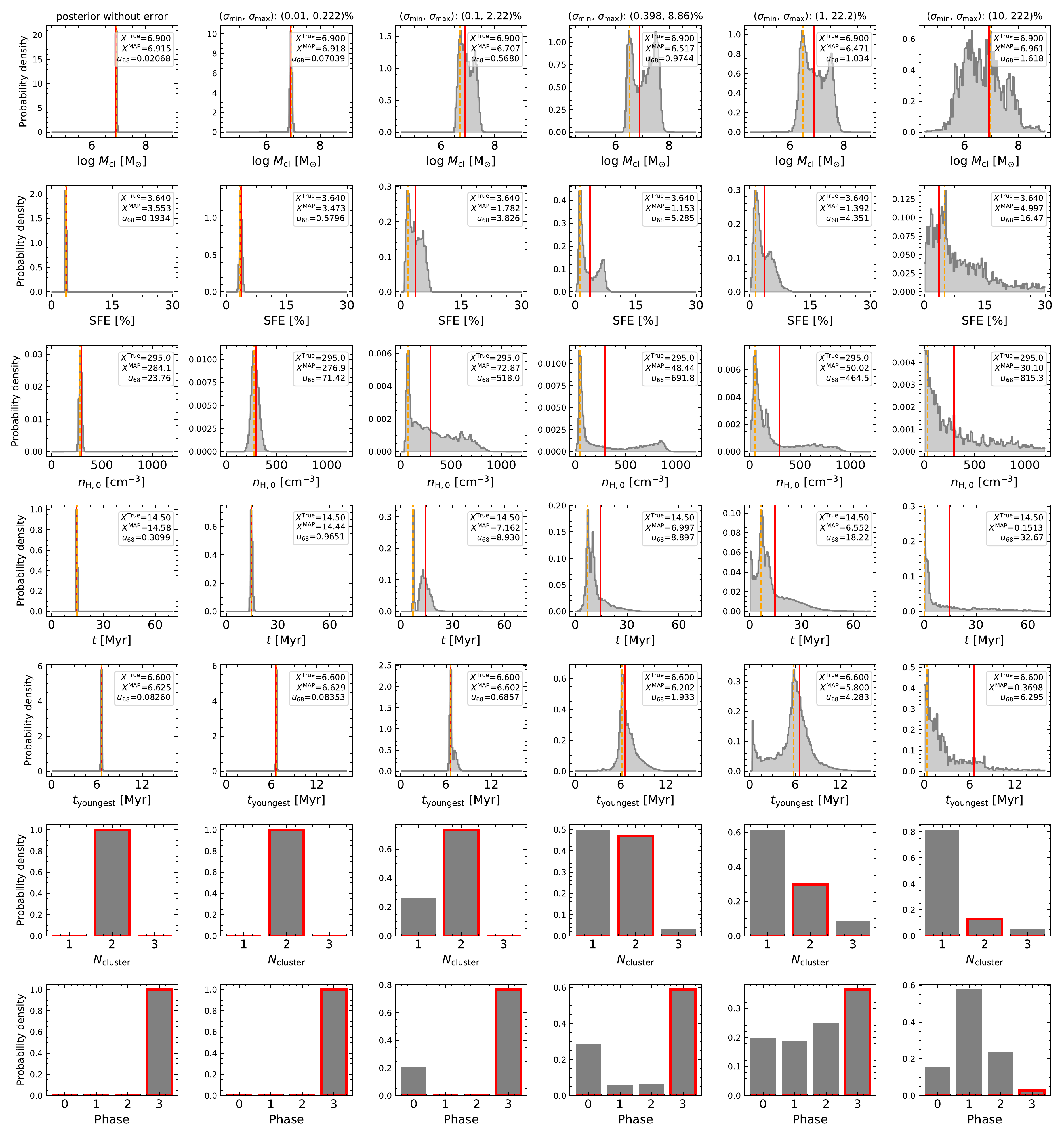}  
    \caption{Posterior distributions of the third example model. Colour codes and lines are the same as in Figure~\ref{fig:post_lerr1}.
     } \label{fig:post_lerr3}
\end{figure*}

In order to investigate the characteristics of the posterior distribution expected from Figure~\ref{fig:violin} and to examine the change of the posterior distribution depending on the luminosity error in detail, we select three from the 100 test models as examples. For these three models, we show the posterior distribution of each parameter for five different minimum luminosity errors (0.01, 0.1, 10$^{-0.4}$, 1, and 10\%, respectively) as well as the posterior distribution without luminosity error in Figures~\ref{fig:post_lerr1},~\ref{fig:post_lerr2}, and ~\ref{fig:post_lerr3} respectively. 

In Figure~\ref{fig:post_lerr1}, we show the posterior distributions of the first model which is the same model shown in the first column of Figure~\ref{fig:showcase_hist}. When we do not apply luminosity errors (i.e., $p_{0}(x)$), this model has a common unimodal posterior distribution which is accurate and precise to the true value (the first column in Figure~\ref{fig:showcase_hist} and \ref{fig:post_lerr1}). The brightest emission line of this model is \ha\ and the faintest line is [\mbox{O\,\textsc{i}}] 6300\AA\ with a factor of 12.7 larger uncertainty.
The posterior distributions do not change much at a minimum luminosity error of 0.01\%. From the $u_{68}$ values of each distribution, we notice that the width of the distribution is 2-3 times wider than that of $p_{0}(x)$, but the width is still narrow enough.
Even when the minimum luminosity error increases to 0.1\%, the MAP value is close to the true value although the width of the posterior distribution considerably widens. However, if the minimum luminosity error is larger than 0.1\%, the peak of the distribution moves further away from the true value and the width widens.

In the case of density, cloud age, and the youngest cluster age, the posterior distributions at $\sigma_{\mathrm{min}}$ of 1\%\ and 10\%\ show a skewed shape. As we expected from Figure~\ref{fig:violin}, the distributions have MAP estimates smaller than the true value but exhibit a tail-like shape toward values larger than the true values. 
For cloud age and the youngest cluster age, the posterior distributions show blended multi-modality at a value of 0.1\%\ $\sigma_{\mathrm{min}}$ because they begin to have a number of younger age estimates. As $\sigma_{\mathrm{min}}$ increases the level of skewness of the posterior distribution increases as well. 
Similarly, in the case of \ncl\ and phase, the posterior distributions become more degenerate with increasing luminosity error. But, unlike \ncl, which maintains an accurate prediction even when the luminosity error increases to 10\%, the accuracy of the phase deteriorates as the luminosity error increases.

The second example model is the one in the third column of Figure~\ref{fig:showcase_hist}. 
For this model, the luminosity error of the faintest emission line, [\mbox{O\,\textsc{i}}] 6300\AA, is a factor of 10.4 larger than that of the brightest line, \ha. Even without luminosity error, the posterior of this model has a bimodal distribution arising from the degeneracy in the \ncl\ prediction. The two modes widen as the luminosity error increases and merge at $\sigma_{\mathrm{min}}\lesssim$ 1\%.
In the case of star formation efficiency and \hdenini, the MAP estimate does not change significantly even when the minimum luminosity error increases by 1\%. However, in the case of cloud age and the youngest cluster age, the MAP estimates at a $\sigma_{\mathrm{min}}$ of 1\%\ or larger are significantly different from those at smaller luminosity errors. As the two degenerate modes in $p_{0}$ merge together, the MAP moves to the middle position between the two modes. But as the minimum luminosity error increases by 1\%, the MAP moves to the peak of the newly formed mode located at a significantly younger age range. As the minimum luminosity error increases by 10\%, the MAP is shifted again toward a much younger range. The result that the posterior distribution changes to a skewed distribution is consistent with the result of the first example and the trend revealed in Figure~\ref{fig:violin}.

For the third example, we select a model that undergoes one recollapse phase and consequently contains two generations of stars, or in our terminology has two clusters. Before applying luminosity errors (the first column in Figure~\ref{fig:post_lerr3}), the posterior distribution of this model has similar characteristics to the first example, which has a unimodal distribution and is very close to the true value. The differences are that the third model has two clusters and older age than the first model. The overall change of posterior distributions with increasing luminosity error is similar to that of the other models as well. Especially at a minimum luminosity error of 10\%, the shape of the posterior distribution of cloud age and the youngest cluster age is similar in all three models. On the other hand, the posterior of \mcl\ and star formation efficiency shows a degenerate bimodal distribution at a $\sigma_{\mathrm{min}}$ of 0.1$\sim$1\%.

The interesting point of this model in Figure~\ref{fig:post_lerr3} is the change of the \ncl\ posterior distribution. From the minimum luminosity error of 0.1\%, the degeneracy appears in the \ncl\ prediction, and the fraction of posterior samples that predict an \ncl\ of 1 increases with increasing luminosity error. Finally, the posterior distribution at the largest luminosity error is almost the same as that of the other two models. However, this means that the network performed poorly compared to the other two cases because this model actually has two clusters. 
Thus, we examine the change of the posterior distribution of \ncl\ for other test models that have two or more clusters like this model. We notice that when the minimum luminosity error is large, around 1$\sim$10\%, the posteriors samples that estimate \ncl\ as 1 are dominant regardless of the intrinsic characteristics of the target. The MAP estimates of \ncl\ with a minimum luminosity error larger than 1\%\ are always 1 for all 100 test models. This means that in the previous two cases, the network seemed to predict the number of clusters reliably, but in fact, it just always provides a similar posterior distribution.

We can interpret the change of the posterior distributions of the cloud age, the youngest cluster age, and phase in the same manner. For example, the posterior distribution of phase at a 10\%\ minimum luminosity error is similar in all three models in that the fraction of phase 1 predictions is the highest. We check the MAP estimate of phase for 100 models and find that when the minimum luminosity error is larger than $\sim 3$\%, more than 85\%\ of models have a phase MAP value of 1. Even for models whose true phase is not 1, still more than 85\%\ of models have a phase MAP estimate of 1. 
This means that at least the peak of the phase posterior distribution is always similar regardless of the conditioned observations if the luminosity error is large enough.

We infer that the consistent posterior distribution when the luminosity error is large is influenced by the bias in our training data. As mentioned in Section~\ref{subsec:overall performance}, even when we do not consider luminosity errors, posterior distributions are frequently degenerate and inaccurate in the case of \hii\ regions that either are old or have more than one cluster. This is because the fraction of single clusters or young \hii\ regions is high in our training data so that our network is well trained and performs well for these kinds of models. We suppose that the bias of the training data also affects the posteriors with luminosity errors as well. Our network delivers good predictions for young and single cluster \hii\ regions but frequently provides degenerate solutions for old, multi-cluster \hii\ regions which include young, single cluster posterior samples. If the luminosity errors are large, mock luminosity sets are more likely to encompass the luminosity of diverse \hii\ regions with various characteristics. Therefore the posteriors with a large luminosity error eventually have a higher proportion of younger and single cluster posterior samples. For this reason, we consider that the skewed posterior distributions in the cloud age, the youngest cluster age, and the \ncl\ are influenced by the characteristics of the training data. 

On the other hand, the posterior distribution of phase at a minimum luminosity error of 10\%\ is not similar to the phase distribution of our training data which has the highest fraction at phase 3. Phase posteriors without luminosity errors were not significantly influenced by the bias of the training data and were accurate most times regardless of the conditioned model. The reason why the fraction of phase 1 in the posterior distribution is high when the luminosity error is large is that very young \hii\ regions, younger than 1~Myr, are mostly in phase 1. As the fraction of posterior samples with a very young age increases when the luminosity error is large, the fraction of phase 1 predictions also increases.

To summarize, the posterior distribution gradually widens and shows a skewed shape with inaccurate MAP estimates as the luminosity error increases. However, the network guarantees the reliable MAP prediction at a $\sigma_{\mathrm{min}} \lesssim 0.1$\%\ or $< 1$\% depending on the parameters. This suggests an approximate minimum signal-to-noise level for the brightest emission line, which is typically the \ha\ line, to obtain reliable posterior samples when we apply our tool to the real observations. 
In this study, we ignore any covariance between different lines, e.g., a relation between a blended line and individual components, and produce mock luminosities as randomized as possible. We expect that our network will perform even better despite the same amount of uncertainties if we consider the covariances between lines because it reduces the randomness in mock luminosities.

\section{Discussion}
\label{sec:discussion}

\subsection{Major assumptions inherent in the training data}
\label{subsec:assumptions}
The WARPFIELD-EMP models we used as training data make several simplifying assumptions. In the previous sections, we demonstrated that the network learned the hidden rules in the training data well, but we did not validate our network using real observations beyond the comparisons already published in \citet{Pellegrini+20}. As our synthetic models are not a perfect representation of nature, we should consider the main assumptions of the WARPFIELD-EMP model in interpreting the posterior distribution if we apply the network to real observations.

One of the main approximations made within WARPFIELD is the assumption of spherical symmetry. The effects of small structures such as pillars, or larger three-dimensional inhomogeneities in the cloud are not taken into account in WARPFIELD's model for the evolution of the \hii\ region. Though it has limitations, this approximation dramatically reduces the computational cost of the calculation compared to a correspondingly detailed 3D simulation, allowing us to produce a large number of synthetic models applicable for machine learning techniques.

Another important assumption is that the clouds are isolated. In reality, some star-forming regions are close enough that their evolution may affect each other. In addition, even if nearby regions evolve independently, they may be blended together in observations that have limited spatial resolution. This is particularly an issue in optical surveys of \hii\ regions in nearby galaxies, as even the highest resolution examples of such surveys struggle to reach spatial resolutions better than $\sim 50$~pc in all but the closest galaxies. Our assumption that the clouds are isolated also implies that we do not account for external effects such as the influence of the large-scale galactic potential or contamination of the observations by emission from diffuse ionized gas.

The WARPFIELD models also make use of a highly idealized treatment of star formation.
All of the stars in each cluster are assumed to form instantaneously, i.e., there is no gradual star formation. Moreover, in cases where the cloud re-collapses and forms a new cluster, we assume that the star formation efficiency is the same as for the original cluster. This assumption is made purely on the grounds of simplicity and it is unclear how well it matches what happens in reality.

WARPFIELD also makes several assumptions that are commonly used in other feedback models. For example, we assume that the shell surrounding the inner bubble is in quasi-hydrostatic equilibrium. The WARPFIELD cloud is in virial equilibrium so that there is no partial gravitational collapse in the cloud. We also apply photoionization and chemical equilibrium to model the cluster and cloud evolution. The mass of stars in the cluster follows the Kroupa initial mass function~\citep{Kroupa01} with an upper stellar mass limit of 120~\msol. As the time-dependent evolution of the cluster is calculated from STARBURST99~\citep{Leitherer+99, Leitherer+14}, we accordingly accepted the physical assumptions used in it.

In addition to the above, we also made assumptions about several parameters when constructing the WARPFIELD-EMP database, resulting in additional constraints. As shown in Figure~\ref{fig:distr_param}, the range of each parameter in the training data is restricted. We confined the range of \mcl, SFE, and \hdenini\ when we randomly sampled the initial clouds and limited the maximum age of the cloud to 30~Myr. On top of these, we assumed a constant radial profile for the initial cloud density and did not account for the effects of the magnetic field and turbulent pressure. As the network is trained on the models within the database, care must be taken when applying our network to objects whose conditions are significantly different from our database models.

\subsection{Effect of noise augmentation on the network performance}
\label{subsec:smoothing}

\renewcommand{\arraystretch}{1.25}
\begin{table*}
    \caption{Comparison of the network performance between two networks: the main network introduced in the paper to which we applied noise augmentation on \ncl\ and phase (Network 1), and the network trained without any noise augmentation (Network 2). The first value in each item shows the performance of Network 2 and the second value shows the performance of Network 1 which is the same value shown in Table~\ref{table:overview}.
    \label{table:smoothing}
    }
    \begin{tabular}{ l  c  c  c  c  c  c   c } 
        \hline
        \hline
        Performance measure & log \mcl\ & SFE & \hdenini\ & $t$ & $t_{\text{youngest}}$ & \ncl\ & Phase \\
        (Network 2 / Network 1)     & log [M$_{\odot}$] & [\%]	& [cm$^{-3}$]	& [Myr]	& [Myr]	\\
        \hline
        $e\mathrm{_{cal}^{med}}$ & 4.7 / 0.44 & 4.9 / 0.26 & 6.7 / 0.87 & 6.2 / 1.3 & 7.4 / 1.1 & 18 / 2.3 & 29 / 0.12 \\
        
        $u\mathrm{_{68}^{med}}$ ($\hat{x}$) & 0.056 / 0.028 & 0.32 / 0.095 & 0.48 / 0.17 & 0.58 / 0.018 & 0.035 / 0.0097 & 0.00088 / 0.15 & 0.00035 / 0.086 \\
        
        $u\mathrm{_{68}^{med}}$ (${x}$) & 0.031 / 0.015 & 0.74 / 0.22 & 54 / 19 & 4.9 / 0.15 & 0.23 / 0.064 & 0.00065 / 0.11 & 0.0004 / 0.1  \\
        
        RMSE ($\hat{x}$) & 0.093 / 0.082 & 0.35 / 0.25 & 0.46 / 0.28 & 0.81 / 0.56 & 0.16 / 0.12 & 0.98 / 0.64 & 0.22 / 0.17  \\
        
        RMSE ($x$) & 0.051 / 0.045 & 0.8 / 0.58 & 51 / 32 & 6.8 / 4.7 & 1.1 / 0.79 & 0.72 / 0.47 & 0.25 / 0.2  \\
    
        \hline
        \hline
    \end{tabular} 
\end{table*}

We mentioned in Section~\ref{subsubsec:preprocess} that smoothing out the discretized parameter distribution by adding a small amount of artificial noise not only helps the network converge in training, but also improves the prediction performance of the network. For this reason, we added Gaussian noise with a standard deviation of 0.05 to smooth out the distribution of \ncl\ and phase to train the main network introduced so far in this paper.
To assess the effect of the smoothing process on network performance, we compare the network used in this study with the network trained without smoothing the distribution of \ncl\ and phase. For convenience, we refer to the former network as Network 1 and refer to the latter network as Network 2. We trained Network 2 with the exact same settings as Network 1 except for smoothing.

To compare the performance of the two networks, we evaluate Network 2 in the same way as described in Section~\ref{subsec:training evaluation}. We sample posteriors for the entire test set by using Network 2 and measure MAP values for each posterior distribution. We first plot the predicted posteriors or MAP estimates against the true values in Figures \ref{fig:all_post_N2} and \ref{fig:all_map_N2}. The direct comparison with the results of Network 1 (Figures~\ref{fig:all_post} and \ref{fig:all_map}) confirms that the prediction performance is improved by applying smoothing. This holds not only for the two parameters to which we applied smoothing, but also for the other five parameters. 
Especially for the star formation efficiency, we find that the arrow-shaped structure is more prominent in Figure~\ref{fig:all_post_N2} than in Figure~\ref{fig:all_post}. We interpret this structure as the degeneracy revealed in the posterior distribution. The more conspicuous V-shaped structure shown in Figure~\ref{fig:all_post_N2} reflects that degeneracy is more frequently observed in the posterior distributions of Network 2 and it is also common in the overall star formation efficiency range.

To compare the two networks more quantitatively, we evaluate the performance of Network 2 in the same way as Network 1 and provide the result in  Table~\ref{table:smoothing}. The first entry in each column is the evaluation result of Network 2 and the second value after the slash is the result of Network 1, note that this last entry to the number provided in Table~\ref{table:overview}. This clearly shows that Network 1 with smoothing applied has significantly improved performance compared to Network 2 without smoothing.
Network 2 has larger median calibration errors in all parameters. Specifically, the calibration errors of \ncl\ and phase reach 18\%\ and 29\%, respectively, which are significantly larger than those of Network 1. Although the error of \ncl\ is the largest in Network 1 as well, it is less than 2.5\%.
From the $u^{\text{med}}_{68}$ representing the average width of posterior distributions, the smoothing process decreases the width of the posterior distribution for five parameters (\mcl, SFE, \hdenini, cloud age, and the youngest cluster age). The posterior distributions of \ncl\ and phase widen by smoothing but this is an expected result. By changing the quantized values to a continuous, wider distribution in training, the network learns to avoid narrow and delta-function like distributions and chooses wider but less degenerate posterior distributions. In addition, the increased width is still sufficiently small compared to the sampling interval of 1. 
Last, the RMSE values also demonstrate that the accuracy of Network 2 is inferior to that of Network 1.

These results prove that augmenting discontinuous parameter distributions with artificial noise can improve the overall performance of the network. By smoothing out the distribution of \ncl\ and phase, the accuracy of \ncl\ is improved, reducing the frequency of degeneracy in the predicted posterior distributions.

\section{Summary}
\label{sec:summary}
In this paper, we introduce the novel method of applying a conditional invertible neural network (cINN) to predict the fundamental physical parameters of \hii\ regions from spectral observations.
When solving the inverse problem to infer the underlying physical parameters (\textbf{x}) from observational data (\textbf{y}), intrinsic degeneracies make the solution  ambiguous. During the forward process, which translates the parameters into observations, inevitable information loss occurs so that different physical systems are mapped onto identical observations. By introducing the latent variables (\textbf{z}) which capture the information loss, a cINN learns the bijective forward mapping between \textbf{x} and \textbf{z} conditioned on \textbf{y}, $\textbf{z} = f(\textbf{x}; \textbf{y})$. The invertibility of the cINN architecture automatically provides the inverse mapping  $\textbf{x} = f^{-1}(\textbf{z}; \textbf{y})$. Once the network is trained through the forward process, we can produce the posterior distributions of \textbf{x} conditioned on \textbf{y} ($p(\textbf{x}|\textbf{y})$) by sampling the latent variables.

As it is difficult to collect the enormous amount of data required for network training, we used a database of synthetic \hii\ region models produced by the WARPFIELD-EMP pipeline~\citep{Pellegrini+20}. WARPFIELD-EMP evolves an isolated massive star-forming cloud using the 1D stellar feedback modelling code WARPFIELD~\citep{Rahner+17} and calculates several observable quantities for the evolving \hii\ region, such as the luminosities of various emission lines, by processing the WARPFIELD output with CLOUDY~\citep{Ferland+17} and POLARIS~\citep{Reissl+16}. The first WARPFIELD-EMP database introduced in \cite{Pellegrini+20} successfully mimicked the BPT diagram of \hii\ regions observed in NGC628~\citep{Rousseau-Nepton+18} but the number of models in that database and the sampling interval of each parameter was not enough to train a network. In this paper, we introduced a new, extended database that consists of 505,748 \hii\ region models evolved from  10,000 randomly sampled initial clouds and used this new database to train and evaluate the network.

The network introduced in this paper is originally designed for the SDSS-V LVM survey, but it can easily be adapted to data from other instruments or telescopes. Using the luminosity of 12 optical emission lines (Table~\ref{table:emission lines}) observable within the wavelength coverage of LVM, our network predicts seven physical parameters of the \hii\ region: initial mass of the star-forming cloud \mcl, star formation efficiency, initial cloud density \hdenini, age of the cloud $t$ which means the age of the first generation stars, age of the youngest cluster (i.e., age of the youngest generation of stars in the system) $t_{\rm youngest}$, number of clusters (i.e., distinct stellar populations) \ncl, and evolutionary phase of the cloud (see also Table~\ref{table:parameters}). We trained the network using 80\%\ of the database and used the remaining 20\%\ to evaluate the trained network with various methods. 
We validated the network performance with WARPFIELD-EMP synthetic models, focusing on the learning ability of the cINN architecture. An application of our newly developed tool to the analysis of real observational data of \hii\ regions from various large-scale surveys will be presented in follow-up studies. 
Our main results of testing the network performance are the following:

\begin{enumerate}
    \item The trained network is able to predict the posterior distribution very fast and efficiently. On average, the cINN can predict posterior distributions for 170 observations per second, sampling each posterior 4096 times with an NVIDIA GeForce RTX 2080 Ti graphic card. With the same graphic card, training of the network takes only a few hours (2--6 hours in most cases) depending on the adopted hyperparameters of the network configuration.

    \item Our network predicts each physical parameter very accurately and precisely. The posteriors commonly show a clear, unimodal distribution with narrow width around the true value. We evaluated the overall performance using three different methods (median calibration error, uncertainty at 68\%\ confidence interval, and the RMSE between the MAP estimate and the true value) and confirmed that our network typically predicts the physical parameters of \hii\ regions very well and accurately. The most difficult properties to determine precisely are the number of clusters and the age of the cloud.

    \item In some cases, the posteriors are degenerate, showing a multimodal posterior distribution. This degeneracy worsens the performance in terms of parameter prediction. However, degenerate posteriors are not incorrect or wrong, instead, they are physically valid alternatives that satisfy the same observational constraints. The network understands the hidden rules in the training data and suggests physically reasonable possibilities. Distinguishing between these possibilities may require additional data currently not considered in our network (e.g.\ broad-band optical or infrared luminosities).

    \item The main source of degeneracy that occurs in our network is caused by multiple star clusters (stellar generations) in a cloud leading to similar emission properties. If the posteriors of \ncl\ are degenerate and exhibit multiple peaks, the other parameters are prone to have degenerate posterior distributions as well. In particular, the cloud age is highly sensitive to variations in \ncl. We confirm that more than 90\%\ of the test models have non-degenerate posterior distributions for other parameters if the \ncl\ prediction is not degenerate.

    \item The performance of the network varies with the characteristics of the observed target. As mentioned above, for clusters with multiple stellar populations (\ncl$ >1$) it is more difficult to get the right value of \ncl\ and the correct cloud age $t$. Also, our network performs better for young \hii\ regions than for old \hii\ regions. This is because multicluster or old \hii\ regions are more likely to have degenerate posterior distributions. Even if the posterior provides an accurate estimate of the cloud age, the posterior distribution is wider if the estimated age is older. When the posterior distribution of the cloud age is multimodal, the younger mode is usually narrower than the older mode. So the MAP estimate is frequently dominated by the peak of the younger mode even if the fraction of posteriors in the older mode is higher. 
    The poorer performance for multicluster \hii\ regions or old \hii\ regions can in part be attributed to the biased parameter distributions in our training data.

    \item We validated the network performance by comparing the luminosity of the predicted posteriors with the corresponding true luminosity. To get this luminosity information, we selected 100 test models, sampled the posterior 100 times per model, and re-run WARPFIELD-EMP for the corresponding 10,000 posterior samples. This final test confirmed that the re-simulated luminosity of the posterior samples is close to the true luminosity with an average offset less than $3\times10^{-3}$dex and scatter of 0.11~dex.
   
\end{enumerate}

The initial evaluation of the network did not take into account the non-negligible uncertainties present in observations of real \hii\ regions. Therefore, in the latter part of the paper, we introduced a Monte Carlo-based method of sampling the posterior distributions that takes luminosity errors into account, and tested the performance of the network for a wide range signal-to-noise levels, i.e., for different luminosity errors. For this, we added a simple Poisson noise model to the original training data generated from WARPFIELD-EMP. And we normalize our approach to the adopted uncertainty in the brightest emission line, which is typically \ha\ in real data. The influence of observational errors on posterior distributions can be summarized as followed.

\begin{enumerate}
    \item[(vii)] The width of the posterior distribution gradually increases as a function of the luminosity error. However, the MAP estimates remain accurate until the error of the brightest emission line (i.e., minimum luminosity error, $\sigma_{\mathrm{min}}$) increases up to 0.1\%. 
    
    \item[(viii)] If the minimum luminosity error is 1\%\ or more, the posterior distributions show a skewed shape with a MAP estimate smaller than the true value and a long tail stretched to larger values than the true value. The skewness of the posterior distribution becomes more pronounced with increasing luminosity error. 
    
    \item[(ix)] If the luminosity error is too large ($\sigma_{\mathrm{min}}\sim10$\%), the network provides similar posterior distributions regardless of the input observations. Specifically, the posterior distributions for five non-discretized parameters (\mcl, SFE, \hdenini, $t$, and $t_{\rm youngest}$) are significantly skewed and as wide as the entire parameter ranges in the training data. For discretized two parameters (\ncl\ and phase), the MAP estimates are determined as the same value with \ncl\ as 1 and Phase as 1 in most cases.
\end{enumerate}

Overall the results of this study demonstrate that the cINN is a time-efficient and powerful tool to predict the fundamental parameters from the observations. We confirmed that with enough training data, the cINN learns the hidden rules within the training data well and can provide accurate predictions of the physical parameters of \hii\ regions from observations of key diagnostic emission lines.

\section*{Acknowledgements}
This work is supported by the Deutsche Forschungsgemeinschaft (DFG, German Research Foundation) under Germany’s Excellence Strategy EXC-2181/1 - 390900948 (the Heidelberg STRUCTURES Cluster of Excellence). We also acknowledge financial support from the European Research Council (ERC) via the ERC Synergy Grant "ECOGAL: Understanding our Galactic ecosystem -- From the disk of the Milky Way to the formation sites of stars and planets" (grant 855130), and we thank for financial support from DFG via the collaborative research center (SFB 881, Project-ID 138713538) ”The Milky Way System” (subprojects A1, B1, B2, and B8). The group makes use of computing resources provided by the state of Baden-W\"{u}rttemberg through bwHPC and the German Research Foundation (DFG) through grant INST 35/1134-1 FUGG. Data are stored at SDS@hd supported by the Ministry of Science, Research and the Arts Baden-Württemberg (MWK) and DFG through grant INST 35/1314-1 FUGG. We also acknowledge computing time provided to the group by the Leibniz Computing Center of the Bavarian Academy of Sciences and Humanities via project pr74nu. 


\section*{Data Availability}
The code FrEIA used in this article as the framework of cINNs is available at \href{url}{https://github.com/VLL-HD/FrEIA}.

The database of WARPFIELD-EMP models and the code WARPFIELD-EMP used in this article will be shared on reasonable request to the corresponding author with the permission of Eric Pellegrini.

The data outcomes underlying this article will be shared on reasonable request to the corresponding author.




\bibliographystyle{mnras}




\appendix

\section{Supplemental materials}
\begin{figure*}
    \includegraphics[width=2\columnwidth]{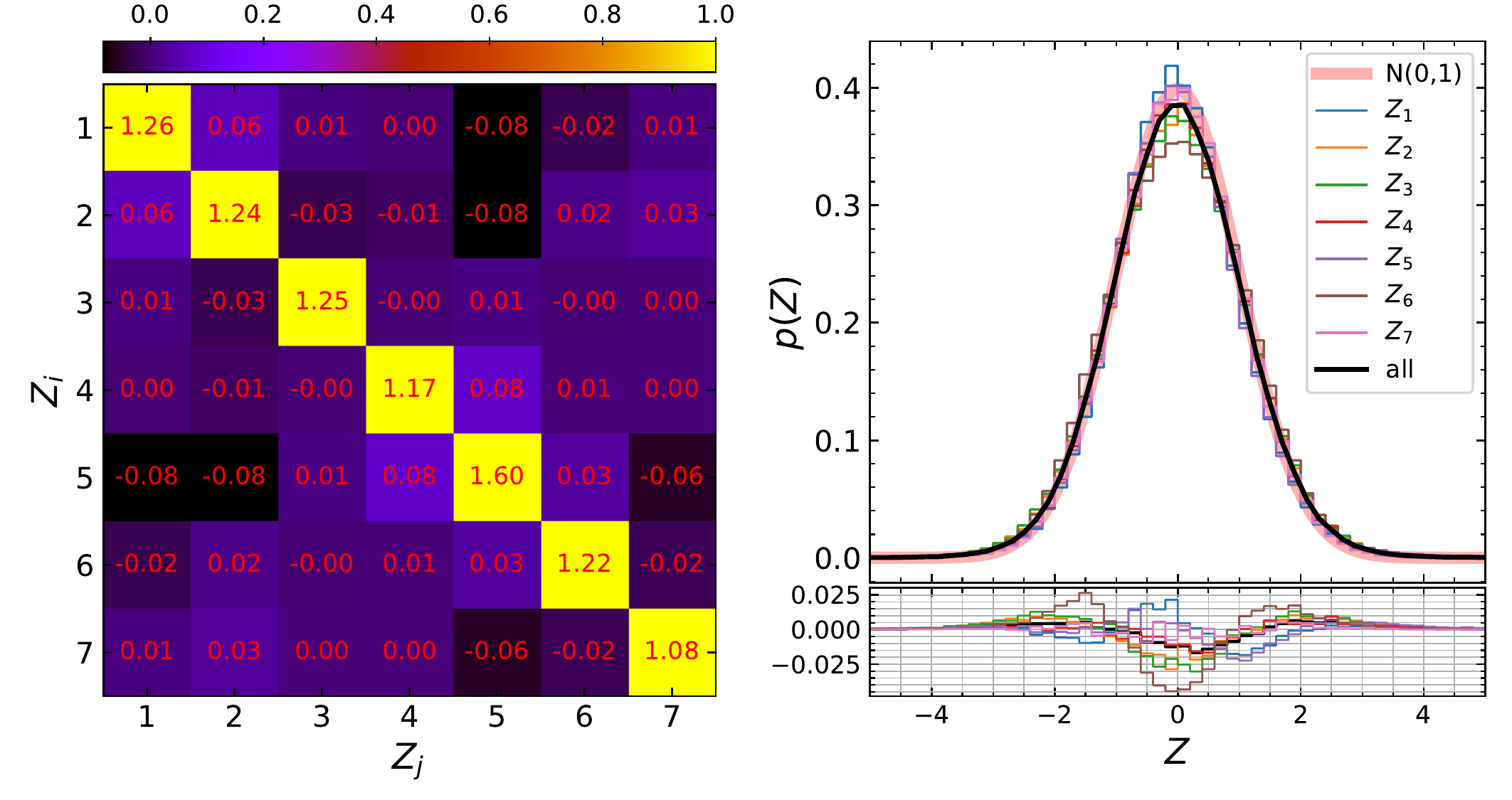}
    \caption{The covariance matrix of the latent variables (left) and distributions of each latent variable (right) as evaluated on 101,149 test set models. In the right panel, the black line shows the distribution of all seven latent variables in one line and the red line indicates the standard normal distribution. The small panel below shows residuals between distributions of latent variables and the standard normal distribution. If the network is well-trained, the covariance matrix is close to the unit matrix and the distribution of latent variable follows the standard normal distribution.}
    \label{fig:z_cov_pdf}
\end{figure*}

The following are supplementary figures of training evaluation of our main network (Section~\ref{subsec:training evaluation}) and the performance of Network 2 without noise augmentation (Section~\ref{subsec:smoothing}).
Figure~\ref{fig:z_cov_pdf} shows the covariance matrix and the probability distributions of latent variables, which are obtained from the forward process of the network for the entire test set ($\mathbf{Z_{test}}$). As mentioned in Sections~\ref{subsec:evaluation methods} and \ref{subsec:training evaluation}, latent variables should follow the prescribed Gaussian normal distribution and their covariance matrix should be close to a unit matrix if the network is well-trained. Though the covariance matrix and probability distributions are not perfectly equal to the desired results, residuals are small enough to conclude that the network converged to a good solution.

\begin{figure*}
	\includegraphics[width=2\columnwidth]{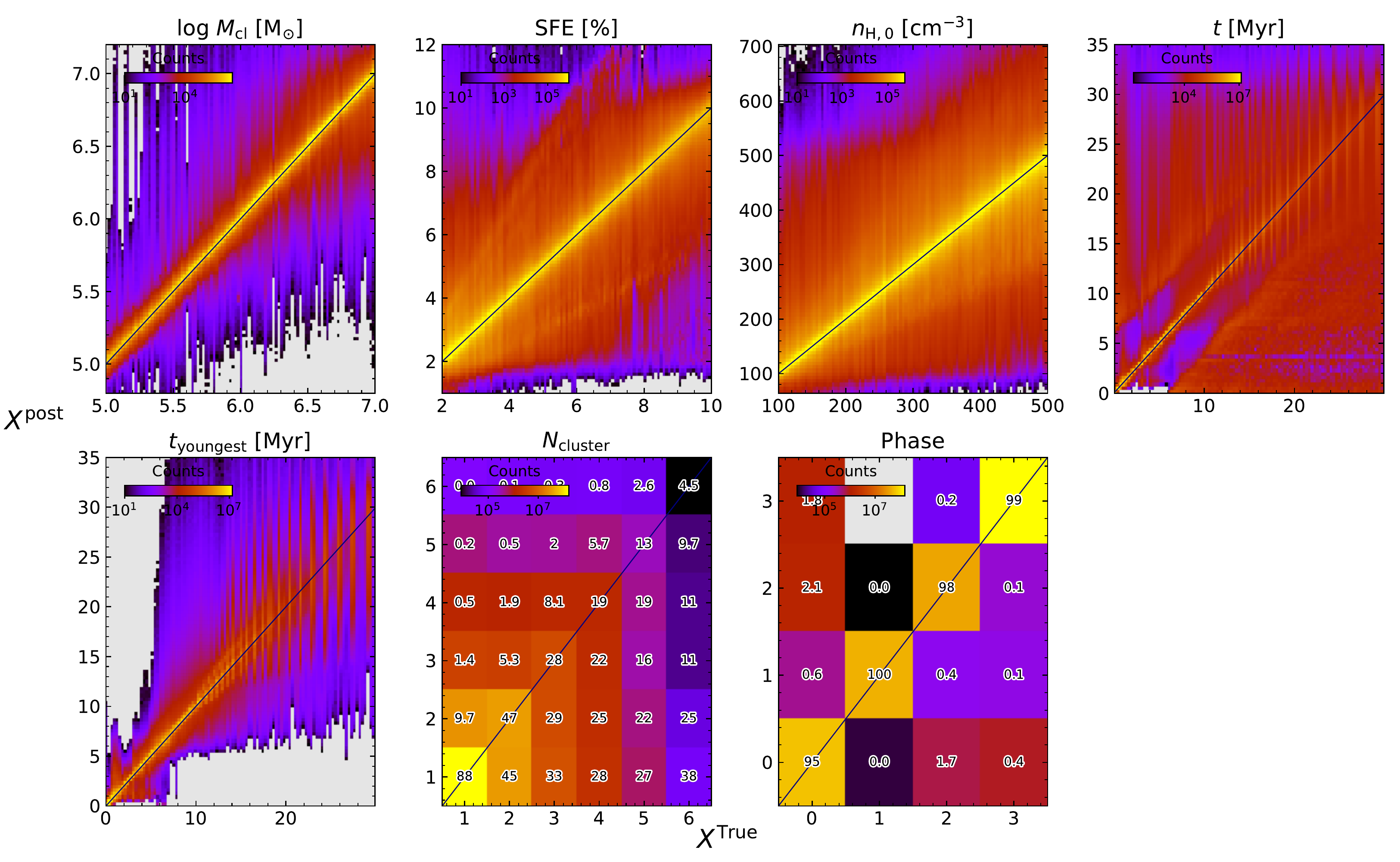}
    \caption{Comparison of true parameter values and all posteriors obtained from the cINN model trained without noise augmentation (Network 2), using the entire 101,149 test models. Figure~\ref{fig:all_post} has a similar but better result which is obtained from our main cINN model, trained with noise augmentation. Colour code indicates the counts of the 2D histogram.} 
    \label{fig:all_post_N2}
\end{figure*}

\begin{figure*}
	\includegraphics[width=2\columnwidth]{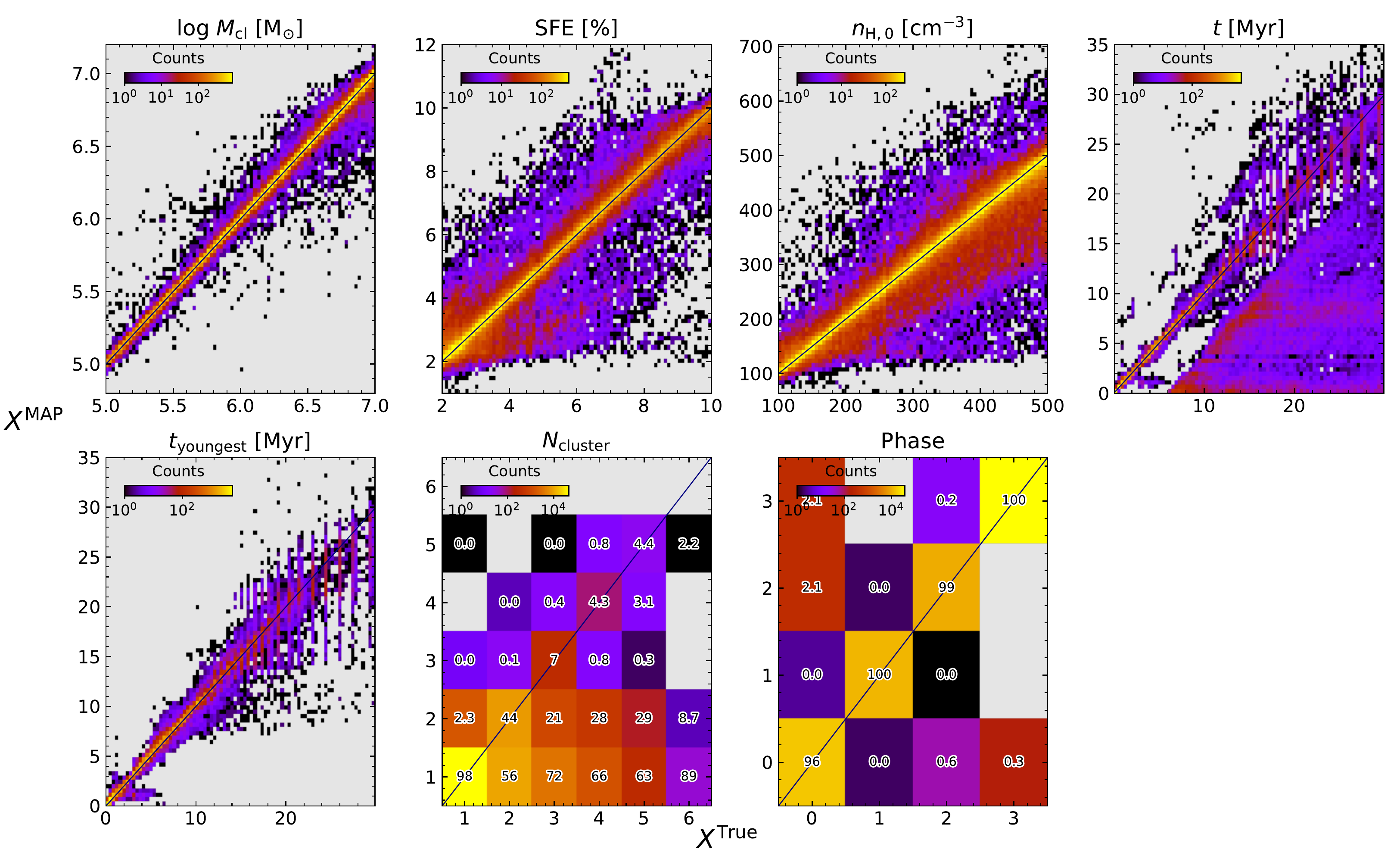}
    \caption{Comparison of true parameter values and MAP estimates obtained from the cINN model trained without noise augmentation using all 101,149 test models. Figure~\ref{fig:all_map} shows the similar result obtained from our main cINN model. Colour code is the same as in Figure~\ref{fig:all_map}.} 
    \label{fig:all_map_N2}
\end{figure*}

Figures~\ref{fig:all_post_N2} and \ref{fig:all_map_N2} show the prediction performance of Network 2 introduced in Section~\ref{subsec:smoothing}. We did not augment the distributions of \ncl\ and phase with artificial noise in training Network 2. After drawing 4096 posterior samples for each model in the entire test set, we compare the true values of the models with all posterior estimates or with the MAP estimates as the representative in Figure~\ref{fig:all_post_N2} and Figure~\ref{fig:all_map_N2}, respectively. As already discussed in the previous section, adding artificial noise to smooth out the discretized parameter distribution improves the overall prediction performance of the network.

\section{Fitting and determining peaks of the posterior distribution}
\label{appendix-B}

\begin{figure*}
	\includegraphics[width=2\columnwidth]{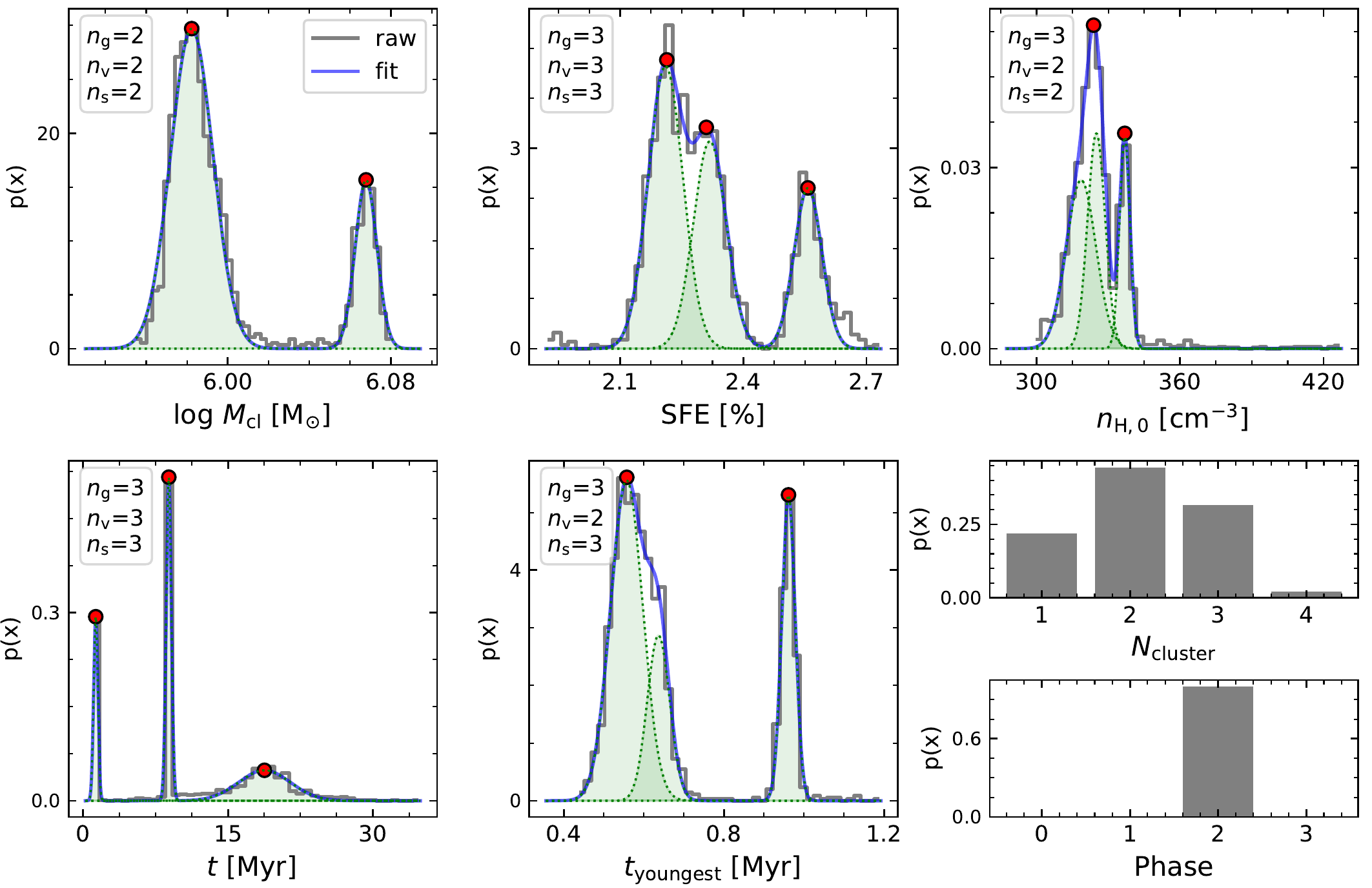}
    \caption{An example of fitting posterior distributions for five parameters. The posterior distribution (grey histogram) is fitted with multiple Gaussian functions. The blue solid line represents the fit result and the green shades with dotted edges represent individual Gaussian components of the fitting curve. We do not fit for \ncl\ and phase.    
    The number of modes in the posterior distribution measured according to the three different definitions of the peak of the mode is listed on the upper left side of each panel: the number of Gaussian components used for the fitting ($n_{\text{g}}$), the number of visible peaks ($n_{\text{v}}$), and the number of separated peaks ($n_{\text{s}}$). The visible peaks of each posterior distribution are denoted by red circles.
    } 
    \label{fig:fit_posterior}
\end{figure*}

In this appendix, we provide the example of the posterior fitting result and the histogram of the number of peaks using different definitions of the peak. We present in Figure~\ref{fig:fit_posterior} the results of fitting the posterior distribution of one model in the test set. As mentioned in Section~\ref{sec:degenerate}, we fit the distribution using up to 6 Gaussian functions, and the number of Gaussian functions used varies even in one observation. In addition to the final fitting curve (blue solid line), individual Gaussian components used in the fitting are shown with green dotted lines. 
Considering the posterior distribution as a multimodal distribution, we provide the number of peaks according to the three different definitions of the peak of the mode. In Figure~\ref{fig:fit_posterior}, the number of Gaussian components ($n_{\text{g}}$), the number of visible peaks (i.e., the number of red circles, $n_{\text{v}}$), and the number of separated peaks ($n_{\text{s}}$) are all the same for \mcl, star formation efficiency, and the cloud age. The case of \hdenini\ shows that even if the distribution is fitted with two Gaussian components, it is considered as one visible peak or one separated peak. On the other hand, the case of the youngest cluster age is where two Gaussian components are considered as one visible peak but two separated peaks depending on the distance between two Gaussian components.

\begin{figure*}
	\includegraphics[width=1.9\columnwidth]{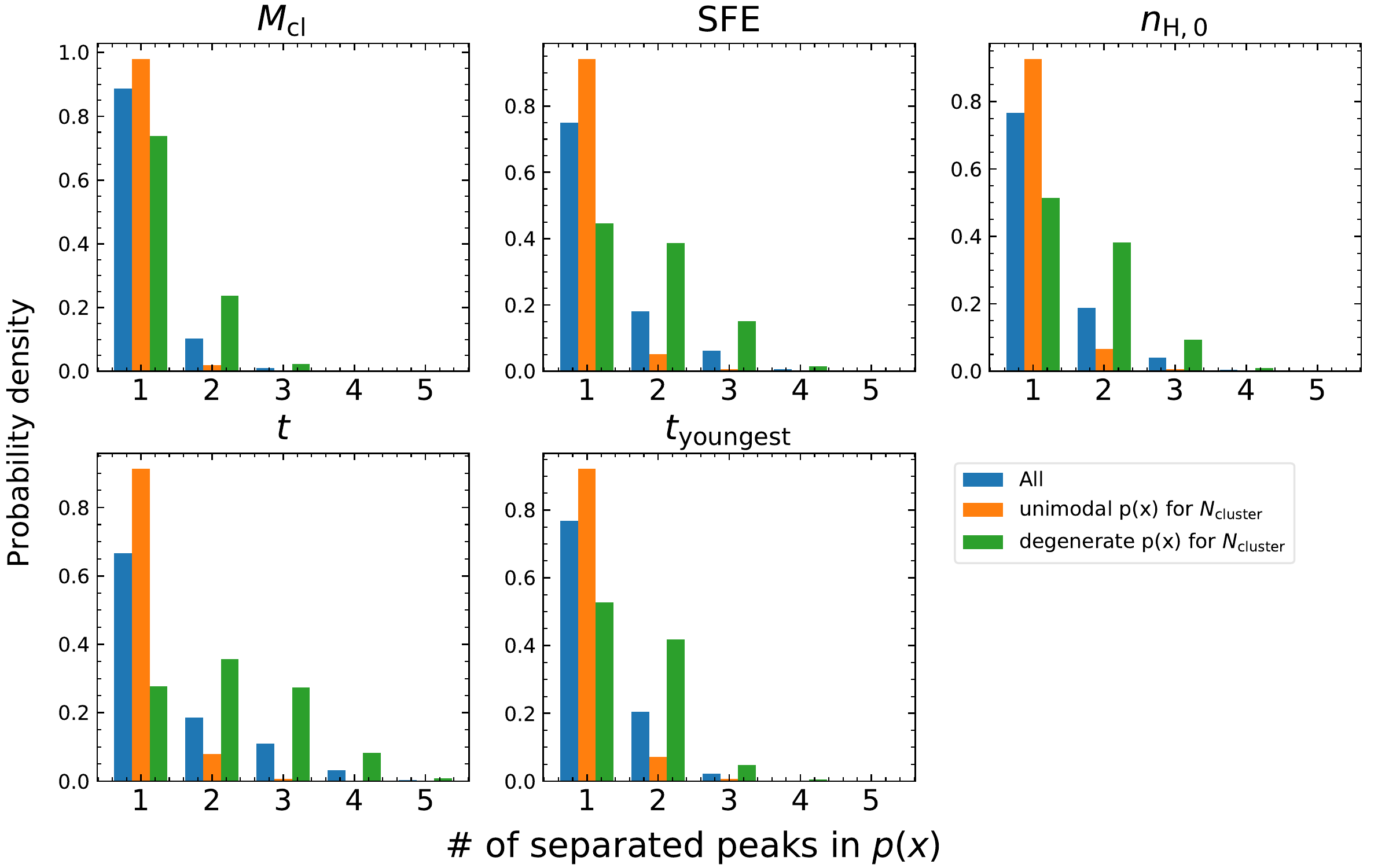}
    \caption{Density histograms of the number of separated peaks in the posterior distributions for five parameters. The colour codes are the same as in Figure~\ref{fig:vpeak_hist}: results of all models in the test set (blue), results of models whose \ncl\ posterior distributions are not degenerate (orange), and results of models whose \ncl\ posterior distributions are degenerate (green).} 
    \label{fig:speak_hist}
\end{figure*}

\begin{figure*}
	\includegraphics[width=1.9\columnwidth]{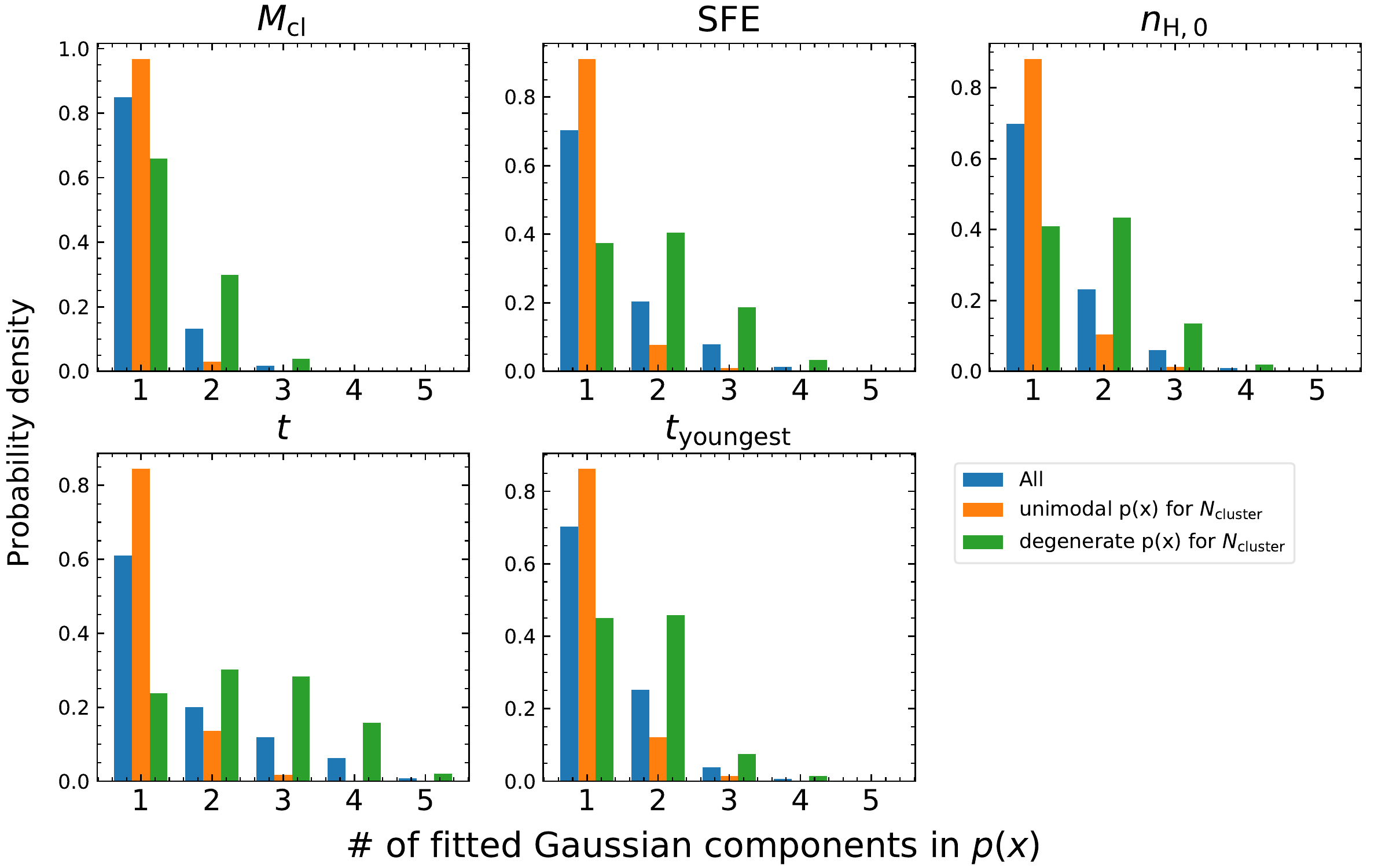}
    \caption{Density histograms of the number of Gaussian components used in the fit of posterior distributions for five parameters. The colour codes are the same as in Figure~\ref{fig:vpeak_hist} and Figure~\ref{fig:speak_hist}.
    } 
    \label{fig:gpeak_hist}
\end{figure*}

In Figures~\ref{fig:speak_hist} and \ref{fig:gpeak_hist}, we present the histogram of the number of peaks using the separated peaks and the number of Gaussian components, respectively. We exclude the result of phase because the number of peaks in the phase posterior distributions does not depend on the definition of the peak so that it is the same as shown in Figure~\ref{fig:vpeak_hist}. These results differ from those obtained using the visible peaks in detail, but the overall trends are similar to Figure~\ref{fig:vpeak_hist}, including the influence of the degenerate \ncl\ prediction on the degeneracy in the posterior distribution of other parameters.

\bsp	
\label{lastpage}
\end{document}